\documentclass[longauth]{aaEC}
\usepackage{amsmath,amssymb,amsfonts}
\usepackage{graphicx}
\usepackage{txfonts}
\usepackage{enumerate} % advanced enumerate environment
\usepackage{colordvi} % for color texthttps://fr.overleaf.com/project/5f92c102fc740c0001d90f85
\usepackage{bm}% bold math
\usepackage{epsfig}
\usepackage{hyperref}
\usepackage[dvipsnames]{xcolor}
\usepackage{color, colortbl}
\usepackage[T1]{fontenc} % if needed
\usepackage{braket}
\usepackage{euclid}
\usepackage{placeins}
\usepackage{multirow}
\usepackage{tabularx}
\usepackage{comment}
\usepackage[switch, mathlines]{lineno}
\usepackage[normalem]{ulem}
\graphicspath{{./Figures/}}

\newcommand\mathcomma{\,,}
\newcommand\mathperiod{\,.}

\newcommand{\orcid}[1]{}
\newcommand{\bfkk}{\mbox{\boldmath$k$}}

\definecolor{amaranth}{rgb}{0.9, 0.17, 0.31}
\definecolor{forestgreen(web)}{rgb}{0.13, 0.55, 0.13}
\definecolor{lavender(web)}{rgb}{0.9, 0.9, 0.98}
\definecolor{cosmiclatte}{rgb}{1.0, 0.97, 0.91}
\definecolor{jonquil}{rgb}{0.98, 0.85, 0.37}
\definecolor{khaki(x11)(lightkhaki)}{rgb}{0.94, 0.9, 0.55}
\definecolor{thistle}{rgb}{0.85, 0.75, 0.85}

\newcommand{\GCsp}{\text{GC}\ensuremath{_\mathrm{sp}}}
\newcommand{\GCph}{\text{GC}\ensuremath{_\mathrm{ph}}}
\newcommand{\Omegam}{\ensuremath{\Omega_{\mathrm{m},0}}}

\newcommand{\lcdm}{\ensuremath{\Lambda\mathrm{CDM}}}

\newcommand{\de}{\mathrm{d}}

\defcitealias{Blanchard:2019oqi}{EC20}
\defcitealias{Blanchard-EP7}{EC20}

\usepackage[nameinlink,noabbrev]{cleveref}
\Crefname{section}{Section}{Sections}
\crefname{section}{Sect.}{Sects.}
\Crefname{figure}{Figure}{Figures}
\crefname{figure}{Fig.}{Figs.}
\Crefname{equation}{Equation}{Equations}
\crefname{equation}{Eq.}{Eqs.}

\makeatletter
\renewcommand*\aa@pageof{, page \thepage{} of \pageref*{LastPage}}
\makeatother

\begin{document}
\title{\Euclid preparation}
\subtitle{Constraining parameterised models of modifications of gravity with the spectroscopic and photometric primary probes} 

\author{Euclid Collaboration: I.~S.~Albuquerque\orcid{0000-0002-5369-1226}\inst{\ref{aff1}}
\and N.~Frusciante\orcid{0000-0002-7375-1230}\inst{\ref{aff2}}
\and Z.~Sakr\orcid{0000-0002-4823-3757}\thanks{\email{zsakr@irap.omp.eu}}\inst{\ref{aff3},\ref{aff4},\ref{aff5}}
\and S.~Srinivasan\orcid{0000-0003-1539-3276}\inst{\ref{aff6}}
\and L.~Atayde\orcid{0000-0001-6373-9193}\inst{\ref{aff1}}
\and B.~Bose\orcid{0000-0003-1965-8614}\inst{\ref{aff7}}
\and V.~F.~Cardone\inst{\ref{aff8},\ref{aff9}}
\and S.~Casas\orcid{0000-0002-4751-5138}\inst{\ref{aff10},\ref{aff11}}
\and M.~Martinelli\orcid{0000-0002-6943-7732}\inst{\ref{aff8},\ref{aff9}}
\and J.~Noller\orcid{0000-0003-2210-775X}\inst{\ref{aff12},\ref{aff11}}
\and E.~M.~Teixeira\orcid{0000-0001-7417-0780}\inst{\ref{aff13}}
\and D.~B.~Thomas\inst{\ref{aff14}}
\and I.~Tutusaus\orcid{0000-0002-3199-0399}\inst{\ref{aff4}}
\and M.~Cataneo\orcid{0000-0002-7992-0656}\inst{\ref{aff15},\ref{aff16}}
\and K.~Koyama\orcid{0000-0001-6727-6915}\inst{\ref{aff11}}
\and L.~Lombriser\inst{\ref{aff17}}
\and F.~Pace\orcid{0000-0001-8039-0480}\inst{\ref{aff18},\ref{aff19},\ref{aff20}}
\and A.~Silvestri\orcid{0000-0001-6904-5061}\inst{\ref{aff21}}
\and N.~Aghanim\orcid{0000-0002-6688-8992}\inst{\ref{aff22}}
\and A.~Amara\inst{\ref{aff23}}
\and S.~Andreon\orcid{0000-0002-2041-8784}\inst{\ref{aff24}}
\and N.~Auricchio\orcid{0000-0003-4444-8651}\inst{\ref{aff25}}
\and C.~Baccigalupi\orcid{0000-0002-8211-1630}\inst{\ref{aff26},\ref{aff27},\ref{aff28},\ref{aff29}}
\and M.~Baldi\orcid{0000-0003-4145-1943}\inst{\ref{aff30},\ref{aff25},\ref{aff31}}
\and S.~Bardelli\orcid{0000-0002-8900-0298}\inst{\ref{aff25}}
\and A.~Biviano\orcid{0000-0002-0857-0732}\inst{\ref{aff27},\ref{aff26}}
\and D.~Bonino\orcid{0000-0002-3336-9977}\inst{\ref{aff20}}
\and E.~Branchini\orcid{0000-0002-0808-6908}\inst{\ref{aff32},\ref{aff33},\ref{aff24}}
\and M.~Brescia\orcid{0000-0001-9506-5680}\inst{\ref{aff34},\ref{aff35},\ref{aff36}}
\and J.~Brinchmann\orcid{0000-0003-4359-8797}\inst{\ref{aff37},\ref{aff38}}
\and S.~Camera\orcid{0000-0003-3399-3574}\inst{\ref{aff18},\ref{aff19},\ref{aff20}}
\and G.~Ca\~nas-Herrera\orcid{0000-0003-2796-2149}\inst{\ref{aff39},\ref{aff21}}
\and V.~Capobianco\orcid{0000-0002-3309-7692}\inst{\ref{aff20}}
\and C.~Carbone\orcid{0000-0003-0125-3563}\inst{\ref{aff40}}
\and J.~Carretero\orcid{0000-0002-3130-0204}\inst{\ref{aff41},\ref{aff42}}
\and M.~Castellano\orcid{0000-0001-9875-8263}\inst{\ref{aff8}}
\and G.~Castignani\orcid{0000-0001-6831-0687}\inst{\ref{aff25}}
\and S.~Cavuoti\orcid{0000-0002-3787-4196}\inst{\ref{aff35},\ref{aff36}}
\and K.~C.~Chambers\orcid{0000-0001-6965-7789}\inst{\ref{aff43}}
\and A.~Cimatti\inst{\ref{aff44}}
\and C.~Colodro-Conde\inst{\ref{aff45}}
\and G.~Congedo\orcid{0000-0003-2508-0046}\inst{\ref{aff7}}
\and C.~J.~Conselice\orcid{0000-0003-1949-7638}\inst{\ref{aff14}}
\and L.~Conversi\orcid{0000-0002-6710-8476}\inst{\ref{aff46},\ref{aff47}}
\and Y.~Copin\orcid{0000-0002-5317-7518}\inst{\ref{aff48}}
\and L.~Corcione\orcid{0000-0002-6497-5881}\inst{\ref{aff20}}
\and F.~Courbin\orcid{0000-0003-0758-6510}\inst{\ref{aff49},\ref{aff50},\ref{aff51}}
\and H.~M.~Courtois\orcid{0000-0003-0509-1776}\inst{\ref{aff52}}
\and A.~Da~Silva\orcid{0000-0002-6385-1609}\inst{\ref{aff53},\ref{aff1}}
\and H.~Degaudenzi\orcid{0000-0002-5887-6799}\inst{\ref{aff54}}
\and S.~de~la~Torre\inst{\ref{aff55}}
\and G.~De~Lucia\orcid{0000-0002-6220-9104}\inst{\ref{aff27}}
\and A.~M.~Di~Giorgio\orcid{0000-0002-4767-2360}\inst{\ref{aff56}}
\and H.~Dole\orcid{0000-0002-9767-3839}\inst{\ref{aff22}}
\and F.~Dubath\orcid{0000-0002-6533-2810}\inst{\ref{aff54}}
\and C.~A.~J.~Duncan\orcid{0009-0003-3573-0791}\inst{\ref{aff14}}
\and X.~Dupac\inst{\ref{aff47}}
\and S.~Dusini\orcid{0000-0002-1128-0664}\inst{\ref{aff57}}
\and A.~Ealet\orcid{0000-0003-3070-014X}\inst{\ref{aff48}}
\and S.~Escoffier\orcid{0000-0002-2847-7498}\inst{\ref{aff58}}
\and M.~Farina\orcid{0000-0002-3089-7846}\inst{\ref{aff56}}
\and S.~Farrens\orcid{0000-0002-9594-9387}\inst{\ref{aff59}}
\and F.~Faustini\orcid{0000-0001-6274-5145}\inst{\ref{aff60},\ref{aff8}}
\and S.~Ferriol\inst{\ref{aff48}}
\and F.~Finelli\orcid{0000-0002-6694-3269}\inst{\ref{aff25},\ref{aff61}}
\and P.~Fosalba\orcid{0000-0002-1510-5214}\inst{\ref{aff62},\ref{aff63}}
\and S.~Fotopoulou\orcid{0000-0002-9686-254X}\inst{\ref{aff64}}
\and M.~Frailis\orcid{0000-0002-7400-2135}\inst{\ref{aff27}}
\and E.~Franceschi\orcid{0000-0002-0585-6591}\inst{\ref{aff25}}
\and M.~Fumana\orcid{0000-0001-6787-5950}\inst{\ref{aff40}}
\and S.~Galeotta\orcid{0000-0002-3748-5115}\inst{\ref{aff27}}
\and B.~Gillis\orcid{0000-0002-4478-1270}\inst{\ref{aff7}}
\and C.~Giocoli\orcid{0000-0002-9590-7961}\inst{\ref{aff25},\ref{aff31}}
\and J.~Gracia-Carpio\inst{\ref{aff65}}
\and A.~Grazian\orcid{0000-0002-5688-0663}\inst{\ref{aff66}}
\and F.~Grupp\inst{\ref{aff65},\ref{aff6}}
\and L.~Guzzo\orcid{0000-0001-8264-5192}\inst{\ref{aff67},\ref{aff24},\ref{aff68}}
\and S.~V.~H.~Haugan\orcid{0000-0001-9648-7260}\inst{\ref{aff69}}
\and W.~Holmes\inst{\ref{aff70}}
\and F.~Hormuth\inst{\ref{aff71}}
\and A.~Hornstrup\orcid{0000-0002-3363-0936}\inst{\ref{aff72},\ref{aff73}}
\and P.~Hudelot\inst{\ref{aff74}}
\and S.~Ili\'c\orcid{0000-0003-4285-9086}\inst{\ref{aff75},\ref{aff4}}
\and K.~Jahnke\orcid{0000-0003-3804-2137}\inst{\ref{aff76}}
\and M.~Jhabvala\inst{\ref{aff77}}
\and B.~Joachimi\orcid{0000-0001-7494-1303}\inst{\ref{aff12}}
\and E.~Keih\"anen\orcid{0000-0003-1804-7715}\inst{\ref{aff78}}
\and S.~Kermiche\orcid{0000-0002-0302-5735}\inst{\ref{aff58}}
\and A.~Kiessling\orcid{0000-0002-2590-1273}\inst{\ref{aff70}}
\and M.~Kilbinger\orcid{0000-0001-9513-7138}\inst{\ref{aff59}}
\and B.~Kubik\orcid{0009-0006-5823-4880}\inst{\ref{aff48}}
\and M.~Kunz\orcid{0000-0002-3052-7394}\inst{\ref{aff17}}
\and H.~Kurki-Suonio\orcid{0000-0002-4618-3063}\inst{\ref{aff79},\ref{aff80}}
\and A.~M.~C.~Le~Brun\orcid{0000-0002-0936-4594}\inst{\ref{aff81}}
\and S.~Ligori\orcid{0000-0003-4172-4606}\inst{\ref{aff20}}
\and P.~B.~Lilje\orcid{0000-0003-4324-7794}\inst{\ref{aff69}}
\and V.~Lindholm\orcid{0000-0003-2317-5471}\inst{\ref{aff79},\ref{aff80}}
\and I.~Lloro\orcid{0000-0001-5966-1434}\inst{\ref{aff82}}
\and G.~Mainetti\orcid{0000-0003-2384-2377}\inst{\ref{aff83}}
\and D.~Maino\inst{\ref{aff67},\ref{aff40},\ref{aff68}}
\and E.~Maiorano\orcid{0000-0003-2593-4355}\inst{\ref{aff25}}
\and O.~Mansutti\orcid{0000-0001-5758-4658}\inst{\ref{aff27}}
\and O.~Marggraf\orcid{0000-0001-7242-3852}\inst{\ref{aff16}}
\and K.~Markovic\orcid{0000-0001-6764-073X}\inst{\ref{aff70}}
\and N.~Martinet\orcid{0000-0003-2786-7790}\inst{\ref{aff55}}
\and F.~Marulli\orcid{0000-0002-8850-0303}\inst{\ref{aff84},\ref{aff25},\ref{aff31}}
\and R.~Massey\orcid{0000-0002-6085-3780}\inst{\ref{aff85}}
\and E.~Medinaceli\orcid{0000-0002-4040-7783}\inst{\ref{aff25}}
\and S.~Mei\orcid{0000-0002-2849-559X}\inst{\ref{aff86}}
\and Y.~Mellier\inst{\ref{aff87},\ref{aff74}}
\and M.~Meneghetti\orcid{0000-0003-1225-7084}\inst{\ref{aff25},\ref{aff31}}
\and E.~Merlin\orcid{0000-0001-6870-8900}\inst{\ref{aff8}}
\and G.~Meylan\inst{\ref{aff49}}
\and A.~Mora\orcid{0000-0002-1922-8529}\inst{\ref{aff88}}
\and M.~Moresco\orcid{0000-0002-7616-7136}\inst{\ref{aff84},\ref{aff25}}
\and L.~Moscardini\orcid{0000-0002-3473-6716}\inst{\ref{aff84},\ref{aff25},\ref{aff31}}
\and R.~C.~Nichol\orcid{0000-0003-0939-6518}\inst{\ref{aff23}}
\and S.-M.~Niemi\inst{\ref{aff39}}
\and J.~W.~Nightingale\orcid{0000-0002-8987-7401}\inst{\ref{aff89}}
\and C.~Padilla\orcid{0000-0001-7951-0166}\inst{\ref{aff90}}
\and S.~Paltani\orcid{0000-0002-8108-9179}\inst{\ref{aff54}}
\and F.~Pasian\orcid{0000-0002-4869-3227}\inst{\ref{aff27}}
\and W.~J.~Percival\orcid{0000-0002-0644-5727}\inst{\ref{aff91},\ref{aff92},\ref{aff93}}
\and V.~Pettorino\inst{\ref{aff39}}
\and S.~Pires\orcid{0000-0002-0249-2104}\inst{\ref{aff59}}
\and G.~Polenta\orcid{0000-0003-4067-9196}\inst{\ref{aff60}}
\and M.~Poncet\inst{\ref{aff94}}
\and L.~A.~Popa\inst{\ref{aff95}}
\and L.~Pozzetti\orcid{0000-0001-7085-0412}\inst{\ref{aff25}}
\and F.~Raison\orcid{0000-0002-7819-6918}\inst{\ref{aff65}}
\and R.~Rebolo\inst{\ref{aff45},\ref{aff96},\ref{aff97}}
\and A.~Renzi\orcid{0000-0001-9856-1970}\inst{\ref{aff98},\ref{aff57}}
\and J.~Rhodes\orcid{0000-0002-4485-8549}\inst{\ref{aff70}}
\and G.~Riccio\inst{\ref{aff35}}
\and E.~Romelli\orcid{0000-0003-3069-9222}\inst{\ref{aff27}}
\and M.~Roncarelli\orcid{0000-0001-9587-7822}\inst{\ref{aff25}}
\and R.~Saglia\orcid{0000-0003-0378-7032}\inst{\ref{aff6},\ref{aff65}}
\and D.~Sapone\orcid{0000-0001-7089-4503}\inst{\ref{aff99}}
\and B.~Sartoris\orcid{0000-0003-1337-5269}\inst{\ref{aff6},\ref{aff27}}
\and J.~A.~Schewtschenko\orcid{0000-0002-4913-6393}\inst{\ref{aff7}}
\and M.~Schirmer\orcid{0000-0003-2568-9994}\inst{\ref{aff76}}
\and T.~Schrabback\orcid{0000-0002-6987-7834}\inst{\ref{aff100}}
\and A.~Secroun\orcid{0000-0003-0505-3710}\inst{\ref{aff58}}
\and E.~Sefusatti\orcid{0000-0003-0473-1567}\inst{\ref{aff27},\ref{aff26},\ref{aff28}}
\and G.~Seidel\orcid{0000-0003-2907-353X}\inst{\ref{aff76}}
\and S.~Serrano\orcid{0000-0002-0211-2861}\inst{\ref{aff62},\ref{aff101},\ref{aff63}}
\and C.~Sirignano\orcid{0000-0002-0995-7146}\inst{\ref{aff98},\ref{aff57}}
\and G.~Sirri\orcid{0000-0003-2626-2853}\inst{\ref{aff31}}
\and L.~Stanco\orcid{0000-0002-9706-5104}\inst{\ref{aff57}}
\and J.~Steinwagner\orcid{0000-0001-7443-1047}\inst{\ref{aff65}}
\and P.~Tallada-Cresp\'{i}\orcid{0000-0002-1336-8328}\inst{\ref{aff41},\ref{aff42}}
\and A.~N.~Taylor\inst{\ref{aff7}}
\and I.~Tereno\inst{\ref{aff53},\ref{aff102}}
\and N.~Tessore\orcid{0000-0002-9696-7931}\inst{\ref{aff12}}
\and S.~Toft\orcid{0000-0003-3631-7176}\inst{\ref{aff103},\ref{aff104}}
\and R.~Toledo-Moreo\orcid{0000-0002-2997-4859}\inst{\ref{aff105}}
\and F.~Torradeflot\orcid{0000-0003-1160-1517}\inst{\ref{aff42},\ref{aff41}}
\and E.~A.~Valentijn\inst{\ref{aff106}}
\and L.~Valenziano\orcid{0000-0002-1170-0104}\inst{\ref{aff25},\ref{aff61}}
\and J.~Valiviita\orcid{0000-0001-6225-3693}\inst{\ref{aff79},\ref{aff80}}
\and T.~Vassallo\orcid{0000-0001-6512-6358}\inst{\ref{aff6},\ref{aff27}}
\and G.~Verdoes~Kleijn\orcid{0000-0001-5803-2580}\inst{\ref{aff106}}
\and A.~Veropalumbo\orcid{0000-0003-2387-1194}\inst{\ref{aff24},\ref{aff33},\ref{aff107}}
\and Y.~Wang\orcid{0000-0002-4749-2984}\inst{\ref{aff108}}
\and J.~Weller\orcid{0000-0002-8282-2010}\inst{\ref{aff6},\ref{aff65}}
\and G.~Zamorani\orcid{0000-0002-2318-301X}\inst{\ref{aff25}}
\and E.~Zucca\orcid{0000-0002-5845-8132}\inst{\ref{aff25}}
\and E.~Bozzo\orcid{0000-0002-8201-1525}\inst{\ref{aff54}}
\and C.~Burigana\orcid{0000-0002-3005-5796}\inst{\ref{aff109},\ref{aff61}}
\and M.~Calabrese\orcid{0000-0002-2637-2422}\inst{\ref{aff110},\ref{aff40}}
\and D.~Di~Ferdinando\inst{\ref{aff31}}
\and J.~A.~Escartin~Vigo\inst{\ref{aff65}}
\and S.~Matthew\orcid{0000-0001-8448-1697}\inst{\ref{aff7}}
\and N.~Mauri\orcid{0000-0001-8196-1548}\inst{\ref{aff44},\ref{aff31}}
\and A.~Pezzotta\orcid{0000-0003-0726-2268}\inst{\ref{aff65}}
\and M.~P\"ontinen\orcid{0000-0001-5442-2530}\inst{\ref{aff79}}
\and C.~Porciani\orcid{0000-0002-7797-2508}\inst{\ref{aff16}}
\and V.~Scottez\inst{\ref{aff87},\ref{aff111}}
\and M.~Tenti\orcid{0000-0002-4254-5901}\inst{\ref{aff31}}
\and M.~Viel\orcid{0000-0002-2642-5707}\inst{\ref{aff26},\ref{aff27},\ref{aff29},\ref{aff28},\ref{aff112}}
\and M.~Wiesmann\orcid{0009-0000-8199-5860}\inst{\ref{aff69}}
\and Y.~Akrami\orcid{0000-0002-2407-7956}\inst{\ref{aff113},\ref{aff114}}
\and V.~Allevato\orcid{0000-0001-7232-5152}\inst{\ref{aff35}}
\and S.~Alvi\orcid{0000-0001-5779-8568}\inst{\ref{aff115}}
\and S.~Anselmi\orcid{0000-0002-3579-9583}\inst{\ref{aff57},\ref{aff98},\ref{aff81}}
\and M.~Archidiacono\orcid{0000-0003-4952-9012}\inst{\ref{aff67},\ref{aff68}}
\and F.~Atrio-Barandela\orcid{0000-0002-2130-2513}\inst{\ref{aff116}}
\and A.~Balaguera-Antolinez\orcid{0000-0001-5028-3035}\inst{\ref{aff45},\ref{aff96}}
\and M.~Ballardini\orcid{0000-0003-4481-3559}\inst{\ref{aff115},\ref{aff25},\ref{aff117}}
\and D.~Bertacca\orcid{0000-0002-2490-7139}\inst{\ref{aff98},\ref{aff66},\ref{aff57}}
\and A.~Blanchard\orcid{0000-0001-8555-9003}\inst{\ref{aff4}}
\and L.~Blot\orcid{0000-0002-9622-7167}\inst{\ref{aff118},\ref{aff81}}
\and S.~Borgani\orcid{0000-0001-6151-6439}\inst{\ref{aff119},\ref{aff26},\ref{aff27},\ref{aff28},\ref{aff112}}
\and M.~L.~Brown\orcid{0000-0002-0370-8077}\inst{\ref{aff14}}
\and S.~Bruton\orcid{0000-0002-6503-5218}\inst{\ref{aff120}}
\and R.~Cabanac\orcid{0000-0001-6679-2600}\inst{\ref{aff4}}
\and A.~Calabro\orcid{0000-0003-2536-1614}\inst{\ref{aff8}}
\and B.~Camacho~Quevedo\orcid{0000-0002-8789-4232}\inst{\ref{aff62},\ref{aff63}}
\and A.~Cappi\inst{\ref{aff25},\ref{aff121}}
\and F.~Caro\inst{\ref{aff8}}
\and C.~S.~Carvalho\inst{\ref{aff102}}
\and T.~Castro\orcid{0000-0002-6292-3228}\inst{\ref{aff27},\ref{aff28},\ref{aff26},\ref{aff112}}
\and F.~Cogato\orcid{0000-0003-4632-6113}\inst{\ref{aff84},\ref{aff25}}
\and A.~R.~Cooray\orcid{0000-0002-3892-0190}\inst{\ref{aff122}}
\and S.~Davini\orcid{0000-0003-3269-1718}\inst{\ref{aff33}}
\and G.~Desprez\orcid{0000-0001-8325-1742}\inst{\ref{aff123}}
\and A.~D\'iaz-S\'anchez\orcid{0000-0003-0748-4768}\inst{\ref{aff124}}
\and S.~Di~Domizio\orcid{0000-0003-2863-5895}\inst{\ref{aff32},\ref{aff33}}
\and A.~G.~Ferrari\orcid{0009-0005-5266-4110}\inst{\ref{aff31}}
\and I.~Ferrero\orcid{0000-0002-1295-1132}\inst{\ref{aff69}}
\and A.~Finoguenov\orcid{0000-0002-4606-5403}\inst{\ref{aff79}}
\and A.~Fontana\orcid{0000-0003-3820-2823}\inst{\ref{aff8}}
\and F.~Fornari\orcid{0000-0003-2979-6738}\inst{\ref{aff61}}
\and L.~Gabarra\orcid{0000-0002-8486-8856}\inst{\ref{aff125}}
\and K.~Ganga\orcid{0000-0001-8159-8208}\inst{\ref{aff86}}
\and J.~Garc\'ia-Bellido\orcid{0000-0002-9370-8360}\inst{\ref{aff113}}
\and T.~Gasparetto\orcid{0000-0002-7913-4866}\inst{\ref{aff27}}
\and V.~Gautard\inst{\ref{aff126}}
\and E.~Gaztanaga\orcid{0000-0001-9632-0815}\inst{\ref{aff63},\ref{aff62},\ref{aff11}}
\and F.~Giacomini\orcid{0000-0002-3129-2814}\inst{\ref{aff31}}
\and F.~Gianotti\orcid{0000-0003-4666-119X}\inst{\ref{aff25}}
\and G.~Gozaliasl\orcid{0000-0002-0236-919X}\inst{\ref{aff127}}
\and A.~Gregorio\orcid{0000-0003-4028-8785}\inst{\ref{aff119},\ref{aff27},\ref{aff28}}
\and M.~Guidi\orcid{0000-0001-9408-1101}\inst{\ref{aff30},\ref{aff25}}
\and C.~M.~Gutierrez\orcid{0000-0001-7854-783X}\inst{\ref{aff128}}
\and S.~Hemmati\orcid{0000-0003-2226-5395}\inst{\ref{aff129}}
\and H.~Hildebrandt\orcid{0000-0002-9814-3338}\inst{\ref{aff15}}
\and J.~Hjorth\orcid{0000-0002-4571-2306}\inst{\ref{aff130}}
\and A.~Jimenez~Mu\~noz\orcid{0009-0004-5252-185X}\inst{\ref{aff131}}
\and J.~J.~E.~Kajava\orcid{0000-0002-3010-8333}\inst{\ref{aff132},\ref{aff133}}
\and Y.~Kang\orcid{0009-0000-8588-7250}\inst{\ref{aff54}}
\and V.~Kansal\orcid{0000-0002-4008-6078}\inst{\ref{aff134},\ref{aff135}}
\and D.~Karagiannis\orcid{0000-0002-4927-0816}\inst{\ref{aff136},\ref{aff137}}
\and C.~C.~Kirkpatrick\inst{\ref{aff78}}
\and S.~Kruk\orcid{0000-0001-8010-8879}\inst{\ref{aff47}}
\and M.~Lattanzi\orcid{0000-0003-1059-2532}\inst{\ref{aff117}}
\and S.~Lee\orcid{0000-0002-8289-740X}\inst{\ref{aff70}}
\and J.~Le~Graet\orcid{0000-0001-6523-7971}\inst{\ref{aff58}}
\and L.~Legrand\orcid{0000-0003-0610-5252}\inst{\ref{aff138},\ref{aff139}}
\and M.~Lembo\orcid{0000-0002-5271-5070}\inst{\ref{aff115},\ref{aff117}}
\and J.~Lesgourgues\orcid{0000-0001-7627-353X}\inst{\ref{aff10}}
\and T.~I.~Liaudat\orcid{0000-0002-9104-314X}\inst{\ref{aff140}}
\and A.~Loureiro\orcid{0000-0002-4371-0876}\inst{\ref{aff141},\ref{aff142}}
\and J.~Macias-Perez\orcid{0000-0002-5385-2763}\inst{\ref{aff131}}
\and M.~Magliocchetti\orcid{0000-0001-9158-4838}\inst{\ref{aff56}}
\and F.~Mannucci\orcid{0000-0002-4803-2381}\inst{\ref{aff143}}
\and R.~Maoli\orcid{0000-0002-6065-3025}\inst{\ref{aff144},\ref{aff8}}
\and J.~Mart\'{i}n-Fleitas\orcid{0000-0002-8594-569X}\inst{\ref{aff88}}
\and C.~J.~A.~P.~Martins\orcid{0000-0002-4886-9261}\inst{\ref{aff145},\ref{aff37}}
\and L.~Maurin\orcid{0000-0002-8406-0857}\inst{\ref{aff22}}
\and R.~B.~Metcalf\orcid{0000-0003-3167-2574}\inst{\ref{aff84},\ref{aff25}}
\and M.~Migliaccio\inst{\ref{aff146},\ref{aff147}}
\and M.~Miluzio\inst{\ref{aff47},\ref{aff148}}
\and P.~Monaco\orcid{0000-0003-2083-7564}\inst{\ref{aff119},\ref{aff27},\ref{aff28},\ref{aff26}}
\and C.~Moretti\orcid{0000-0003-3314-8936}\inst{\ref{aff29},\ref{aff112},\ref{aff27},\ref{aff26},\ref{aff28}}
\and G.~Morgante\inst{\ref{aff25}}
\and C.~Murray\inst{\ref{aff86}}
\and S.~Nadathur\orcid{0000-0001-9070-3102}\inst{\ref{aff11}}
\and K.~Naidoo\orcid{0000-0002-9182-1802}\inst{\ref{aff12}}
\and Nicholas~A.~Walton\orcid{0000-0003-3983-8778}\inst{\ref{aff149}}
\and K.~Paterson\orcid{0000-0001-8340-3486}\inst{\ref{aff76}}
\and L.~Patrizii\inst{\ref{aff31}}
\and A.~Pisani\orcid{0000-0002-6146-4437}\inst{\ref{aff58},\ref{aff150}}
\and V.~Popa\orcid{0000-0002-9118-8330}\inst{\ref{aff95}}
\and D.~Potter\orcid{0000-0002-0757-5195}\inst{\ref{aff151}}
\and I.~Risso\orcid{0000-0003-2525-7761}\inst{\ref{aff152}}
\and P.-F.~Rocci\inst{\ref{aff22}}
\and M.~Sahl\'en\orcid{0000-0003-0973-4804}\inst{\ref{aff153}}
\and E.~Sarpa\orcid{0000-0002-1256-655X}\inst{\ref{aff29},\ref{aff112},\ref{aff28}}
\and A.~Schneider\orcid{0000-0001-7055-8104}\inst{\ref{aff151}}
\and M.~Schultheis\inst{\ref{aff121}}
\and D.~Sciotti\orcid{0009-0008-4519-2620}\inst{\ref{aff8},\ref{aff9}}
\and E.~Sellentin\inst{\ref{aff154},\ref{aff155}}
\and M.~Sereno\orcid{0000-0003-0302-0325}\inst{\ref{aff25},\ref{aff31}}
\and L.~C.~Smith\orcid{0000-0002-3259-2771}\inst{\ref{aff149}}
\and K.~Tanidis\inst{\ref{aff125}}
\and C.~Tao\orcid{0000-0001-7961-8177}\inst{\ref{aff58}}
\and G.~Testera\inst{\ref{aff33}}
\and R.~Teyssier\orcid{0000-0001-7689-0933}\inst{\ref{aff150}}
\and S.~Tosi\orcid{0000-0002-7275-9193}\inst{\ref{aff32},\ref{aff33}}
\and A.~Troja\orcid{0000-0003-0239-4595}\inst{\ref{aff98},\ref{aff57}}
\and M.~Tucci\inst{\ref{aff54}}
\and C.~Valieri\inst{\ref{aff31}}
\and D.~Vergani\orcid{0000-0003-0898-2216}\inst{\ref{aff25}}
\and F.~Vernizzi\orcid{0000-0003-3426-2802}\inst{\ref{aff156}}
\and G.~Verza\orcid{0000-0002-1886-8348}\inst{\ref{aff157},\ref{aff158}}
\and P.~Vielzeuf\orcid{0000-0003-2035-9339}\inst{\ref{aff58}}}
										   
%%%% please do not edit the affiliation list -- contact ECEB Bureau for changes
\institute{Instituto de Astrof\'isica e Ci\^encias do Espa\c{c}o, Faculdade de Ci\^encias, Universidade de Lisboa, Campo Grande, 1749-016 Lisboa, Portugal\label{aff1}
\and
Dipartimento di Fisica "E. Pancini", Universita degli Studi di Napoli Federico II, Via Cinthia 6, 80126, Napoli, Italy\label{aff2}
\and
Institut f\"ur Theoretische Physik, University of Heidelberg, Philosophenweg 16, 69120 Heidelberg, Germany\label{aff3}
\and
Institut de Recherche en Astrophysique et Plan\'etologie (IRAP), Universit\'e de Toulouse, CNRS, UPS, CNES, 14 Av. Edouard Belin, 31400 Toulouse, France\label{aff4}
\and
Universit\'e St Joseph; Faculty of Sciences, Beirut, Lebanon\label{aff5}
\and
Universit\"ats-Sternwarte M\"unchen, Fakult\"at f\"ur Physik, Ludwig-Maximilians-Universit\"at M\"unchen, Scheinerstrasse 1, 81679 M\"unchen, Germany\label{aff6}
\and
Institute for Astronomy, University of Edinburgh, Royal Observatory, Blackford Hill, Edinburgh EH9 3HJ, UK\label{aff7}
\and
INAF-Osservatorio Astronomico di Roma, Via Frascati 33, 00078 Monteporzio Catone, Italy\label{aff8}
\and
INFN-Sezione di Roma, Piazzale Aldo Moro, 2 - c/o Dipartimento di Fisica, Edificio G. Marconi, 00185 Roma, Italy\label{aff9}
\and
Institute for Theoretical Particle Physics and Cosmology (TTK), RWTH Aachen University, 52056 Aachen, Germany\label{aff10}
\and
Institute of Cosmology and Gravitation, University of Portsmouth, Portsmouth PO1 3FX, UK\label{aff11}
\and
Department of Physics and Astronomy, University College London, Gower Street, London WC1E 6BT, UK\label{aff12}
\and
Laboratoire univers et particules de Montpellier, Universit\'e de Montpellier, CNRS, 34090 Montpellier, France\label{aff13}
\and
Jodrell Bank Centre for Astrophysics, Department of Physics and Astronomy, University of Manchester, Oxford Road, Manchester M13 9PL, UK\label{aff14}
\and
Ruhr University Bochum, Faculty of Physics and Astronomy, Astronomical Institute (AIRUB), German Centre for Cosmological Lensing (GCCL), 44780 Bochum, Germany\label{aff15}
\and
Universit\"at Bonn, Argelander-Institut f\"ur Astronomie, Auf dem H\"ugel 71, 53121 Bonn, Germany\label{aff16}
\and
Universit\'e de Gen\`eve, D\'epartement de Physique Th\'eorique and Centre for Astroparticle Physics, 24 quai Ernest-Ansermet, CH-1211 Gen\`eve 4, Switzerland\label{aff17}
\and
Dipartimento di Fisica, Universit\`a degli Studi di Torino, Via P. Giuria 1, 10125 Torino, Italy\label{aff18}
\and
INFN-Sezione di Torino, Via P. Giuria 1, 10125 Torino, Italy\label{aff19}
\and
INAF-Osservatorio Astrofisico di Torino, Via Osservatorio 20, 10025 Pino Torinese (TO), Italy\label{aff20}
\and
Institute Lorentz, Leiden University, Niels Bohrweg 2, 2333 CA Leiden, The Netherlands\label{aff21}
\and
Universit\'e Paris-Saclay, CNRS, Institut d'astrophysique spatiale, 91405, Orsay, France\label{aff22}
\and
School of Mathematics and Physics, University of Surrey, Guildford, Surrey, GU2 7XH, UK\label{aff23}
\and
INAF-Osservatorio Astronomico di Brera, Via Brera 28, 20122 Milano, Italy\label{aff24}
\and
INAF-Osservatorio di Astrofisica e Scienza dello Spazio di Bologna, Via Piero Gobetti 93/3, 40129 Bologna, Italy\label{aff25}
\and
IFPU, Institute for Fundamental Physics of the Universe, via Beirut 2, 34151 Trieste, Italy\label{aff26}
\and
INAF-Osservatorio Astronomico di Trieste, Via G. B. Tiepolo 11, 34143 Trieste, Italy\label{aff27}
\and
INFN, Sezione di Trieste, Via Valerio 2, 34127 Trieste TS, Italy\label{aff28}
\and
SISSA, International School for Advanced Studies, Via Bonomea 265, 34136 Trieste TS, Italy\label{aff29}
\and
Dipartimento di Fisica e Astronomia, Universit\`a di Bologna, Via Gobetti 93/2, 40129 Bologna, Italy\label{aff30}
\and
INFN-Sezione di Bologna, Viale Berti Pichat 6/2, 40127 Bologna, Italy\label{aff31}
\and
Dipartimento di Fisica, Universit\`a di Genova, Via Dodecaneso 33, 16146, Genova, Italy\label{aff32}
\and
INFN-Sezione di Genova, Via Dodecaneso 33, 16146, Genova, Italy\label{aff33}
\and
Department of Physics "E. Pancini", University Federico II, Via Cinthia 6, 80126, Napoli, Italy\label{aff34}
\and
INAF-Osservatorio Astronomico di Capodimonte, Via Moiariello 16, 80131 Napoli, Italy\label{aff35}
\and
INFN section of Naples, Via Cinthia 6, 80126, Napoli, Italy\label{aff36}
\and
Instituto de Astrof\'isica e Ci\^encias do Espa\c{c}o, Universidade do Porto, CAUP, Rua das Estrelas, PT4150-762 Porto, Portugal\label{aff37}
\and
Faculdade de Ci\^encias da Universidade do Porto, Rua do Campo de Alegre, 4150-007 Porto, Portugal\label{aff38}
\and
European Space Agency/ESTEC, Keplerlaan 1, 2201 AZ Noordwijk, The Netherlands\label{aff39}
\and
INAF-IASF Milano, Via Alfonso Corti 12, 20133 Milano, Italy\label{aff40}
\and
Centro de Investigaciones Energ\'eticas, Medioambientales y Tecnol\'ogicas (CIEMAT), Avenida Complutense 40, 28040 Madrid, Spain\label{aff41}
\and
Port d'Informaci\'{o} Cient\'{i}fica, Campus UAB, C. Albareda s/n, 08193 Bellaterra (Barcelona), Spain\label{aff42}
\and
Institute for Astronomy, University of Hawaii, 2680 Woodlawn Drive, Honolulu, HI 96822, USA\label{aff43}
\and
Dipartimento di Fisica e Astronomia "Augusto Righi" - Alma Mater Studiorum Universit\`a di Bologna, Viale Berti Pichat 6/2, 40127 Bologna, Italy\label{aff44}
\and
Instituto de Astrof\'{\i}sica de Canarias, V\'{\i}a L\'actea, 38205 La Laguna, Tenerife, Spain\label{aff45}
\and
European Space Agency/ESRIN, Largo Galileo Galilei 1, 00044 Frascati, Roma, Italy\label{aff46}
\and
ESAC/ESA, Camino Bajo del Castillo, s/n., Urb. Villafranca del Castillo, 28692 Villanueva de la Ca\~nada, Madrid, Spain\label{aff47}
\and
Universit\'e Claude Bernard Lyon 1, CNRS/IN2P3, IP2I Lyon, UMR 5822, Villeurbanne, F-69100, France\label{aff48}
\and
Institute of Physics, Laboratory of Astrophysics, Ecole Polytechnique F\'ed\'erale de Lausanne (EPFL), Observatoire de Sauverny, 1290 Versoix, Switzerland\label{aff49}
\and
Institut de Ci\`{e}ncies del Cosmos (ICCUB), Universitat de Barcelona (IEEC-UB), Mart\'{i} i Franqu\`{e}s 1, 08028 Barcelona, Spain\label{aff50}
\and
Instituci\'o Catalana de Recerca i Estudis Avan\c{c}ats (ICREA), Passeig de Llu\'{\i}s Companys 23, 08010 Barcelona, Spain\label{aff51}
\and
UCB Lyon 1, CNRS/IN2P3, IUF, IP2I Lyon, 4 rue Enrico Fermi, 69622 Villeurbanne, France\label{aff52}
\and
Departamento de F\'isica, Faculdade de Ci\^encias, Universidade de Lisboa, Edif\'icio C8, Campo Grande, PT1749-016 Lisboa, Portugal\label{aff53}
\and
Department of Astronomy, University of Geneva, ch. d'Ecogia 16, 1290 Versoix, Switzerland\label{aff54}
\and
Aix-Marseille Universit\'e, CNRS, CNES, LAM, Marseille, France\label{aff55}
\and
INAF-Istituto di Astrofisica e Planetologia Spaziali, via del Fosso del Cavaliere, 100, 00100 Roma, Italy\label{aff56}
\and
INFN-Padova, Via Marzolo 8, 35131 Padova, Italy\label{aff57}
\and
Aix-Marseille Universit\'e, CNRS/IN2P3, CPPM, Marseille, France\label{aff58}
\and
Universit\'e Paris-Saclay, Universit\'e Paris Cit\'e, CEA, CNRS, AIM, 91191, Gif-sur-Yvette, France\label{aff59}
\and
Space Science Data Center, Italian Space Agency, via del Politecnico snc, 00133 Roma, Italy\label{aff60}
\and
INFN-Bologna, Via Irnerio 46, 40126 Bologna, Italy\label{aff61}
\and
Institut d'Estudis Espacials de Catalunya (IEEC),  Edifici RDIT, Campus UPC, 08860 Castelldefels, Barcelona, Spain\label{aff62}
\and
Institute of Space Sciences (ICE, CSIC), Campus UAB, Carrer de Can Magrans, s/n, 08193 Barcelona, Spain\label{aff63}
\and
School of Physics, HH Wills Physics Laboratory, University of Bristol, Tyndall Avenue, Bristol, BS8 1TL, UK\label{aff64}
\and
Max Planck Institute for Extraterrestrial Physics, Giessenbachstr. 1, 85748 Garching, Germany\label{aff65}
\and
INAF-Osservatorio Astronomico di Padova, Via dell'Osservatorio 5, 35122 Padova, Italy\label{aff66}
\and
Dipartimento di Fisica "Aldo Pontremoli", Universit\`a degli Studi di Milano, Via Celoria 16, 20133 Milano, Italy\label{aff67}
\and
INFN-Sezione di Milano, Via Celoria 16, 20133 Milano, Italy\label{aff68}
\and
Institute of Theoretical Astrophysics, University of Oslo, P.O. Box 1029 Blindern, 0315 Oslo, Norway\label{aff69}
\and
Jet Propulsion Laboratory, California Institute of Technology, 4800 Oak Grove Drive, Pasadena, CA, 91109, USA\label{aff70}
\and
Felix Hormuth Engineering, Goethestr. 17, 69181 Leimen, Germany\label{aff71}
\and
Technical University of Denmark, Elektrovej 327, 2800 Kgs. Lyngby, Denmark\label{aff72}
\and
Cosmic Dawn Center (DAWN), Denmark\label{aff73}
\and
Institut d'Astrophysique de Paris, UMR 7095, CNRS, and Sorbonne Universit\'e, 98 bis boulevard Arago, 75014 Paris, France\label{aff74}
\and
Universit\'e Paris-Saclay, CNRS/IN2P3, IJCLab, 91405 Orsay, France\label{aff75}
\and
Max-Planck-Institut f\"ur Astronomie, K\"onigstuhl 17, 69117 Heidelberg, Germany\label{aff76}
\and
NASA Goddard Space Flight Center, Greenbelt, MD 20771, USA\label{aff77}
\and
Department of Physics and Helsinki Institute of Physics, Gustaf H\"allstr\"omin katu 2, 00014 University of Helsinki, Finland\label{aff78}
\and
Department of Physics, P.O. Box 64, 00014 University of Helsinki, Finland\label{aff79}
\and
Helsinki Institute of Physics, Gustaf H{\"a}llstr{\"o}min katu 2, University of Helsinki, Helsinki, Finland\label{aff80}
\and
Laboratoire Univers et Th\'eorie, Observatoire de Paris, Universit\'e PSL, Universit\'e Paris Cit\'e, CNRS, 92190 Meudon, France\label{aff81}
\and
SKA Observatory, Jodrell Bank, Lower Withington, Macclesfield, Cheshire SK11 9FT, UK\label{aff82}
\and
Centre de Calcul de l'IN2P3/CNRS, 21 avenue Pierre de Coubertin 69627 Villeurbanne Cedex, France\label{aff83}
\and
Dipartimento di Fisica e Astronomia "Augusto Righi" - Alma Mater Studiorum Universit\`a di Bologna, via Piero Gobetti 93/2, 40129 Bologna, Italy\label{aff84}
\and
Department of Physics, Institute for Computational Cosmology, Durham University, South Road, Durham, DH1 3LE, UK\label{aff85}
\and
Universit\'e Paris Cit\'e, CNRS, Astroparticule et Cosmologie, 75013 Paris, France\label{aff86}
\and
Institut d'Astrophysique de Paris, 98bis Boulevard Arago, 75014, Paris, France\label{aff87}
\and
Aurora Technology for European Space Agency (ESA), Camino bajo del Castillo, s/n, Urbanizacion Villafranca del Castillo, Villanueva de la Ca\~nada, 28692 Madrid, Spain\label{aff88}
\and
School of Mathematics, Statistics and Physics, Newcastle University, Herschel Building, Newcastle-upon-Tyne, NE1 7RU, UK\label{aff89}
\and
Institut de F\'{i}sica d'Altes Energies (IFAE), The Barcelona Institute of Science and Technology, Campus UAB, 08193 Bellaterra (Barcelona), Spain\label{aff90}
\and
Waterloo Centre for Astrophysics, University of Waterloo, Waterloo, Ontario N2L 3G1, Canada\label{aff91}
\and
Department of Physics and Astronomy, University of Waterloo, Waterloo, Ontario N2L 3G1, Canada\label{aff92}
\and
Perimeter Institute for Theoretical Physics, Waterloo, Ontario N2L 2Y5, Canada\label{aff93}
\and
Centre National d'Etudes Spatiales -- Centre spatial de Toulouse, 18 avenue Edouard Belin, 31401 Toulouse Cedex 9, France\label{aff94}
\and
Institute of Space Science, Str. Atomistilor, nr. 409 M\u{a}gurele, Ilfov, 077125, Romania\label{aff95}
\and
Universidad de La Laguna, Departamento de Astrof\'{\i}sica, 38206 La Laguna, Tenerife, Spain\label{aff96}
\and
Consejo Superior de Investigaciones Cientificas, Calle Serrano 117, 28006 Madrid, Spain\label{aff97}
\and
Dipartimento di Fisica e Astronomia "G. Galilei", Universit\`a di Padova, Via Marzolo 8, 35131 Padova, Italy\label{aff98}
\and
Departamento de F\'isica, FCFM, Universidad de Chile, Blanco Encalada 2008, Santiago, Chile\label{aff99}
\and
Universit\"at Innsbruck, Institut f\"ur Astro- und Teilchenphysik, Technikerstr. 25/8, 6020 Innsbruck, Austria\label{aff100}
\and
Satlantis, University Science Park, Sede Bld 48940, Leioa-Bilbao, Spain\label{aff101}
\and
Instituto de Astrof\'isica e Ci\^encias do Espa\c{c}o, Faculdade de Ci\^encias, Universidade de Lisboa, Tapada da Ajuda, 1349-018 Lisboa, Portugal\label{aff102}
\and
Cosmic Dawn Center (DAWN)\label{aff103}
\and
Niels Bohr Institute, University of Copenhagen, Jagtvej 128, 2200 Copenhagen, Denmark\label{aff104}
\and
Universidad Polit\'ecnica de Cartagena, Departamento de Electr\'onica y Tecnolog\'ia de Computadoras,  Plaza del Hospital 1, 30202 Cartagena, Spain\label{aff105}
\and
Kapteyn Astronomical Institute, University of Groningen, PO Box 800, 9700 AV Groningen, The Netherlands\label{aff106}
\and
Dipartimento di Fisica, Universit\`a degli studi di Genova, and INFN-Sezione di Genova, via Dodecaneso 33, 16146, Genova, Italy\label{aff107}
\and
Infrared Processing and Analysis Center, California Institute of Technology, Pasadena, CA 91125, USA\label{aff108}
\and
INAF, Istituto di Radioastronomia, Via Piero Gobetti 101, 40129 Bologna, Italy\label{aff109}
\and
Astronomical Observatory of the Autonomous Region of the Aosta Valley (OAVdA), Loc. Lignan 39, I-11020, Nus (Aosta Valley), Italy\label{aff110}
\and
ICL, Junia, Universit\'e Catholique de Lille, LITL, 59000 Lille, France\label{aff111}
\and
ICSC - Centro Nazionale di Ricerca in High Performance Computing, Big Data e Quantum Computing, Via Magnanelli 2, Bologna, Italy\label{aff112}
\and
Instituto de F\'isica Te\'orica UAM-CSIC, Campus de Cantoblanco, 28049 Madrid, Spain\label{aff113}
\and
CERCA/ISO, Department of Physics, Case Western Reserve University, 10900 Euclid Avenue, Cleveland, OH 44106, USA\label{aff114}
\and
Dipartimento di Fisica e Scienze della Terra, Universit\`a degli Studi di Ferrara, Via Giuseppe Saragat 1, 44122 Ferrara, Italy\label{aff115}
\and
Departamento de F{\'\i}sica Fundamental. Universidad de Salamanca. Plaza de la Merced s/n. 37008 Salamanca, Spain\label{aff116}
\and
Istituto Nazionale di Fisica Nucleare, Sezione di Ferrara, Via Giuseppe Saragat 1, 44122 Ferrara, Italy\label{aff117}
\and
Center for Data-Driven Discovery, Kavli IPMU (WPI), UTIAS, The University of Tokyo, Kashiwa, Chiba 277-8583, Japan\label{aff118}
\and
Dipartimento di Fisica - Sezione di Astronomia, Universit\`a di Trieste, Via Tiepolo 11, 34131 Trieste, Italy\label{aff119}
\and
California Institute of Technology, 1200 E California Blvd, Pasadena, CA 91125, USA\label{aff120}
\and
Universit\'e C\^{o}te d'Azur, Observatoire de la C\^{o}te d'Azur, CNRS, Laboratoire Lagrange, Bd de l'Observatoire, CS 34229, 06304 Nice cedex 4, France\label{aff121}
\and
Department of Physics \& Astronomy, University of California Irvine, Irvine CA 92697, USA\label{aff122}
\and
Department of Astronomy \& Physics and Institute for Computational Astrophysics, Saint Mary's University, 923 Robie Street, Halifax, Nova Scotia, B3H 3C3, Canada\label{aff123}
\and
Departamento F\'isica Aplicada, Universidad Polit\'ecnica de Cartagena, Campus Muralla del Mar, 30202 Cartagena, Murcia, Spain\label{aff124}
\and
Department of Physics, Oxford University, Keble Road, Oxford OX1 3RH, UK\label{aff125}
\and
CEA Saclay, DFR/IRFU, Service d'Astrophysique, Bat. 709, 91191 Gif-sur-Yvette, France\label{aff126}
\and
Department of Computer Science, Aalto University, PO Box 15400, Espoo, FI-00 076, Finland\label{aff127}
\and
Instituto de Astrof\'\i sica de Canarias, c/ Via Lactea s/n, La Laguna 38200, Spain. Departamento de Astrof\'\i sica de la Universidad de La Laguna, Avda. Francisco Sanchez, La Laguna, 38200, Spain\label{aff128}
\and
Caltech/IPAC, 1200 E. California Blvd., Pasadena, CA 91125, USA\label{aff129}
\and
DARK, Niels Bohr Institute, University of Copenhagen, Jagtvej 155, 2200 Copenhagen, Denmark\label{aff130}
\and
Univ. Grenoble Alpes, CNRS, Grenoble INP, LPSC-IN2P3, 53, Avenue des Martyrs, 38000, Grenoble, France\label{aff131}
\and
Department of Physics and Astronomy, Vesilinnantie 5, 20014 University of Turku, Finland\label{aff132}
\and
Serco for European Space Agency (ESA), Camino bajo del Castillo, s/n, Urbanizacion Villafranca del Castillo, Villanueva de la Ca\~nada, 28692 Madrid, Spain\label{aff133}
\and
ARC Centre of Excellence for Dark Matter Particle Physics, Melbourne, Australia\label{aff134}
\and
Centre for Astrophysics \& Supercomputing, Swinburne University of Technology,  Hawthorn, Victoria 3122, Australia\label{aff135}
\and
School of Physics and Astronomy, Queen Mary University of London, Mile End Road, London E1 4NS, UK\label{aff136}
\and
Department of Physics and Astronomy, University of the Western Cape, Bellville, Cape Town, 7535, South Africa\label{aff137}
\and
DAMTP, Centre for Mathematical Sciences, Wilberforce Road, Cambridge CB3 0WA, UK\label{aff138}
\and
Kavli Institute for Cosmology Cambridge, Madingley Road, Cambridge, CB3 0HA, UK\label{aff139}
\and
IRFU, CEA, Universit\'e Paris-Saclay 91191 Gif-sur-Yvette Cedex, France\label{aff140}
\and
Oskar Klein Centre for Cosmoparticle Physics, Department of Physics, Stockholm University, Stockholm, SE-106 91, Sweden\label{aff141}
\and
Astrophysics Group, Blackett Laboratory, Imperial College London, London SW7 2AZ, UK\label{aff142}
\and
INAF-Osservatorio Astrofisico di Arcetri, Largo E. Fermi 5, 50125, Firenze, Italy\label{aff143}
\and
Dipartimento di Fisica, Sapienza Universit\`a di Roma, Piazzale Aldo Moro 2, 00185 Roma, Italy\label{aff144}
\and
Centro de Astrof\'{\i}sica da Universidade do Porto, Rua das Estrelas, 4150-762 Porto, Portugal\label{aff145}
\and
Dipartimento di Fisica, Universit\`a di Roma Tor Vergata, Via della Ricerca Scientifica 1, Roma, Italy\label{aff146}
\and
INFN, Sezione di Roma 2, Via della Ricerca Scientifica 1, Roma, Italy\label{aff147}
\and
HE Space for European Space Agency (ESA), Camino bajo del Castillo, s/n, Urbanizacion Villafranca del Castillo, Villanueva de la Ca\~nada, 28692 Madrid, Spain\label{aff148}
\and
Institute of Astronomy, University of Cambridge, Madingley Road, Cambridge CB3 0HA, UK\label{aff149}
\and
Department of Astrophysical Sciences, Peyton Hall, Princeton University, Princeton, NJ 08544, USA\label{aff150}
\and
Department of Astrophysics, University of Zurich, Winterthurerstrasse 190, 8057 Zurich, Switzerland\label{aff151}
\and
INAF-Osservatorio Astronomico di Brera, Via Brera 28, 20122 Milano, Italy, and INFN-Sezione di Genova, Via Dodecaneso 33, 16146, Genova, Italy\label{aff152}
\and
Theoretical astrophysics, Department of Physics and Astronomy, Uppsala University, Box 516, 751 37 Uppsala, Sweden\label{aff153}
\and
Mathematical Institute, University of Leiden, Einsteinweg 55, 2333 CA Leiden, The Netherlands\label{aff154}
\and
Leiden Observatory, Leiden University, Einsteinweg 55, 2333 CC Leiden, The Netherlands\label{aff155}
\and
Institut de Physique Th\'eorique, CEA, CNRS, Universit\'e Paris-Saclay 91191 Gif-sur-Yvette Cedex, France\label{aff156}
\and
Center for Cosmology and Particle Physics, Department of Physics, New York University, New York, NY 10003, USA\label{aff157}
\and
Center for Computational Astrophysics, Flatiron Institute, 162 5th Avenue, 10010, New York, NY, USA\label{aff158}}

\date{\today}

\authorrunning{Euclid Collaboration: I. S. Albuquerque et al.}

\titlerunning{\Euclid: forecasts on parameterised modified gravity.}

\abstract
{The \Euclid mission has the potential to understand the fundamental physical nature of late-time cosmic acceleration and, as such, of deviations from the standard cosmological model, $\lcdm$. 
In this paper, we focus on model-independent methods to modify the evolution of scalar perturbations at linear scales. 
We consider two powerful and convenient approaches: the first is based on the two phenomenological modified gravity (PMG) parameters, $\mu_{\rm mg}$ and $\Sigma_{\rm mg}$, which are phenomenologically connected to the clustering of matter and weak lensing, respectively; and the second is the effective field theory (EFT) of dark energy and modified gravity, which we use to parameterise the braiding function, $\alpha_{\rm B}$, which defines the mixing between the metric and the dark energy field typical of Galileon theories.}
{We discuss the predictions from spectroscopic and photometric primary probes by \Euclid on the cosmological parameters and a given set of additional parameters featuring the PMG and EFT models.}
{We use the Fisher matrix method applied to spectroscopic galaxy clustering (\GCsp), weak lensing (WL), photometric galaxy clustering (\GCph), and cross-correlation (XC) between \GCph\ and WL. For the modelling of photometric predictions on nonlinear scales, we use the halo model reaction approach to cover two limiting cases for the screening mechanism: the unscreened (US) case, for which the screening mechanism is not present; and the super-screened (SS) case, which assumes strong screening. 
We also assume scale cuts to account for our uncertainties in the modelling of nonlinear perturbation evolution. We choose a time-dependent form for $\{\mu_{\rm mg},\Sigma_{\rm mg}\}$, with two fiducial sets of values for the corresponding model parameters at the present time, $\{\bar{\mu}_0,\bar{\Sigma}_0\}$, and two forms for $\alpha_{\rm B}$, with one fiducial set of values for each of the model parameters, $\alpha_{\rm B,0}$ and $\{\alpha_{\rm B,0},m\}$.} 
{At the 68.3\% confidence level, the percentage relative errors obtained with \Euclid alone and our conservative settings for the full combination of probes for the US case are: for $\{\bar{\mu}_0,\bar{\Sigma}_0\}$, with a $\lcdm$ fiducial, \{$23.3\%$, $2.6\%$\}; for a fiducial  $\{\bar{\mu}_0,\bar{\Sigma}_0\}$=$\{0.5,0.5\}$ we obtain {$36.2\%$, $2.7\%$}; for $\alpha_{\rm B,0}$ whose fiducial is $0.2$, we have $31.1\%$ and for $\{\alpha_{\rm B,0},m\}$ with fiducial $\{0.9,2.4\}$ we have $11.6\%$ and $11.8\%$. The constraints we obtain with the SS prescription provide similar values. We also compute the constraints for different combinations of probes to assess their standalone and complementary constraining power.}
{}


\maketitle

%%%%%%%%%%%%%%%%%%%%%%
\section{Introduction} \label{sec:intro}
%%%%%%%%%%%%%%%%%%%%%%

The discovery of a late-time acceleration of the Universe's expansion has stimulated the study of modifications to general relativity (GR) on cosmological scales. There are many distinct modified gravity (MG) models proposed, and forecasts for \Euclid's ability to constrain some of the most representative ones have already been made \citep{Casas23a,Frusciante23,EP-Bose}. However, the landscape of possible remaining viable theories is still significant, so it is difficult to test all of them exhaustively. Therefore, the primary purpose of the present work is to constrain deviations from the standard cosmological model (\lcdm) based on parameterised approaches that could encapsulate a larger class of models. Parameterised tests of gravity are motivated more generally as a null test of the \lcdm\, paradigm and as precision tests of GR on cosmological scales \citep[analogous to those carried out on small scales, see for example][]{2014LRR....17....4W}. 
However, to constrain parameterised models of dark energy (DE) and MG, we inevitably need to specify the selection of parameterisations that are being considered. In this paper, we will focus on the following two approaches.

\paragraph{Phenomenological Modified Gravity (PMG) Parameterisation:} a parameterisation of the gravitational couplings $\{\mu_{\rm mg}, \Sigma_{\rm mg}\}$ describing the modifications of the Poisson equations that link the Bardeen potentials $\Phi$ and $\Psi$ \citep{Bardeen:1980kt} to the matter density contrast \citep{Zhang:2007nk, Amendola:2007rr,Pogosian:2010tj, Planck:2015bue}, respectively. More precisely, $\mu_{\rm mg}$ is linked to the Poisson equation for $\Psi$. This Bardeen potential controls the evolution of non-relativistic particles and, hence, the cosmological clustering of matter. In fact, it quantifies modifications to the effective strength of gravity as felt by matter. On the other hand, $\Sigma_{\rm mg}$ plays an analogous role in the equation of lensing or Weyl potential, $\Phi + \Psi$. This potential controls the evolution of massless, relativistic particles and, hence, of light. \lcdm\, is recovered when $\mu_{\rm mg}$ and $\Sigma_{\rm mg}$ are equal to unity, implying that deviations away from this limit in either of these parameters would indicate evidence of new physics. 
Note that, by construction, both $\mu_{\rm mg}$ and $\Sigma_{\rm mg}$ are closely related to observable effects, so they are expected to be well constrained by large-scale structure surveys such as \Euclid. While we choose specific functional forms for $\mu_{\rm mg}$ and $\Sigma_{\rm mg}$ (see \cref{Sec:modelpotential} for these choices), we note that in principle these quantities may be treated as model independent in the sense that they capture observational degrees of freedom. In the literature, work has been done investigating redshift binning for $\mu_{\rm mg}$ and $\Sigma_{\rm mg}$ \citep{Casas:2017eob, Srinivasan:2021gib, Srinivasan:2024} and also AI-aided reconstruction as a function of redshift \citep{Ruiz_Zapatero_2022}. Essentially, one can think of the PMG approach as an effective observational data-driven null test of \lcdm. 

\paragraph{The `Effective Field Theory of DE and MG' (EFT):} 
in contrast to more phenomenological parameterisations, the EFT is theoretically constructed using a small set of fundamental assumptions and organising principles about the underlying physics \citep[see, e.g.][for a review]{2013JCAP...02..032G,2013JCAP...08..010B,Frusciante:2019xia}. This approach encompasses a wide range of MG models, with one extra scalar degree of freedom that can be specified through a set of time-dependent functions,
called EFT functions, appearing in the action and multiplying the operators compatible with the residual symmetries of unbroken spatial diffeomorphisms. Generalisations of the EFT beyond its first formulation exist -- see, for example, \cite{Gleyzes:2013ooa}, \cite{Gao:2014soa}, and \cite{Frusciante:2017nfr} for approaches going to higher derivative orders; \cite{Kase:2014cwa}, \cite{Frusciante:2015maa}, and \cite{Frusciante:2016xoj} for beyond scalar-tensor theories; and \cite{Lagos:2016wyv} and \cite{Heisenberg:2014rta} for constructing analogous frameworks involving additional vector and tensor degrees of freedom.
In this paper, we use the $\alpha$-basis \citep{2014JCAP...07..050B}, in which the EFT functions,  namely $\alpha_i$, encode the general dynamics at the level of linear perturbations of scalar-tensor theories à la Horndeski/Galileon. 
While these effective functions do not have as evident a link to observable effects as the PMG parameterisation, they directly hold the information about physical effects and provide a compelling alternative parameterisation to consider.
Note that a well-defined mapping exists between the PMG and EFT parameterisations within linear theory, with elegant analytic expressions for PMG in terms of the functions $\alpha_i$ emerging in the quasi-static approximation (QSA), whose accuracy is controlled by the speed of sound of scalar perturbations and can also be linked to the proximity of the background evolution to that of \lcdm\, \citep{Sawicki:2015zya}.

A systematic investigation of deviations from \lcdm\, can be captured by constraining either the PMG functions or the EFT functions. It must be stressed that even though these approaches are independent of the underlying model, the constraints on each functions will depend on the choices for their functional forms, and therefore the constraints might be highly sensitive to the selected parameterisation (a parameterisation is differently sensitive to the redshift evolution, according to the time scaling of the model) and eventually to the assumed priors. 
Additionally, when using the PMG approach, one must remember that preserving a connection to the underlying gravity model is possible by using the EFT formulation to compute the constraints. In this case, once a parameterisation of the EFT functions is chosen, one can translate the constraints on the EFT parameters into those for PMG. In contrast, a direct parameterisation of $\{\mu_{\rm mg}, \Sigma_{\rm mg}\}$ will lose connection with a specific theory.
Moreover, in a specific theory, $\mu_{\rm mg}$ and $\Sigma_{\rm mg}$ are not independent, and a direct parameterisation of these functions does not allow them to retain the link they share, thus affecting the constraints. One can overcome this issue by using the EFT approach instead. It is important to note that the PMG approach is designed to parameterise observational degrees of freedom directly and, therefore, its strength is not in distinguishing between models given a measured signal, but in detecting a deviation from \lcdm, if present in the data. However, in both approaches, the time evolution of the functions involved might result in simplified behaviours compared to those obtained from full covariant theories, such as scalar-tensor theories \citep{Linder:2015rcz,Perenon:2015sla,Linder:2016wqw}. The risk is to miss MG signatures or to provide false detections.

Additionally, the use of the EFT approach allows us to impose appropriate stability conditions that guarantee the viability of the chosen model \citep{2013JCAP...08..010B,Piazza:2013pua,Frusciante:2016xoj,DeFelice:2016ucp} and to use observational and experimental priors, such as those on the Hubble parameter \citep{Planck:2018vyg}, the effective Planck mass \citep{Planck:2018vyg} and the speed of gravitational waves \citep{LIGOScientific:2017vwq,LIGOScientific:2017zic}. These conditions have a strong constraining power on the parameters of the model, which can be translated a posteriori into the $\{\mu_{\rm mg}, \Sigma_{\rm mg}\}$ parameters' plane \citep{Raveri:2014cka,Frusciante:2015maa,Salvatelli:2016mgy,Perenon:2019dpc}. 
In contrast, a direct parameterisation of $\{\mu_{\rm mg}, \Sigma_{\rm mg}\}$ does not allow us to impose stability or additional observational priors. On the one hand, it can be seen as a disadvantage, but on the other hand, it provides a good null test of \lcdm, which enables a clear identification of the trends in the data that are inconsistent with \lcdm, but not in line with any predictions available from the EFT model. Usually, the PMG approach also requires a reduced number of free parameters compared to the EFT approach, increasing the power in constraining of data and reducing degeneracies. A further advantage is that it does not need to be based on perturbation theory, so it applies to an arbitrarily large density contrast.

In summary, both approaches are useful for acquiring deeper knowledge about the nature of the gravitational interaction and deriving novel predictions at cosmological scales. Using both gives a better picture of possible modified gravity effects in cosmology than just using one or the other.\\

We expect a significant contribution to the constraints on PMG and EFT from \Euclid, which we quantify in this work. To fully profit from the constraining power of \Euclid \citep[see, e.g.,][]{Blanchard-EP7,Martinelli21a,EuclidSkyOverview}, it is crucial to be able to exploit observations down to small scales, where the linear description of density perturbations breaks down, and hence one has to model nonlinear scales in these extensions of the \lcdm\, model. 
 
Nonlinear scales are particularly tricky for MG theories, where standard methods (notably $N$-body simulations and tools to approximate the output of $N$-body simulations) must be modified, sometimes substantially \citep{Winther:2015wla,Hassani:2020rxd,EP-Adamek}.
There are several options in the literature for handling MG predictions on nonlinear scales, including using \texttt{Halofit} \citep{Smith:2002dz}, fitting formulae \citep{Zhao:2013dza, Winther:2019mus}, nonlinear PPF approaches \citep{Hu:2007pj}, spherical collapse \citep[e.g.,][]{Lombriser:2016zfz}, halo model approaches \citep[see][for a recent review]{Asgari:2023mej}, emulators \citep[e.g.,][]{Ruan:2023mgq}, and approximate numerical simulations \citep{Winther:2017jof,Wright:2022krq,Brando:2023fzu}. 
However, most of these approaches are designed for specific models or classes of models and, therefore, cannot be applied to phenomenological parameterisations such as those we are investigating here. \texttt{Halofit}-based approaches, which can, in principle, work in more generic cases, are conceptually problematic instead, since these tools were designed explicitly for $\Lambda$CDM, and have been shown not to be valid for phenomenological MG cosmologies \citep{Srinivasan:2021gib}. In this work, we use the halo model reaction approach \citep{Cataneo:2018cic}, a flexible semi-analytic framework for beyond-$\Lambda$CDM models based on the halo model, which has the advantage of being suitable for both specific models and more model-independent approaches \citep{Srinivasan:2021gib,Bose:2022vwi}.

This paper is organised as follows. In \cref{sec:models}, we review the phenomenological approaches to PMG and EFT. \Cref{sec:methods} presents the functional forms that we have chosen for the PMG and EFT functions. In \cref{sec:thpred}, we review the theoretical predictions for the primary photometric and spectroscopic probes of \Euclid. In \cref{sec:fisher}, we describe the analysis method and present the fiducial parameters used in the Fisher matrix analysis and the survey specifications. Then, we discuss the results in \cref{sec:results} and conclude in \cref{sec:conclusions}.

%%%%%%%%%%%%%%%%%%%%%%
\section{Parameterised modified gravity models} \label{sec:models}
%%%%%%%%%%%%%%%%%%%%%%

%---------------------
\subsection{Phenomenological approach to the modification of the gravitational couplings} \label{Sec:theorypotential}
%---------------------

We define the line element of the flat Friedmann--Lemaître--Robertson--Walker (FLRW) metric by following the Bardeen formalism \citep{Bardeen:1980kt,1995ApJ...455....7M}: 
\begin{equation}
\label{eq:perturbed_metric}
 \de s^2 = -(1+2\Psi)\,c^2\,\de t^2 + a^2(t)\,(1-2\Phi)\,\delta_{ij} \, \de x^i\,\de x^j\mathcomma
\end{equation}
where $a(t)$ is the scale factor as a function of the cosmic time $t$, $\Psi$ and $\Phi$ are the two dimensionless scalar potentials, and $c$ is the speed of light.
When the matter anisotropic stress becomes negligible in GR, the Bardeen potentials that describe scalar perturbations to the metric are equal, that is, $\Phi=\Psi$. This implies that the lensing potential, $ (\Phi+ \Psi)/2$, describing the geodesics of relativistic particles, is equal to the gravitational potential governing non-relativistic particles, $\Psi$. Extended theories of gravity introduce new degrees of freedom, most typically in the form of non-minimally coupled scalar fields, which generally mediate new interactions. Consequently, the equivalence between $(\Phi+\Psi)/2$ and $\Psi$ is generically broken. By combining weak lensing and galaxy clustering data, \Euclid will search for differences between the lensing and Newtonian ($\Psi$) potentials and thus constrain alternative gravity theories. 
 
In an agnostic approach to such tests, two phenomenological functions are commonly introduced to parameterise possible modifications to the Poisson equations relating the matter density contrast to the lensing and Newtonian potentials. Taking into account the contribution of anisotropic stress from matter, so that the resulting equations also apply at early times, these functions are defined by \citep{Zhang:2007nk, Amendola:2007rr,Pogosian:2010tj, Planck:2015bue}
\begin{align}
-k^2\Psi=&\frac{4\pi\,G_{\sfont{N}}}{c^2} \,a^2\mu_{\rm mg}(a,k)\left[\bar\rho\Delta+3\left(\bar\rho+\bar {p}/c^2\right)\sigma\right]\mathcomma \label{eq:mu}\\ 
%  \label{eq:mu}
-k^2\left(\Phi+\Psi\right)  =& \frac{8\pi\,G_{\sfont{N}}}{c^2}\,a^2\bigg\{\Sigma_{\rm mg}(a,k)\left[\bar\rho\Delta+3\left(\bar\rho+\bar{p}/c^2\right)\sigma\right] \nonumber \\
& -\frac{3}{2}\mu_{\rm mg}(a,k)\left(\bar\rho+\bar{p}/c^2\right)\sigma\bigg\}\mathcomma \label{eq:sigma}
\end{align}
where $G_{\sfont{N}}$ is Newton's gravitational constant, $\bar{\rho} = \bar{\rho}_{\rm m} + \bar{\rho}_{\rm r} $ and $\bar{p} = \bar{p}_{\rm m} + \bar{p}_{\rm r}$ are the total background energy density and pressure, respectively, $\bar{\rho} \Delta = \bar{\rho}_{\rm m} \Delta_{\rm m} + \bar{\rho}_{\rm r} \Delta_{\rm r} $ with $\Delta_{i}$ the comoving density contrast, and $\sigma$ is the anisotropic stress. 
The modified gravitational couplings $\mu_{\rm mg}$ and $\Sigma_{\rm mg}$ are defined to be identical to $1$ in the \lcdm\ limit. More generally, they will be time- and possibly scale-dependent functions that encode any departure of these equations from their \lcdm\, form. 

As an alternative to $\Sigma_{\rm mg}$, it is possible to combine $\mu_{\rm mg}$ with a function that defines the ratio of the two potentials, $\eta=\Phi/\Psi$. The three functions, $\mu_{\rm mg}, \Sigma_{\rm mg}$, and $\eta$, are not independent, and they can be related to each other; once the anisotropic stress of matter becomes negligible, the relation assumes the straightforward expression
\begin{equation}\label{eq:sigderived}
 \Sigma_{\rm mg} = \frac{1}{2}\mu_{\rm mg}(1+\eta)\mathperiod
\end{equation}
Only a phenomenological form for two out of three of these gravitational couplings is necessary to fully describe large-scale structures' dynamics once a background cosmology choice has been made \citep{Pogosian:2010tj}.

The PMG framework has been implemented in \texttt{MGCAMB}~\citep{Zhao:2008bn,Hojjati:2011ix,Zucca:2019xhg} and \texttt{MGCLASS II} \citep{Sakr:2021ylx}, a patch to the Einstein--Boltzmann solvers \texttt{CAMB}~\citep{Lewis:1999bs} and \texttt{CLASS}~\citep{Blas:2011rf}, respectively. The user can choose among the built-in functional forms of $\mu_{\rm mg}$, $\eta$, $\Sigma_{\rm mg}$, and the dark energy equation of state or implement specific choices. \Cref{Sec:modelpotential} presents the functional forms adopted for this paper.

%---------------------
\subsection{Effective Field Theory of Dark Energy and Modified gravity}\label{Sec:theoryEFT}
%---------------------

The EFT of DE/MG provides a model-independent approach capable of encompassing all single-field DE/MG models \citep[see][for a review]{2013JCAP...02..032G,2013JCAP...08..010B,Frusciante:2019xia}. It allows us to describe the evolution of the background and linear perturbations in such a class of models. This work focusses on the description of EFT in the so-called $\alpha$-basis developed by \cite{2014JCAP...07..050B}. This description encodes specific physical properties of the Horndeski formalism \citep{Horndeski:1974wa} in four free coupling functions of time, namely $\alpha_{\rm M}(t)$, $\alpha_{\rm B}(t)$, $\alpha_{\rm K}(t)$ and $\alpha_{\rm T}(t)$. 

In the $\alpha$-basis, the quadratic action of EFT that covers Horndeski models can be written in the Arnowitt--Deser--Misner formalism \citep{Arnowitt:1959ah} as \citep{Gleyzes:2014rba,Pogosian:2016pwr}
\begin{align}\label{alphageneralized}
S=&\int{}{\rm d}^3x \, {\rm d}t\,a^3\,\frac{M^2}{2}\bigg\{\delta N\,\delta_1 \mathcal{R} + 2H\alpha_{\rm B}\,\delta N\,\delta K + \delta K^i_j\,\delta K^j_i\nonumber\\
&- \left(\delta K\right)^2 + \alpha_{\rm K} H^2\left(\delta N\right)^2+\left(1+\alpha_{\rm T}\right)\,\delta_2 \left[\mathcal{R}\,\delta \left( \! \! \sqrt{\zeta} \right) \right]\bigg\}\mathcomma 
\end{align}
where $H \equiv \Dot{a}/a$ is the Hubble rate with a dot standing for the derivative with respect to cosmic time $t$, $M^2$ is the effective Planck mass\footnote{The effective Planck mass represents a generalization of the Planck mass and hence of the gravitational constant in natural units. It is denoted by $M^2$ because it has the dimensions of a mass squared.}, $\delta N$ is the perturbation of the lapse function $N (t,x^i)$, $\mathcal{R}$ is the three-dimensional Ricci scalar, and $K^i_j$ and $K \equiv K^{i}_{i}$ are the extrinsic curvature tensor and its trace, respectively. Additionally, $\delta_2$ refers to taking the expansion at the second order in perturbations, and $\zeta$ is the determinant of the spatial metric $\zeta_{ij}$.

Each $\alpha$-function that composes the above basis then has a specific physical interpretation \citep{2014JCAP...07..050B} as follows:
\begin{itemize}

\item $\alpha_{\rm B}$ is designated as the `braiding function' and defines the kinetic mixing between the metric and the DE field, hence affecting the clustering properties of DE.

\item $\alpha_{\rm K}$ is called `kineticity', and it is relevant for defining the stability space of a given model (the `ghost condition'), but its effect cannot be measured because it is dominated by cosmic variance \citep{Frusciante:2018jzw}.

\item $\alpha_{\rm T}$ is known as the `tensor speed excess', and it defines departures in the GW speed of propagation from the speed of light. It is constrained to be $\alpha_{\rm T} < \mathcal{O}(10^{-16})$ at $z=0$ \citep{LIGOScientific:2017zic}, which has also been considered a strong constraint on the MG theories \citep{Creminelli:2017sry,Baker:2017hug,Ezquiaga:2017ekz}. However, since the energy scales observed by the LVK (LIGO-Virgo-Kagra) collaboration lie very close to the typical cutoff of several MG models, the translation of the $\alpha_{\rm T}$ constraint to a cosmological setting is non-trivial \citep{deRham:2018red}.

\item $\alpha_{\rm M}$ is the `running Planck mass' which can be defined from the previously introduced effective Planck mass, $M^2$, as
\begin{equation}
\alpha_{\rm M}=\frac{1}{H}\frac{{\rm d}\ln M^2}{{\rm d}\ln t}\mathcomma
\end{equation}
thus, a function characterising the evolution rate of the effective Planck mass can introduce modifications to the growth of structures.
\end{itemize} 

The original $\alpha$ basis described here was later generalised to include other classes of theory \citep{Gleyzes:2014qga,Gleyzes:2014rba,Frusciante:2016xoj,Lagos:2017hdr}. 
The EFT formalism also allows one to obtain analytical expressions for the modified gravitational couplings $\mu_{\rm mg}$, $\Sigma_{\rm mg}$, and $\eta$ under the QSA. For the particular cases we will consider, where $\alpha_{\rm T} = 0 = \alpha_{\rm M}$, one has $\mu_{\rm mg} =\Sigma_{\rm mg}$ and the phenomenological function $\Sigma_{\rm mg}$ expressed in terms of the $\alpha$-basis reads \citep{Pogosian:2016pwr}
\begin{equation}
\Sigma_{\rm mg}(a)=1+\frac{\alpha_{\rm B}^2}{2\alpha c_{\rm s}^2} \mathcomma
\end{equation}
with $\alpha = \alpha_{\rm K} + \left( 3/2 \right) \alpha_{\rm B}^2$ and
\begin{equation}
    \alpha c_{\rm s}^2 = 2\left[ \left( 1-\frac{\alpha_{\rm B}}{2} \right) \left(\frac{\alpha_{\rm B}}{2}-\frac{\dot{H}}{H^2} \right)+\frac{\dot{\alpha}_{\rm B}}{2H}-\frac{3}{2}\Omega_{\rm m}(a)\right] \mathcomma
\end{equation}
where $\Omega_{\rm m}(a)$ is the energy density of the matter components and $c_{\rm s}^2$ the sound speed of the perturbations in units of the speed of light.

The EFT formalism described in this section has been implemented in Einstein--Boltzmann codes, such as \texttt{EFTCAMB} \citep{Hu:2013twa}, \texttt{HiCLASS} \citep{Zumalacarregui:2016pph} and \texttt{COOP} \citep{Huang:2012mt}, which solve linear perturbation equations. Such modified codes have been validated and shown to agree at the subpercent level \citep{Bellini:2017avd}. \Cref{Sec:modelEFT} defines the parameterisations we choose for the $\alpha$ functions.

%%%%%%%%%%%%%%%%%%%%%%
\section{Choices of the parameterised methods for the modified gravity models} \label{sec:methods}
%%%%%%%%%%%%%%%%%%%%%%

%---------------------
\subsection{Phenomenological approach to the modification of the gravitational couplings} \label{Sec:modelpotential}

%---------------------
For the modified gravitational couplings $\mu_{\rm mg}$ and $\Sigma_{\rm mg}$ we consider a scale-independent parameterisation.
Since \Euclid's main observables are mostly sensitive to the late-time evolution of the background, we choose a late-time parameterisation similar to the one adopted in \citet{DES:2018ufa},
\begin{eqnarray}
\mu_{\rm mg}(z) = 1 + \frac{\Omega_{\rm DE}(z)\, }{ \, \Omega_{\rm DE}(z=0)} \, \mu_0\mathcomma \label{eq:plkmu}\\
\Sigma_{\rm mg}(z) = 1 + \frac{\Omega_{\rm DE}(z)\, }{ \, \Omega_{\rm DE}(z=0)} \, \Sigma_0\mathcomma\label{eq:plksigma}
\end{eqnarray}
where $\mu_0$ and $\Sigma_0$ are constants that control the amplitude of the deviation from GR, and $\Omega_{\rm DE}(z)$ is the DE density parameter that we assume to evolve as in a \lcdm\ background.
In \cref{sec:results} we will show constraints on the present values of the parameters in Eqs.~\eqref{eq:plkmu} and \eqref{eq:plksigma}, which we can relate to the MG parameters as
\begin{equation}
\bar{\mu}_0=\mu \,(z=0)=1+\mu_0\mathcomma \quad \bar{\Sigma}_0=\Sigma\,(z=0)=1+\Sigma_0 \mathperiod
\end{equation}

In this work, we consider two fiducial sets of values for the $\{\mu_0,\Sigma_0\}$ parameters: one coinciding with the GR limit, which we will refer to as PMG-1; and a second one for which the fiducial values are far from GR, dubbed PMG-2. Details will be given in \cref{sec:fisher}.

\citet{Planck:2018vyg} provided constraints on the $\mu_{\rm mg}(z)$ and $\Sigma_{\rm mg}(z)$ functions also in combination with external data, which, converted to our formalism\footnote{It must be pointed out that in \citet{Planck:2018vyg} the $\mu_{\rm mg}(z)$ and $\eta(z)$ functions are the independent ones, while $\Sigma_{\rm mg}(z)$ is obtained via Eq.~\eqref{eq:sigderived}.}, are $\mu_0=-0.07^{+0.19}_{-0.32}$ and $\Sigma_0=0.018^{+0.059}_{-0.048}$. \citet{DES:2022ccp} reported instead a bound of $\mu_0=-0.4\pm0.4$ and $\Sigma_0=-0.06^{+0.09}_{-0.10}$, inferred from measurements from the Dark Energy Survey's first three years of observations, together with external data that include redshift space distortions (RSDs), baryonic acoustic oscillations (BAOs) and supernovae (SNe) measurements.  

This parameterisation has also been used to obtain forecast constraints for upcoming experiments. As an example, \citet{Casas:2022vik} exploited the spectroscopic and continuum observables from the SKA Observatory (SKAO), finding that $\mu_0$ and $\Sigma_0$ can be constrained, respectively, at the $2.7\%$ and $1.8\%$ levels. A discussion of other possible constraints, achievable with upcoming surveys, on this and other parameterisations, can be found in \citet{Martinelli:2021hir}.

We implemented the parameterisation used by \citet{DES:2018ufa} in \texttt{MGCLASS II} \citep{Sakr:2021ylx} and in a private independent internally modified version of the \texttt{CLASS} code. We compared the predictions of the linear matter power spectrum from both codes and found sub-percent level agreement.

%---------------------
\subsection{Effective Field Theory modified gravity models}\label{Sec:modelEFT}

%---------------------
For a general scalar-tensor model, it might not be possible to specify a generic analytical expression for the equation of state $w(a)$ or for the Hubble function $H$ and for the perturbation functions $\alpha_{i}$. For this reason, it is often convenient to assume simple parameterisations, where the $\alpha_{i}$ functions are proportional to the energy density of the scalar field $\Omega_{\phi}$ \citep{Bellini:2015xja,Gleyzes:2015rua,Alonso:2016suf,Linder:2016wqw,Gleyzes:2017kpi,Noller:2018wyv}, to the scale factor $a$ \citep{Gleyzes:2017kpi,Noller:2018wyv,Denissenya:2018mqs}, or a power of $a$, that is, $a^n$, with $n$ constant \citep{Gleyzes:2017kpi,Noller:2018wyv,Denissenya:2018mqs,Lombriser:2018olq}. The exponent can be the same for all functions or chosen separately for each. Even though these functional forms are elementary, they can still be helpful for a glimpse into the physical evolution of some observables under the action of a DE component or modifications to the gravitational sector.
Regarding the background expansion, it is customary to assume a given model for it, choosing either a \lcdm\ background or a varying equation of state for DE, usually parameterised through the $w_0$ and $w_a$ parameters following the Chevalier--Polarski--Linder \citep[CPL,][]{Chevallier:2000qy,Linder:2002et} parameterisation
\begin{equation}\label{eq:cpl}
    w(a) = w_0+w_a(1-a)\mathperiod
\end{equation}

In this work, we will consider $\alpha_{\rm M}=0=\alpha_{\rm T}$, the latter being in agreement with results from gravitational wave observations of the neutron star merger GW170817 \citep{LIGOScientific:2017adf,LIGOScientific:2017zic,LIGOScientific:2017vwq}. The choice of $\alpha_{\rm M}=0$ is motivated by the fact that we want to reduce the degeneracy between the MG parameters. In the literature, it is common to use the relation $\alpha_{\rm M}=\beta \alpha_{\rm B}$ where $\beta$ can be, for example, $\beta=-1$ \citep{Planck:2015bue,Planck:2018vyg} for conformally coupled models such as $f(R)$-gravity, Brans-Dicke, and chameleon theories, or $\beta=-1/2$ \citep{Linder:2018jil,Brush:2018dhg} for No Slip Gravity models. In this work, we decided to investigate a different class of theories since previous work by the \Euclid collaboration has already investigated a similar class of theories \citep{Casas23a,Frusciante23}. Additionally, the $\alpha_{\rm K}$ function cannot be constrained by data at leading order \citep{Bellini:2015xja,Frusciante:2018jzw}. Therefore, we will fix it to a constant value. Thus, we are left with only one free function, $\alpha_{\rm B}$, and the choice of background evolution, $H$. When choosing a background evolution different from that of $\Lambda$CDM we decided to keep the parameters of the DE equation of state fixed. This means that in our analysis we do not provide forecasts for those parameters because we are interested in constraining the modifications to the gravitational interaction that affect only the linear perturbations. We present our choices in the following subsections.

\paragraph{\textit{Linear scale factor $\alpha_{\rm B}$ parameterisation on a \lcdm\ background (EFT-1)}:}
we will assume 
\begin{equation}
 \alpha_{\rm B}=\alpha_{\rm B,0}\,a\,,\label{eq:EFT1}
\end{equation}
with $\alpha_{\rm B,0}$ being a constant and we also consider a \lcdm\ background. We use $\alpha_{\rm K,0}=10$, as the fiducial value. Let us stress that $\alpha_{\rm K}$ cannot be constrained by the data \citep{Bellini:2015xja,Frusciante:2018jzw} and does not affect the constraints on the other parameters as shown in \cite{Bellini:2015xja}. However, $\alpha_{\rm K,0}$ changes the viable parameter space \citep{Frusciante:2018jzw} because it enters in the definition of the ghost condition. In this work, we selected $\alpha_{\rm K,0}=10$ to closely follow \cite{Traykova:2021hbr} and to allow for a large parameter space. 
For this scenario, constraints were obtained by \cite{Noller:2018wyv}, whose analysis we repeat here using the cosmic microwave background (CMB) updated to Planck 2018 \citep{Planck:2018vyg}, BAO, RSD and the matter power spectrum from SDSS DR4 luminous red galaxies (LRG), but fixing $\alpha_{\rm M,0}=0$. Considering stability conditions,  in particular imposing the speed of sound of the extra degree of freedom to be positive, implies $\alpha_{\rm B,0} \geq 0$.
In this case, we find that the constraints on $\alpha_{\rm B,0}$ are $\alpha_{\rm B,0} = 0.09^{+0.04}_{-0.05}$. This posterior for $\alpha_{\rm B,0}$ is mildly non-Gaussian and the above limits encompass $68\%$ of cases, so we take these bounds as a proxy to be compared with forecasted $1\,\sigma$ bounds later.
Note how these bounds differ from previous constraints found when marginalising over $\alpha_{\rm M,0}$ (as opposed to fixing its value, as we do here) -- see the parameterisation II of \cite{Noller:2018wyv}, where $\alpha_{\rm B,0}<1.3$ was found, and \cite{Seraille:2024beb}, where $\alpha_{\rm B,0}<0.6$ was obtained by using galaxy auto- and cross-correlations (especially as related to the integrated Sachs-Wolfe effect) in addition to the above data sets. The tighter bounds found in our analysis here can be explained by the strong degeneracy between the two parameters (and hence the strong effect of fixing $\alpha_{\rm M,0}$), as we see from Fig.~1 in \cite{Noller:2018wyv} or \cite{Noller:2020afd}.

The choice of the  $\alpha$-parameterisation in Eq.~\eqref{eq:EFT1} is implemented in both \texttt{EFTCAMB} and \texttt{hi\_class}. The validation of the implementations was extensively explored in \cite{Bellini:2017avd}, with the two codes showing sub-percent level agreement. This paper uses the linear predictions produced with \texttt{EFTCAMB}.

\paragraph{\textit{A shift-symmetric scalar-tensor model on a $w_0w_a$ background (EFT-2):}}
we will assume the parameterisation proposed in \citet{Traykova:2021hbr} to encompass shift-symmetric scalar-tensor theories \citep{Babichev:2011iz,Pirtskhalava:2015nla,Finelli:2018upr}, which is
\begin{equation}
 \alpha_{\rm B} = \alpha_{\rm B,0}\left(\frac{H_0}{H}\right)^{4/m}\mathcomma\label{eq:EFT2}
\end{equation}
with $m$ and $\alpha_{\rm B,0}$ being free constants and $H_0=H(z=0)$. For the background, it is shown in \cite{Traykova:2021hbr} that the equation of state can be assumed to follow the CPL parameterisation of Eq.~\eqref{eq:cpl}.

As fiducial values, we assume $\alpha_{\rm K,0}=10$, $\alpha_{\rm B,0}=0.9$, and $m=2.4$. The value for $m$ is the best fit obtained by \cite{Traykova:2021hbr};  for $\alpha_{\rm B,0}$ we consider the value that corresponds to the upper limit at 68\%, as found in their analysis.
For the background equation of state, we use the best fit values $w_0 = -0.97$ and $w_a = -0.11$, and we hold them fixed. Accordingly, the equation of state would necessarily undergo phantom crossing at -1.

To constrain the values of the free parameters, \cite{Traykova:2021hbr} used the auto- and cross-correlation of temperature and polarisation fluctuations and the lensing potential together with BAO and RSD measurements. In addition, the authors also considered SNIa data for two different cases, one in which the cosmological constant is absent, that is, when the scalar field contributes all of the dark energy component, and another in which the cosmological constant is present and, therefore, the dark energy component is given by the combined contributions of the scalar field and the cosmological constant. Here we only consider the case with $\Lambda = 0$, for which the authors found $\alpha_{\rm B,0} = 0.6 \pm 0.3$ and $m = 2.4 \pm 0.4$ at the $68\%$ confidence level. It follows that $\alpha_{\rm B,0} = 0$ is found to be excluded at $2\sigma$ (see e.g.  Fig. 11 in \cite{Traykova:2021hbr}. 

We implemented the parameterisation in Eq.~\eqref{eq:EFT2} in the \texttt{EFTCAMB} and \texttt{hi\_class} codes. We compared the predictions of the linear matter power spectrum up to $k = 8 \, h$ Mpc$^{-1}$ from both codes and reached sub-percent level ($\lesssim 0.25 \%$) agreement. We then used the theoretical predictions obtained with \texttt{EFTCAMB} for this work.

%%%%%%%%%%%%%%%%%%%%%%
\section{Theoretical predictions for \Euclid observables} \label{sec:thpred}

In this analysis, we follow the procedure presented in \citet{Blanchard-EP7} to compute the theoretical predictions for the different observables with \Euclid specifications. Given that we consider extended models beyond the baseline cases considered in~\citet{Blanchard-EP7}, we also rely on the modifications described in~\citet{Casas23a} and \citet{Frusciante23}. Given that all the extended models considered in this analysis are scale independent, we closely follow the latter and refer the interested reader to it for the technical details. However, we briefly describe the main steps for self-consistency in the following.
%---------------------
\subsection{Spectroscopic survey}\label{sec:spect}
%---------------------
For the spectroscopic survey, the main observable considered in this analysis is the observed galaxy power spectrum, which can be expressed as
\begin{multline}
P_\text{obs}(k_\text{ref}, \mu_{\theta,\text{ref}} ;z) = 
\frac{1}{q_\perp^2(z)\, q_\parallel(z)} %AP
\left\{\frac{\left[b\sigma_8(z)+f\sigma_8(z)\mu_{\theta}^2\right]^2}{1+ \Big[ f(z)\ k\ \mu_{\theta}\ \sigma_{\rm p}(z) \Big]^2 } \right\} 
\\
\times \frac{P_\text{dw}(k,\mu_{\theta};z)}{\sigma_8^2(z)}  %nonlinear damping
F_z(k,\mu_{\theta};z) %z-error
+ P_\text{s}(z) \mathcomma % shot noise
\label{eq:GC:pk-ext}
\end{multline}
where $\mu_{\theta}$ is the cosine of the angle between the wavevector $\bm{k}$ and the line-of-sight direction. The quantities on the right-hand side are functions of their counterparts for a reference cosmology, that is, $k \equiv k(k_{\rm ref})$, $\mu_{\theta} \equiv \mu_{\theta,\text{ref}}$, which are transformed due to the Alcock--Paczynski effect \citep{1979Natur.281..358A}. The prefactor with $q_{\perp}$ and $q_{\parallel}$ also accounts for the same effect on the overall $P_\text{obs}$, see \cite{Blanchard-EP7} for the explicit formula. The term within curly brackets contains the redshift-space distortion contribution correcting for the `fingers-of-God' effect with the $\sigma_{\rm p}$ nuisance parameters, one for each redshift bin. The terms $b\sigma_8$ and $f\sigma_8$ are the product of the galaxy bias $b$ and the growth rate $f$ and the normalisation of the matter power spectrum $\sigma_8$. The $P_{\rm dw}(k,\mu_{\theta};z)$ term accounts for the smearing of the BAO features with
\begin{equation}
    P_{\rm dw}(k,\mu_{\theta};z)=P_{\delta\delta}^{\rm lin}(k;z)\text{e}^{-\sigma_{\rm v}k^2} + P_{\rm nw}(k;z)\left(1-\text{e}^{-\sigma_{\rm v}k^2}\right)\mathcomma
    \label{eq:GC-BAOdamp}
\end{equation}
where $P_{\delta\delta}^{\rm lin}(k;z)$ stands for the linear matter power spectrum, $P_{\rm nw}(k;z)$ is a power spectrum with only the broadband shape (no BAO feature), and the $\sigma_{\rm v}$ nuisance parameters, one for each redshift bin, are evaluated along with $\sigma_{\rm p}$ at the fiducial following:
\begin{align}
\sigma^2_{\rm v}(z, \mu_{\theta}) &= \frac{1}{6\pi^2}\int\de k\, P_{\delta\delta}(k,z)\left\{1 - \mu_{\theta}^2 + \mu_{\theta}^2\left[1+f(k,z)\right]^2\right\},\\
\sigma_{\rm p}^2(z) &= \frac{1}{6\pi^2}\int\de k\, P_{\delta\delta}(k,z)f^2(k,z)\,.
\end{align} 
Finally, the $F_z(k,\mu_{\theta};z)= \text{exp}\left[-k^2\mu_{\theta}^2\sigma_r^2(z)\right]$ terms account for the smearing of the galaxy density field along the line of sight due to the spectroscopic redshift uncertainty, which is given by $\sigma_r(z)=c\,(1+z)\sigma_{0,z}/H(z)$, $\sigma_{0,z}$ is the uncertainty on the measured redshifts, and $P_{\rm s}(z)$ represents any residual shot noise and enters as a nuisance parameter with a fiducial value of zero.

For the inclusion of non-linearities in the spectroscopic observable we are using a phenomenological semi-analytic model, which aims at describing the most dominant nonlinear effects in the redshift-space galaxy power spectrum, those being the fingers-of-God effect in Eq.~(\ref{eq:GC:pk-ext}) and the BAO damping in Eq.~(\ref{eq:GC-BAOdamp}). This model is fully described in \cite{Blanchard-EP7} and has recently been tested against the one-loop effective field theory of large-scale structure in \cite{Linde:2024uzr}.

%---------------------
\subsection{Photometric survey} \label{sec:photo}
%---------------------

The main observables considered for the photometric survey are the harmonic angular power spectra $C_{ij}^{XY}(\ell)$, where $i$ and $j$ refer to two tomographic redshift bins, and $X$ and $Y$ stand for two of the photometric probes. These are either galaxy clustering (GC$_{\rm ph}$) or weak lensing (WL). We note that when combining the two of them, we effectively include the third probe into the analysis, that is, galaxy-galaxy lensing or simply the cross-correlation (XC). In the following, we will refer to the entire combination of these photometric observables as 3\texttimes2\,pt. statistics. 

Within the Limber approximation \citep{{Limber1953}}, and assuming vanishing spatial curvature (that is a vanishing curvature parameter, $\Omega_K=0$), the angular power spectra are given by
\begin{equation}
    C_{ij}^{XY}(\ell)=c\int_{z_{\rm min}}^{z_{\rm max}}\text{d}z\frac{W_i^X(z)\,W_j^Y(z)}{H(z)\,r^2(z)}P_{\delta\delta}(k_{\ell},z)\mathcomma
\end{equation}
where $k_{\ell}=(\ell+1/2)/r(z)$, with $r(z)$ representing the comoving distance as a function of redshift, and $P_{\delta\delta}(k_{\ell},z)$ stands for the nonlinear matter power spectrum evaluated at a wavenumber $k_{\ell}$ and redshift $z$. The kernels, or window functions, are given by
\begin{align}
    W_i^{\rm G}(k,z)=& b_i(k,z)\frac{n_i(z)}{\bar{n}}\frac{H(z)}{c}\mathcomma\\
    W_i^{\rm L}(k,z) =& \frac{3}{2}\Omegam \frac{H_0^2}{c^2}(1+z)\,r(z)\,\Sigma_{\rm mg}(z)\nonumber\\
    & \times \int_z^{z_{\rm max}}{\de z'\frac{n_i(z')}{\bar{n}_i}\frac{r(z') - r(z)}{r(z')}} + W^{\rm IA}_i(k,z)\mathcomma
\end{align}
for GC$_{\rm ph}$ and WL, respectively. The ratio $n_i(z)/\bar{n}$ represents the normalised galaxy distribution as a function of redshift, while $b_i(k,z)$ is the galaxy bias in the $i$-th tomographic bin. The galaxy bias is assumed to be constant within each redshift bin, with fiducial values $b_i = \sqrt{1 + \bar{z}_i}$, and $\bar{z}_i$ the bin centre, but we note that when adding very small scales into the analysis, as is the case here, a more detailed modelling of the galaxy bias is required such as through the use of perturbation theory, which introduces a nonlinear and nonlocal galaxy bias \citetext{\citealp{Paranjape:2013wna,Lazeyras:2015lgp,BOSS:2016off}, or more recently, \citealt{Tutusaus20}}. This would add some degeneracy between the MG and the bias parameters. However, perturbation theory breaks down at our scales, and dedicated simulations for our models at smaller scales are still lacking to calibrate the bias. Therefore, we choose to model the bias in each redshift bin as a free parameter and rely on the inclusion of weak lensing information and the XC between the lensing and the galaxy clustering probes to simultaneously obtain constraints on the cosmological and model parameters, while also auto-calibrating the galaxy bias itself.
We also note that the effects of extended models beyond $\Lambda$CDM will be encapsulated in $\Sigma_{\rm mg}(z)$ for the lensing potential, in $H(z)$ and $r(z)$ for the background, and in the matter power spectrum $P_{\delta \delta}(k_{\ell},z)$. The intrinsic alignment contribution to the weak lensing kernel enters through the $W_i^{\rm IA}(k,z)$ term. Following \cite{Blanchard-EP7}, \cite{Casas23a}, and \cite{Frusciante23} we consider the extended nonlinear alignment model with three free parameters: the amplitude of intrinsic alignments, $A_{\rm IA}$; their redshift evolution, $\eta_{\rm IA}$; and their luminosity dependence, $\beta_{\rm IA}$.

%---------------------
\subsection{Nonlinear modelling for \texorpdfstring{3\texttimes2\,pt. statistics}{3x2}}\label{sec:nonlinear-WL}

As opposed to spectroscopic galaxy clustering, photometric probes require the theoretical modelling to extend deeply into the nonlinear regime of the matter power spectrum.
In the following subsections, we comment on our prescriptions for EFT and PMG. 
In some MG models, it was shown that baryonic effects and MG effects can be treated independently \citep{Arnold:2019zup}, and we can use the baryon models developed in $\Lambda$CDM to take into account baryonic effects. However, it is not clear whether this assumption is valid in the case of cosmologies based on phenomenological approaches. In our paper, we will use conservative scale cuts to address the uncertainties in nonlinear modelling in MG models, and baryonic effects are not important at the scales we consider.

For our forecasts, we use the halo-model reaction approach \citep{Cataneo:2018cic}, which has been compared in detail against $N$-body simulations for both specific models \citep{Cataneo:2018cic,Cataneo:2019fjp,Bose:2021mkz,Adamek23} and more phenomenological cases \citep{Srinivasan:2021gib,Carrilho:2021rqo,Parimbelli:2022pmr,Srinivasan:2024}. We briefly summarise the halo-model reaction approach in \cref{sec:nl-hmrapproach}. 

We also study the impact on nonlinear predictions caused by the possible existence of screening, which is a mechanism that removes the effects of modified gravity in particular environments \citep[see, e.g.,][for recent reviews]{Koyama:2015vza,Brax:2021wcv}. 
Since the forecasts in this work are focused on constraining more phenomenological or parameterised forms of modified gravity, we extend the halo-model reaction approach to cover two limiting cases, namely the unscreened (US) and super-screened (SS) cases. The former assumes that no screening is operating, allowing for the maximum deviation from GR. In contrast, the latter assumes that the screening is strong enough to thoroughly screen single halos from modified gravity effects. In practice, this means that the halo model power spectrum's `1-halo-term' is given by the $\Lambda$CDM prediction. The vast majority of modified gravity phenomenology should fall between these two cases; we explain the cases in more detail in \cref{sec:ss&us}.
In the case of EFT, assuming that there are no new parameters from higher-order interactions \citep{Cusin:2017wjg}, it is possible to model nonlinearities consistently, for example, by using the spherical collapse model \citep{Albuquerque:2024hwv}  or computing the effective Newton constant \citep{Brando:2023fzu}. In this paper, to account for the uncertainties due to higher-order interactions, we consider two extreme cases for screening. The consistent nonlinear prediction falls between these two cases. This analysis could be used to inform scale cuts in future work and to exclude these uncertainties.

In terms of the choice of parametrised approaches, one of the advantages of the PMG parametrisation is that, although originally defined perturbatively, it can be generalised to large density contrasts \citep{Thomas:2020duj,Hassani:2019wed}. Thus, the definition of $\mu_{\rm mg}$ and $\Sigma_{\rm mg}$ is valid on all cosmological scales, although some care is necessary if one wants to also include the effect of screening mechanisms \citep{Thomas:2020duj}.
There is another approach to parameterised modified gravity that is fundamentally defined to include nonlinear scales \citep{Sanghai:2016tbi,Clifton:2018cef}. 
However, the predictions of this approach for nonlinear scales have yet to be computed, so we do not use it here.

The situation is more complicated when focussing on the EFT approach. Here, the model-independent parameters are coefficients of a perturbative expansion of invariants at the level of the action, and a finite number of coefficients is sufficient to describe the linear theory. It is as yet unclear how to extend this to the fully nonlinear regime (where an arbitrary density contrast is allowed), and how many coefficients (finite or infinite) may be required in this case. Instead, we simply truncate the perturbative expansion at the linear level and assume that the correspondence between PMG and EFT at the linear level can be extended to the nonlinear regime.

%.....................
\subsubsection{Halo-model reaction} \label{sec:nl-hmrapproach}
%.....................

We employ the halo model and 1-loop perturbation theory approach of \cite{Cataneo:2018cic}, which models the nonlinear matter power spectrum as 
\begin{equation}
 P_{\rm NL}(k,z) = \mathcal{R}(k,z) \, P^{\rm pseudo}_{\rm NL}(k,z)\mathperiod
 \label{eq:nonlinpk}
\end{equation}
Here, $P^{\rm pseudo}_{\rm NL}(k,z)$ is the pseudo-model power spectrum, which is defined as the nonlinear spectrum in a $\Lambda$CDM universe, but with initial conditions such that its linear clustering matches the beyond-$\Lambda$CDM universe at some target redshift, $z$. This guarantees that the halo mass functions in the target and pseudo-model universes are similar, yielding a smoother transition of the power spectrum between the inter- and intra-halo regimes. The pseudo-model power spectrum can be modelled using nonlinear formulas, such as \texttt{HMCode} \citep{Mead:2015yca,Mead:2016zqy,Mead:2020vgs}, which take as input a specified linear power spectrum and allow the user to specify the linear clustering while keeping the nonlinear specifications fixed to $\Lambda$CDM, according to the pseudo-model spectrum definition. 

The second component is the halo-model reaction, $\mathcal{R}(k,z)$. This gives the corrections to the pseudo spectrum coming from nonlinear beyond-$\Lambda$CDM physics. We can write this as \citep{Cataneo:2019fjp,Bose:2021mkz,Bose:2022vwi} 
\begin{equation}
    \mathcal{R}(k)=\frac{\left(1-f_{\nu}\right)^{2} P_{\mathrm{hm}}^{(\mathrm{cb})}(k)+2 f_{\nu}\left(1-f_{\nu}\right) P_{\mathrm{hm}}^{(\mathrm{cb} \nu)}(k)+f_{\nu}^{2} P_{\mathrm{L}}^{(\nu)}(k)}{ P_{\rm hm}^{\rm pseudo}(k,z)} \, ,
    \label{eq:reaction}
\end{equation}
with the subscript ‘hm’ representing the halo model, ${\rm (m)} \coloneqq (cb+\nu) $, with cb for CDM and baryons, $\nu$ for massive neutrinos, and $f_\nu = \Omega_{\nu,0}/\Omega_{\rm m,0}$ being the energy density fraction of massive neutrinos at $z=0$. The matter density fraction is given by $\Omega_{\rm m}(a)=\Omega_{\rm{m},0} H_0^2/\left[a^3 H^2(a)\right]$. Massive neutrino effects are linearly modelled via a weighted sum of the nonlinear cb halo model and linear massive neutrino spectra \citep{Agarwal:2010mt,Adamek23}. Here we fix the neutrino mass to $0.06~{\rm eV}$. 

The components of the reaction are given by 
\begin{align}
    P_{\mathrm{hm}}^{(\mathrm{cb} \nu)}(k) \approx &\, \sqrt{P_{\mathrm{hm}}^{(\mathrm{cb})}(k) \, P_{\mathrm{L}}^{(\nu)}(k)} \, , \\ 
    P_{\mathrm{hm}}^{(\mathrm{cb})}(k) =&\, P_{\mathrm{L}}^{(\mathrm{cb})}(k)+P_{\mathrm{1h}}^{(\mathrm{cb})}(k)  \, , \label{eq:1hcb}  \\  
  P_{\rm hm}^{\rm pseudo}(k,z) = &\,   P_{\rm L} (k,z) + P_{\rm 1h}^{\rm pseudo}(k,z)\,. \label{Pk-halos} 
  \end{align}
We note here that we have omitted the 1-loop standard perturbation theory (SPT) corrections found in the original formalism \citep{Cataneo:2018cic}. It has been noted in \cite{Bose:2022vwi} that this correction is likely a small effect in theories that do not induce scale dependency in $\mu_{\rm mg}(k,z)$, which is the case considered in this paper.

The calculation of $\mathcal{R}$ thus only requires the computation of linear and spherical collapse predictions for the overdensity, $\delta$. In particular, these computations require the specification of the Poisson equation, which relates $\delta$ to the Newtonian potential $\Phi$:
\begin{align}
-\left(\frac{c \, k}{a H(a)}\right)^2\Phi_{\rm L} (\bfkk,a)= &
\frac{3 \Omega_{\rm m}(a)}{2} \mu_{\rm mg}(k,a) \, \delta_{\rm L}(\bfkk,a)\, , \label{eq:poisson1} \\ %+ S(\bfkk,a) \, , \label{eq:poisson1} \\ 
-\left(\frac{c \, k}{a H(a)}\right)^2\Phi_{\rm NL} (\bfkk,a)= &
\frac{3 \Omega_{\rm m}(a)}{2} [1+ \mathcal{F}(k,a)] \, \delta_{\rm NL}(\bfkk,a) \, ,
\label{eq:poisson2}
\end{align} 
where L and NL denote `linear' and `nonlinear', respectively. 
$\mathcal{F}$ is the nonlinear modification of the Poisson equation, which should have the property $ \lim_{k \rightarrow 0} \mathcal{F} \rightarrow \mu_{\rm mg}-1$. This work focusses on two limiting cases for the nonlinear power spectrum, US and SS, for which we describe our choices of $\mathcal{F}$ in \cref{sec:ss&us}. We use the {\tt ReACT} code \citep{Bose:2020wch,Bose:2022vwi} to compute Eq.~(\ref{eq:reaction}). 

The halo-model reaction has been shown to provide reasonable accuracy for many MG models. The reaction $\mathcal{R}$ is typically accurate on the order of $2\%$ for $k\leq 3 \, h \, {\rm Mpc}^{-1}$ \citep{Cataneo:2018cic}. The \texttt{HMCode2020} pseudo-spectrum is comparable, being accurate at about $2.5\%$ over the same range \citep{Mead:2020vgs}. This gives roughly a $5\%$ accuracy to the prediction of Eq.~(\ref{eq:nonlinpk}). On the other hand, the $\Lambda$CDM-based \texttt{EuclidEmulator2} \citep{Knabenhans-EP2} is about $1\%$ accurate for $k\leq 10\,h \, {\rm Mpc}^{-1}$. Furthermore, the $5\%$ accuracy for Eq.~(\ref{eq:nonlinpk}) assumes an exact prediction for $\mathcal{F}$ for a choice of cosmology and $\mu_{\rm mg}(k, a)$ in Eq.~(\ref{eq:poisson2}). Since we wish to forecast model-independent constraints, this will inevitably limit the scales that we can reliably use. 

%.....................
\subsubsection{Super-screened and Unscreened cases} \label{sec:ss&us}
\begin{figure*}[ht!]
    \centering
    \begin{tabular}{cc}
    \includegraphics[width=0.45\linewidth]{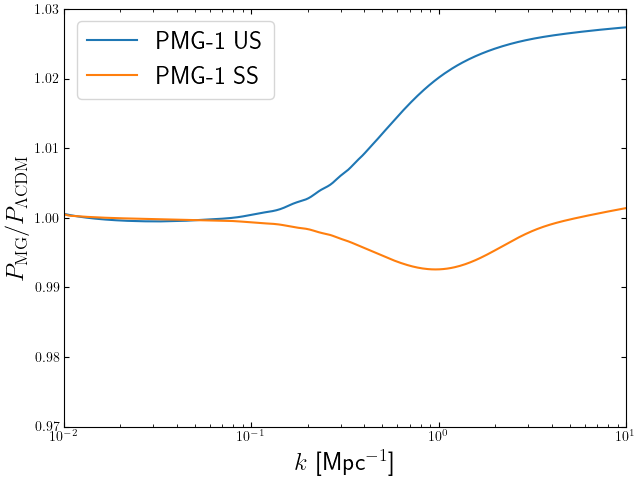}
    \includegraphics[width=0.45\linewidth]{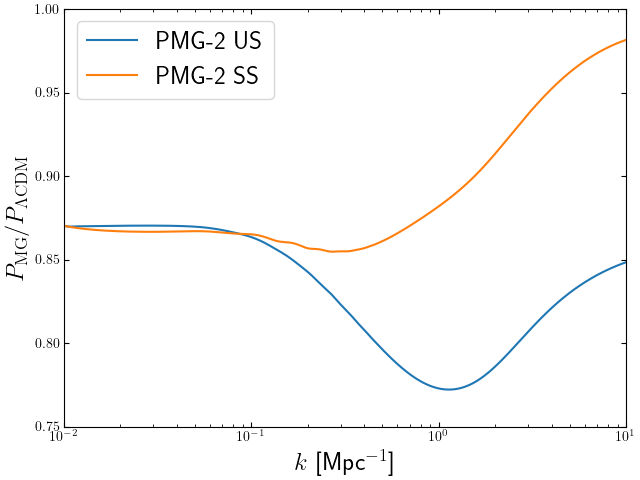}
    \end{tabular}
    \caption{We present the ratio of the modified gravity power spectrum relative to the $\Lambda$CDM power spectrum for the US and SS cases for PMG-1 (left panel) and PMG-2 (right panel) at $z=0$. Note that in the SS case the ratio smoothly goes to 1 at $k>1\,{\rm Mpc}^{-1}$ by construction. Of course, there is no such trend for the US case.} 
    \label{fig:US_vs_SS}
\end{figure*}

%.....................
For the two limiting cases that we consider, the modifications to the \texttt{ReACT} code for both the EFT and PMG approaches are the same.\footnote{Note that in general, the code would not necessarily be modified the same way for these two approaches; see \cite{Srinivasan:2021gib,Srinivasan:2024} for a more detailed discussion of how to incorporate the PMG approach into \texttt{ReACT} in the general case.}
In the US case, we do not consider any screening effects and set $\mathcal{F} = \mu_{\rm mg}(k,a)-1$ in our calculations, so we effectively remove all screening mechanisms. This represents the case where gravity is modified on all scales relevant to our analysis.
In the SS case, all modifications to gravity are screened beyond the linear level, and we return to GR as quickly as possible. To achieve this, we set $\mathcal{F}=0$, but we must also modify the pseudo-spectrum in Eq.~(\ref{eq:nonlinpk}), since it contains nonlinear, MG corrections coming from the modified linear spectrum input, essentially giving a $\Lambda$CDM prediction with a larger $\sigma_8$ than our baseline. As a workaround to this, in the SS case, we have chosen to use the $\Lambda$CDM spectra instead of the pseudo-model ones in Eqs.~(\ref{eq:nonlinpk}) and (\ref{eq:reaction}). As a result, the reaction is given by the ratio of the modified linear power spectrum to $\Lambda$CDM on linear and quasi-linear scales (and is thus not equal to 1). On nonlinear scales, the reaction proceeds smoothly to $\mathcal{R}=1$. This can be seen in Fig.~\ref{fig:US_vs_SS}.

We found good agreement with the parameterised post-Friedmann approach for both the US and SS cases. This nonlinear phenomenological model was motivated in \cite{Hu:2007pj} to parameterise screening in a particular class of MG models. Then it was used in \cite{Casas:2017eob} to perform forecasts on MG models. 

We note that the validation of the \texttt{ReACT} code against $N$-body simulations for the PMG approach, that has been carried out so far \citep{Srinivasan:2021gib,Srinivasan:2024},  applies only to the unscreened limit that we consider here, not to the superscreened limit. In addition, in those works, the validation was carried out for the case where $\mu_{\rm mg}$ is composed of piecewise constant bins in redshift in \cite{Srinivasan:2021gib}. For this forecast, additional simulations were run to validate the specific time-dependent parameterisation. \Cref{fig:kmax_sims} shows some of the details of this validation. The validity of the \texttt{ReACT} approach used here for EFT has not been tested using $N$-body simulations for either of the US or SS cases (although, as noted earlier, there are specific models whose nonlinear phenomenology is captured by \texttt{ReACT}).

%%%%%%%%%%%%%%%%%%%%%%
\section{Analysis method} \label{sec:fisher}
%%%%%%%%%%%%%%%%%%%%%%

This work forecasts the constraints achievable with \Euclid primary probes on the MG models we discuss in \cref{sec:models}. For such a purpose, we exploit the Fisher matrix method, following the approach described in \citet{Blanchard-EP7}, to which we refer the reader for a description of the formalism used to compute the Fisher matrix for the photometric and spectroscopic observables we consider, and to estimate from it the probability distributions of the cosmological parameters. We also follow the same formalism as used in \citet{Blanchard-EP7} when computing the covariance matrices in which it is assumed for the spectroscopic probe that the observed power spectrum follows a Gaussian distribution. The same assumption was also made for the photometric probe, deciding to not implement the non-Gaussian terms since even in \lcdm\, there are large theoretical uncertainties regarding how to model this quantity, both in the quasi-linear and nonlinear regime.

The first ingredient to obtain constraints with this approach is to assume a fiducial cosmological model; that is, we need to specify the values of the parameters that enter the computation of \Euclid observables.
The models we consider have in common the same set of standard parameters, that is, the total and baryonic matter abundances ($\Omega_{\rm m,0}$ and $\Omega_{\rm b,0}$), the dimensionless Hubble parameter ($h = H_0/[100\,\kmsMpc]$), the sum of neutrino masses ($\Sigma m_\nu$, which we keep fixed to 0.06 eV), the spectral index of the matter power spectrum ($n_s$) and the primordial amplitude of matter perturbations ($A_{\rm s}$), for which we choose the fiducial values reported in Table \ref{tab:fidcosmo}.
\begin{table}[h]
    \centering
    \caption{Fiducial values for the standard cosmological parameters common to all the models considered.}
    \label{tab:fidcosmo}
    \begin{tabular}{ccccc}
    \hline
    \rule{0pt}{2ex}
         $\Omega_{\rm m,0}$ & $\Omega_{\rm b,0}$ & $h$ & $n_{\rm s}$ & $A_{\rm s}\times10^9$ \\
         \hline
         \rule{0pt}{2ex}
         0.315 & 0.05 & 0.674 & 0.966 & 2.097 \\ 
    \hline
    \end{tabular}
\end{table}
In addition, the CPL parameters ($w_0$ and $w_a$), specifying the DE equation of state, need to be set, but we do this on a model-by-model basis.

The freely varying standard cosmological parameters, for which we need to compute derivatives of our observables, are
\begin{equation}\label{eq:freestdpars}
  \bm\Theta=\{\Omega_{\rm m,0},\, \Omega_{\rm b,0},\, h,\, n_{\rm s},\, \sigma_8\}\mathcomma
\end{equation}
where $\sigma_8$ is a quantification of matter density fluctuations. Notice that while the parameter varied in all our analyses is $\sigma_8$, we choose to express the fiducial of Table \ref{tab:fidcosmo} in terms of $A_{\rm s}$. The two parameters are related to each other, but, starting from the same value of $A_{\rm s}$, the resulting $\sigma_8$ will depend on the assumed cosmological model. For such a reason, we decide to keep the fiducial value of $A_{\rm s}$ constant for all the cosmologies we investigate, thus implicitly assuming the same inflationary phase of the Universe, and we derive the corresponding value of $\sigma_8$ for each case.

For each model that we specify below, the fiducial value of the parameters will be those of Table \ref{tab:fidcosmo}, those of the model specific parameters considered, and the corresponding value of $\sigma_8$. We also specify the nuisance parameters values used for the intrinsic alignment modelling for the photometric probes in Table \ref{tab:fidIAcosmo}.

\begin{table}[h]
    \centering
    \caption{Fiducial values for the intrinsic alignment model parameters adopted for the photometric probes.}
    \label{tab:fidIAcosmo}
    \begin{tabular}{cccc}
    \hline
    \rule{0pt}{2ex}
          $\mathcal A_{\rm IA}$ & $\eta_{\rm IA}$ & $\beta_{\rm IA}$ & $\mathcal{C}_{IA}$ \\
         \hline
         \rule{0pt}{2ex}
          1.72 & -0.41 & 2.17 & 0.0134  \\ 
    \hline
    \end{tabular}
\end{table}

\paragraph{Phenomenological Modified Gravity}
In the phenomenological approach to MG described in \cref{Sec:modelpotential}, the additional parameters to be added to the standard set are the amplitudes of the deviations from GR in clustering and lensing, that is, $\mu_0$ and $\Sigma_0$. For these, we consider two fiducial models, one coinciding with the GR limit of the parameterisation (PMG-1) and one that deviates from the standard \lcdm\ predictions (PMG-2), where $\mu_0$ and $\Sigma_0$ take values compatible with the results of \citet{DES:2022ccp}. For both cases, we consider a background evolution that mimics that of \lcdm, which allows us to set the CPL parameters.
The fiducial values of these two choices are shown in Table \ref{tab:GCfid}, together with the corresponding values of the $\sigma_8$ parameter.
For both PMG-1 and PMG-2, we consider as free parameters $\mu_0$ and $\Sigma_0$ in addition to the set of Eq.~(\ref{eq:freestdpars}), while we keep $w_0$ and $w_a$ fixed to their fiducial values.

\begin{table}[h]
    \centering
    \caption{Fiducial values for the two model choices in the gravitational couplings approach.}
    \label{tab:GCfid}
    \begin{tabular}{cccccc}
    \hline
    \rule{0pt}{2ex}
        Model & $\mu_{0}$ & $\Sigma_{0}$ & $w_{0}$ & $w_a$ & $\sigma_{8}$ \\
    \hline
    \rule{0pt}{2ex}
        PMG-1 & $0$ & 0 & $-1$ & $0$ & $0.82$\\
        PMG-2 & $-0.5$ & \;\;\,$0.5$ & $-1$ & $0$ & $0.76$\\
    \hline
    \end{tabular}
\end{table}

\paragraph{Effective Field Theory}

We derive forecasts for two particular models, EFT-1 and EFT-2, as described in \cref{Sec:modelEFT}. The former is a simple linear parameterisation of the free function $\alpha_{\rm B}$ on a standard \lcdm\ background, while the latter considers a departure both in the perturbation and background sectors, namely through a shift-symmetric scalar-tensor model parameterisation of $\alpha_{\rm B}$ on a $w_0 w_a$ background. Both models require us to specify as additional parameters $\alpha_{\rm B,0}$ and $\alpha_{\rm K,0}$, together with the CPL parameters, while EFT-2 also requires the parameter $m$. The fiducial values we choose are listed in Table \ref{tab:EFTfid} with each model's corresponding $\sigma_8$ value.
In the analysis of these models, we choose to keep $\alpha_{\rm K,0}$ and the CPL parameters fixed to their fiducial values, while at the same time, we vary the other additional parameters in addition to the set of Eq.~\eqref{eq:freestdpars}.

\begin{table}[h]
    \centering
    \caption{Fiducial values for the two model choices in the EFT approach.}
    \label{tab:EFTfid}
    \begin{tabular}{ccccccc}
    \hline
    \rule{0pt}{2ex}
        Model & $\alpha_{\rm B,0}$ & $\alpha_{\rm K,0}$ & $m$ & $w_{0}$ & $w_a$ & $\sigma_{8}$ \\
    \hline
    \rule{0pt}{2ex}
        EFT-1 & $0.2$ & $10$ & {---} & $-1$ & $0$ & $0.86$\\
        EFT-2 & $0.9$ & $10$ & $2.4$ & $-0.97$ & $-0.11$ & $0.85$\\
    \hline
    \end{tabular}
\end{table}

\subsection{Survey specifications}\label{sec:surveyspecs}

The specifications for the \Euclid survey that we consider, needed to compute the Fisher matrix, are the same as those used in \citet{Blanchard-EP7}. However, we rely on a different definition for the pessimistic and optimistic cases, namely different choices of small-scales cuts to account for our uncertainties in the modelling of nonlinear perturbation evolution. While \citet{Blanchard-EP7} relies on a $k_{\rm max}$ cut for \GCsp\ and an $\ell_{\rm max}$ cut for photometric observables, we choose to implement a particular $k_{\rm max}$ cut for both surveys, that is, we specify values of $k_{\rm max}^{\rm 3\times2pt}$ and $k_{\rm max}^{\GCsp}$.

For \GCsp, we fix the maximum wavenumber to $k_{\rm max}^{\GCsp} = 0.1$ Mpc$^{-1}$, which we consider to be a conservative case, given that most of the contribution to the constraining power of this probe will now come from linear scales.

\begin{figure}
    \centering
    \includegraphics[width = 0.45\textwidth]{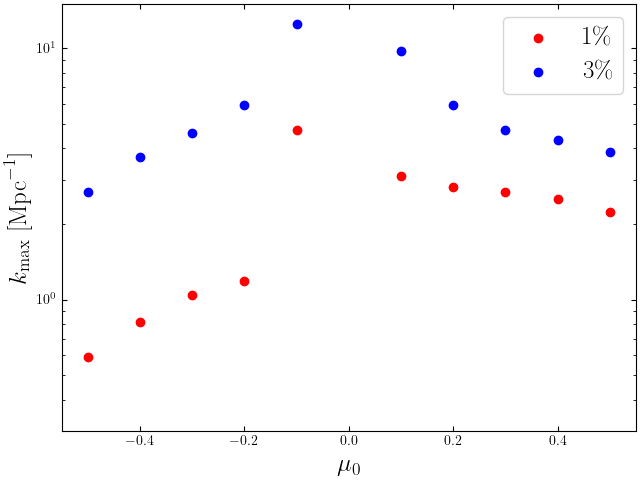}
    \caption{We show the value of the $k_{\rm max}$ parameter inferred from comparing the nonlinear unscreened power spectrum as predicted by \texttt{ReACT} to full $N$-body simulations with the late-time parameterisation at $z=0$ (see text for more details). The data points correspond to the value of $k$ at which the disagreement exceeds 1\% (red) or 3\% (blue). We find that the 3\% threshold is a more robust choice from the point of view of retaining consistency in $k_{\rm max}$ with varying $\mu_{\rm mg}(z)$.}
    \label{fig:kmax_sims}
\end{figure}

To identify the scale below which we cut for the photometric probes, we need to assess the scale down to which the nonlinear matter power spectrum can be accurately computed. To do this, we follow the approach taken in \cite{Srinivasan:2021gib}, where the accuracy of \texttt{ReACT} was tested against modified gravity simulations with a time-dependent $\mu_{\rm mg}$, adapted for the functional form we use here. These are dark-matter only simulations with $1024^3$ particles in a periodic comoving box of $250\,h^{-1}\,{\rm Mpc}$. As shown in appendix A of \cite{Srinivasan:2021gib}, these simulations were shown to have resolved $P(k)$ down to $k\sim 5\,h\,{\rm Mpc}^{-1}$. In this work, we compute the value $k_{\rm max}$ by quantifying the disagreement in the unscreened $P_{\rm NL}(k,z)$ between \texttt{ReACT} and $N$-body simulations with the parameterisation in Eqs.~\eqref{eq:plkmu} and \eqref{eq:plksigma}. To do this, we run a simulation with the corresponding value of $\mu_0$ and measure $P(k)$ at $z=0$. We then run a $\Lambda$CDM simulation with $\sigma_8$ varied such that the linear $P(k)$ at $z=0$ is identical to the modified gravity cosmology we are interested in, and measure $P(k)$ again at $z=0$ (this is the pseudo-cosmology). By taking the ratio of the modified gravity power spectrum relative to the pseudo case, we have the reaction computed from simulations, which we compare to that computed using \texttt{ReACT}. As noted in \cite{Srinivasan:2024}, we find that $k_{\rm max}$ increases with redshift. In other words, we find the smallest value of $k_{\rm max}$ at $z=0$, making our choice for the scale cuts a conservative one. We show the variation in $k_{\rm max}$ as a function of $\mu_0$ in Fig.~\ref{fig:kmax_sims} for different accuracy thresholds, that is, various levels of disagreement between \texttt{ReACT} and the simulations. We see that the 1\% threshold strongly depends on the sign of $\mu_0$. This is because the effect of modifying $\mu_{\rm mg}(z)$ on $P_{\rm NL}(k,z)$ is inherently asymmetric on linear scales, which is compounded on nonlinear scales. We recommend that the $3\%$ threshold be used to infer optimistic constraints on the modified gravity parameters. This choice is further justified by the fact that previous work that achieved benchmark accuracy with \texttt{ReACT} involved the modification of the halo model parameters, such as the halo mass function \citep{Cataneo:2018cic} or the concentration-mass relation \citep{Srinivasan:2024} systematically as a function of $\mu_{\rm mg}(z)$, which is beyond the scope of this paper.

We implement the scale cuts in the Fisher matrix forecasts by progressively removing elements from the data vector (and the associated covariance), converting a given $k_{\rm max}^{\rm 3\times2pt}$ into an $\ell_{\rm max}^{ij}$ estimated as

\begin{equation}
\ell_{\rm max}^{ij} = k_{\rm max}^{\rm 3\times2pt}\,{\rm min}{\left\{r(z_i), r(z_j)\right\}}\,,
\label{eq: ellmaxij}
\end{equation}
with $z_i$ the centre of the $i$-th redshift bin. It is worth to note that this method can greatly reduce the length of the data vector. Indeed, we start from 60 logarithmically spaced bins over the range $(10, 5000)$, so that we have a data vector made by
\begin{equation*}
{\cal{N}}_{{\rm full}} = 60 \times [10 \times (10 + 1)/2 + 10 \times 10 + 10 \times (10 + 1)/2 ] = 12\,600
\end{equation*}
elements, each data vector being the value of $C_{ij}^{XY}(\ell)$ evaluated in the central value of the multipole bin. 
We then remove all the elements with $\ell > 3000$ to mimic the optimistic cut in \citet{Blanchard-EP7}, thus obtaining a data vector with ${\cal{N}}_{\rm IST:F} = 11550$ elements. We then compute our Fisher matrices for $k_{\rm max}^{\rm 3\times2pt}$ values spanning the range $(0.05, 10) \ {\rm Mpc}^{-1}$ going from the linear to strongly nonlinear regimes. Note that, for the smallest $k_{\rm max}^{\rm 3\times2pt}$ value, we end up with a data vector containing only 807 elements, that is, $7\%$ of the optimistic case used in \citet{Blanchard-EP7}. 
We estimate that the constraints will degrade by more than an order of magnitude because of the reduced length of the data vector. The final effect will be further worsened by removing more and more nonlinear scales as $k_{\rm max}^{\rm 3\times2pt}$ becomes smaller, hence reducing the constraining power mainly of WL and galaxy-galaxy lensing (GGL), which, due to the broader kernel, are more sensitive to the nonlinear scales.

As a further remark, we mention that this method of cutting scales is sub-optimal, since, due to the width of the WL kernel, Eq.~(\ref{eq: ellmaxij}) does not ensure that scales with $k > k_{\rm max}^{\rm 3\times2pt}$ do not contribute to the integral giving $C_{ij}^{LL}(\ell)$ even if $\ell \le \ell_{\rm max}^{ij}$. Instead, one should rely on the Bernardeau--Nishimichi--Taruya (BNT) formalism, which ensures a better separation of nonlinear scales by applying a linear transform to the WL kernels \citep{BNT}. 

In this work, we rely on the more straightforward approach described by Eq.~(\ref{eq: ellmaxij}) to mimic the methods used by the currently ongoing Stage III surveys. The difference with the BNT method becomes smaller as $k_{\rm max}^{\rm 3\times2pt}$ becomes larger, so that, at least for part of the range we have considered, we do not expect to be dramatically affected by the use of a less accurate approach.
In the following, we will rely on a baseline choice of $k_{\rm max}^{\rm 3\times2pt} = 0.25$ Mpc$^{-1}$ to draw our main conclusions. This specific value is quite conservative for WL, as the methods we apply for small-scale corrections would allow us to push for higher values (see Fig.~\ref{fig:kmax_sims}). On the other hand, the scale cut we apply is optimistic for \GCph, given that we do not include nonlinear bias in the analysis. In general, we choose $k_{\rm max}^{\rm 3\times2pt} = 0.25$ Mpc$^{-1}$ as a compromise between these two observables. Nevertheless, we will also explore how a change in this assumption will affect the results.

%%%%%%%%%%%%%%%%%%%%%%
\section{Results} \label{sec:results}
%%%%%%%%%%%%%%%%%%%%%%

In this section, we present the results obtained for the PMG and EFT approaches, taking advantage of the methodology described previously. In detail, we show the constraints on cosmological and model parameters assuming conservative choices for $k_{\rm max}^{\GCsp}$ and $k_{\rm max}^{\rm 3\times2pt}$, as outlined in the previous section. We also provide a trend of the constraints achievable on the MG parameters by varying $k_{\rm max}^{\rm 3\times2pt}$.

%--------------------------
\subsection{ Phenomenological Modified Gravity }\label{Sec:Resultspotential}
%--------------------------

As detailed in \cref{Sec:modelpotential}, here we present the forecast results for \Euclid, considering two scenarios, namely PMG-1 and PMG-2. In PMG-1, the fiducial model aligns with the \lcdm\ predictions, while in PMG-2, we assume deviations from GR (see Table \ref{tab:GCfid}).

Table \ref{tab:GCresults} lists the marginalised 68\% confidence level constraints on the free parameters of the analysis. The constraints are expressed as percentages relative to the fiducial values of the parameters that are achievable with the combination of \Euclid primary probes, along with the two approaches to account for MG at nonlinear scales, namely SS and US.

For PMG-1, we observe that \GCsp\ constrains the values of the MG function at the present time, denoted as $\bar{\mu}_0$  to $53.0\%$. Note that galaxy clustering probes offer minimal constraints on $\Sigma_0$ and for this reason we keep this parameter fixed to its fiducial value for the \GCsp\ analyses, while the 3\texttimes2\,pt. statistics provides bounds on $\bar{\mu}_0$ and $\bar{\Sigma}_0$ of respectively $278.5\%$ and $35.0\%$ ($169.4\%$ and $22.2\%$) in the SS (US) case.

In our analysis, we observe that the parameter $\bar{\mu}_0$ is much more constrained by the spectroscopic probe. This trend is expected due to the conservative cutoff at $k_{\rm max}^{\rm 3\times2pt}$, which prevents us from fully benefiting from nonlinear scales for the photometric probes. Even with the pessimistic $k_{\rm max}$ scale cut we still observe more than $100\%$ gain difference between US and SS in the 3\texttimes2\,pt. statistics constraints. This is because, in the latter case, our observables revert to GR in the nonlinear regime, thereby limiting the constraining power of these scales.

\begin{figure*}[ht!]
 \centering
 \begin{tabular}{cc}
 \includegraphics[width=0.45\linewidth]{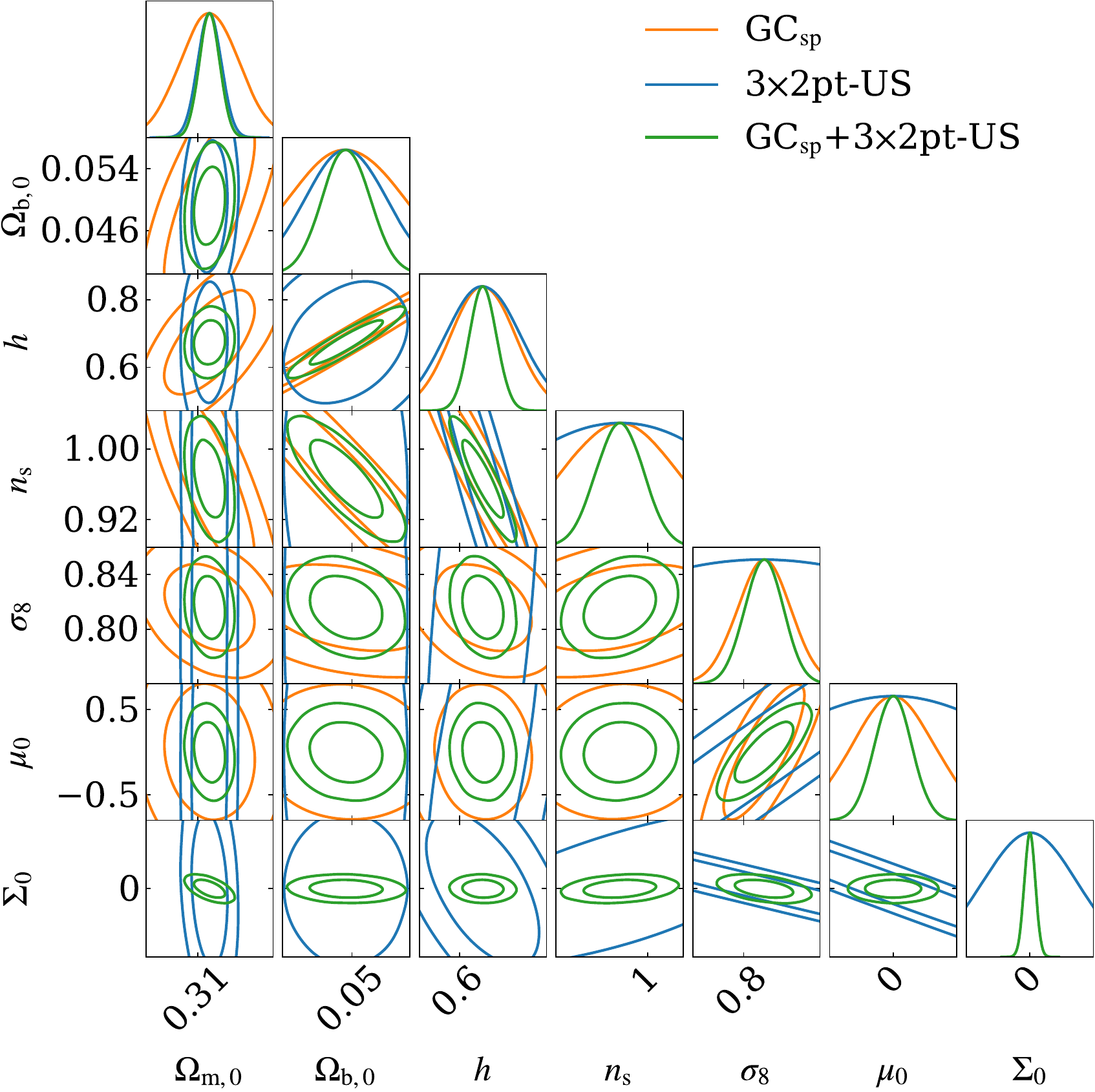}&
 \includegraphics[width=0.45\linewidth]{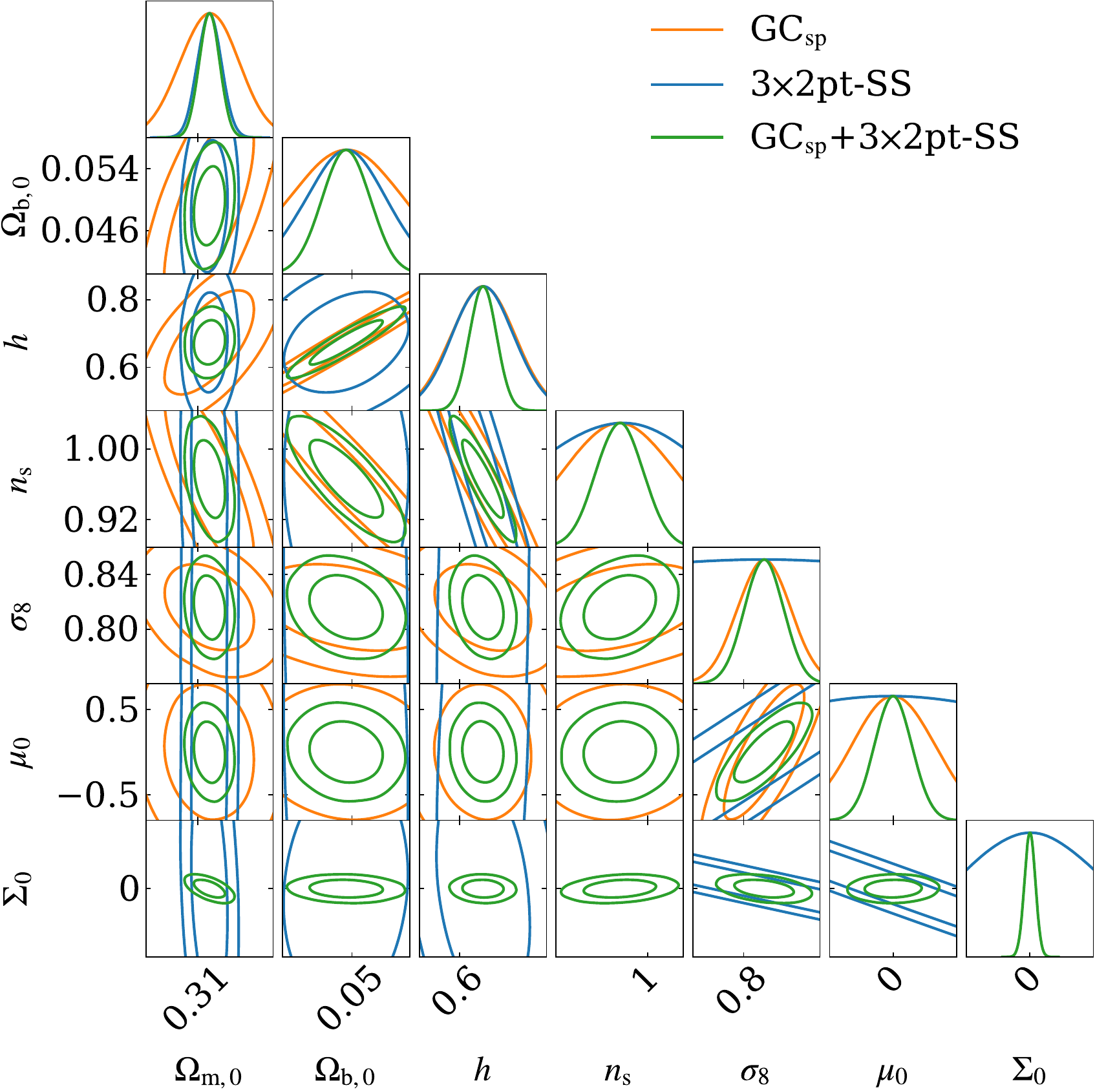}
 \end{tabular}
 
 \caption{68\% and 95\% joint marginal error contours on the cosmological parameters in the phenomenological modified gravity approach for the PMG-1 case, using different probe combinations: \GCsp (orange); 3\texttimes2\,pt. statistics (blue); and \GCsp+3\texttimes2\,pt. statistics (green). Two different nonlinear corrections have also been considered for the photometric probes: US (left panels); and SS (right panels). 
 }
 \label{fig:ellipses-musigma_GC1}
\end{figure*}

Upon combining \Euclid's photometric and spectroscopic surveys, we find improvements for the constraints of $\bar{\mu}_0$ and $\bar{\Sigma}_0$ up to $23.7\%$ and $2.6\%$ ($23.3\%$ and $2.6\%$) for both the SS and US scenarios.  We expect the improvement of the full combination over \GCsp\ to be small, because the constraints from the latter are much stronger than that from the photometric probes. We also observe a significant enhancement in the $\bar{\Sigma}_0$ parameter, despite \GCsp\ not providing constraints. The latter improvement can be attributed to the complementary constraining power of the two surveys, which, when combined, help reduce the impact of nuisance parameters and break the degeneracy between cosmological parameters. We show this effect in Fig.~\ref{fig:ellipses-musigma_GC1}, which shows the marginalised 68\% and 95\% confidence level contours in different combinations of parameters.

\begin{figure*}[ht!]
 \centering
 \begin{tabular}{cc}
 \includegraphics[width=0.45\linewidth]{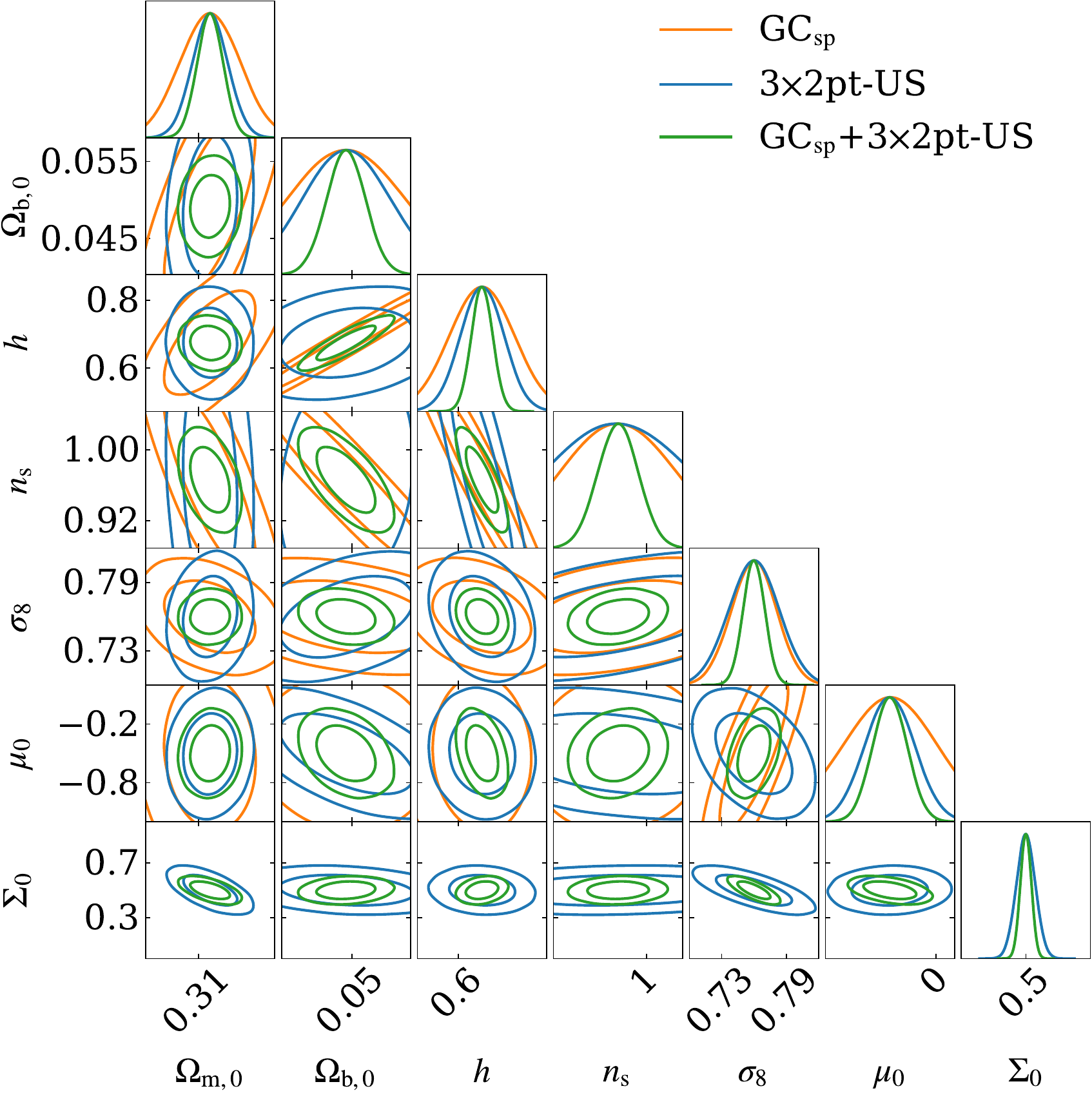}&
 \includegraphics[width=0.45\linewidth]{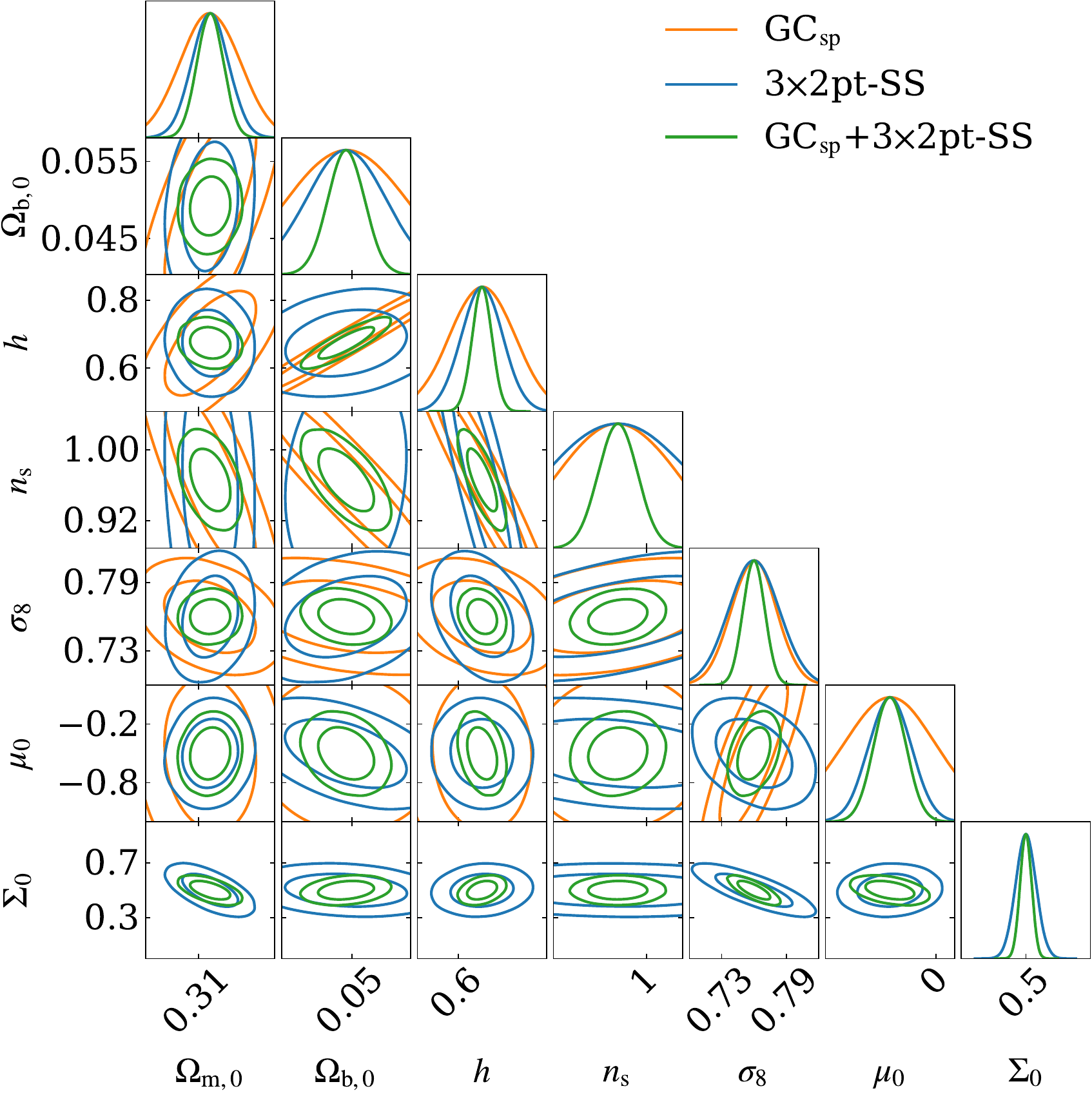}
 \end{tabular}
 
 \caption{68\% and 95\% joint marginal error contours on the cosmological parameters in the phenomenological modified gravity approach for the PMG-2 case, using different probe combinations: \GCsp (orange); 3\texttimes2\,pt. statistics (blue); and \GCsp+3\texttimes2\,pt. statistics (green). Two different nonlinear corrections have also been considered for the photometric probes: US (left panels); and SS (right panels).}
 \label{fig:ellipses-musigma_GC2}
\end{figure*}

We present the results for the PMG-2 case in Table \ref{tab:GCresults} and Fig.~\ref{fig:ellipses-musigma_GC2}. The values of the MG functions at present, $\bar{\mu}_0$ and $\bar{\Sigma}_0$, are constrained by the 3\texttimes2\,pt. statistics to $54.7\%$ and $4.9\%$, respectively, ($46.6\%$ and $5.3\%$) in the SS (US) case. Furthermore, \GCsp\ provides bounds on $\bar{\mu}_0$ of $103.4\%$. Combining \Euclid's photometric and spectroscopic surveys, we find that the bounds on $\bar{\mu}_0$ and $\bar{\Sigma}_0$ improve to $34.4\%$ and $2.8\%$ ($36.2\%$ and $2.7\%$) in the SS (US) case.

These findings follow similar qualitative trends as PMG-1 when comparing constraints from spectroscopic and photometric probes alone and their combinations and when moving from the US to the SS case. The main difference with respect to PMG-1 case lies in the constraints given by 3\texttimes2\,pt. statistics alone, which are tighter in PMG-2. The different fiducials chosen for the $\mu_0$ and $\Sigma_0$ parameters make the observables more sensitive to changes in their values, thus leaving an imprint on the derivatives entering the Fisher matrix analysis. Another difference concerns the constraints on $\sigma_8$, which are nearly unchanged in PMG-2 between US and SS, since, in the latter case, we revert to GR while the fiducial is now far from it.

We also notice increased errors from \GCsp\ alone in the PMG-2 case for some parameters such as $\Omega_{\rm m,0}$, $\sigma_{8}$ or $\bar{\mu}_0$ and this affected the combination with 3\texttimes2\,pt. statistics, but in general errors decrease in PMG-2 with respect to PMG-1. We checked that this is due either to the difference in the fiducial value, making the Fisher values higher, or to changes in the degeneracies compared to PMG-1, with the parameters no longer being correlated with those from the 3\texttimes2\,pt. statistics $\bar{\mu}_0$, as shown in Fig.~\ref{fig:ellipses-musigma_GC2}. As a result, the combination has more ability to break degeneracies between cosmological and modified gravity parameters. This is clearly shown by the orthogonality in the contours between \GCsp\ and 3\texttimes2\,pt. statistics in, e.g., the $\bar{\mu}_0$-$\sigma_8$ case, which is more pronounced in Fig.~\ref{fig:ellipses-musigma_GC2} compared to Fig.~\ref{fig:ellipses-musigma_GC1}.

In terms of improvements compared to current constraints, such as those of the \Planck mission \citep{Planck:2018vyg} or the DES survey \citep{DES:2022ccp} mentioned in \cref{Sec:modelpotential}, our relative errors are, in general, and for probe combinations similar to those presented in the former papers, one order of magnitude smaller for $\bar{\mu}_0$ and $\bar{\Sigma}_0$. Our constraints, although less tight, are also comparable to the predictions from \cite{Casas:2022vik} using other future Stage IV experiments.

\begin{figure*}
    \centering
    \begin{tabular}{ccc}
\includegraphics[width=0.3\textwidth]{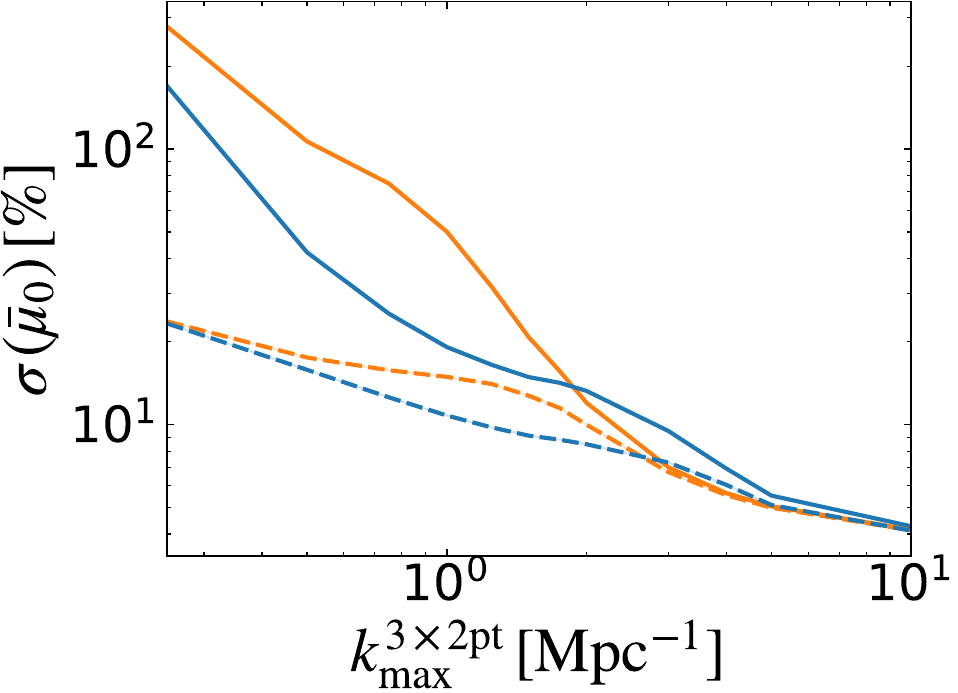} &
\includegraphics[width=0.3\textwidth]{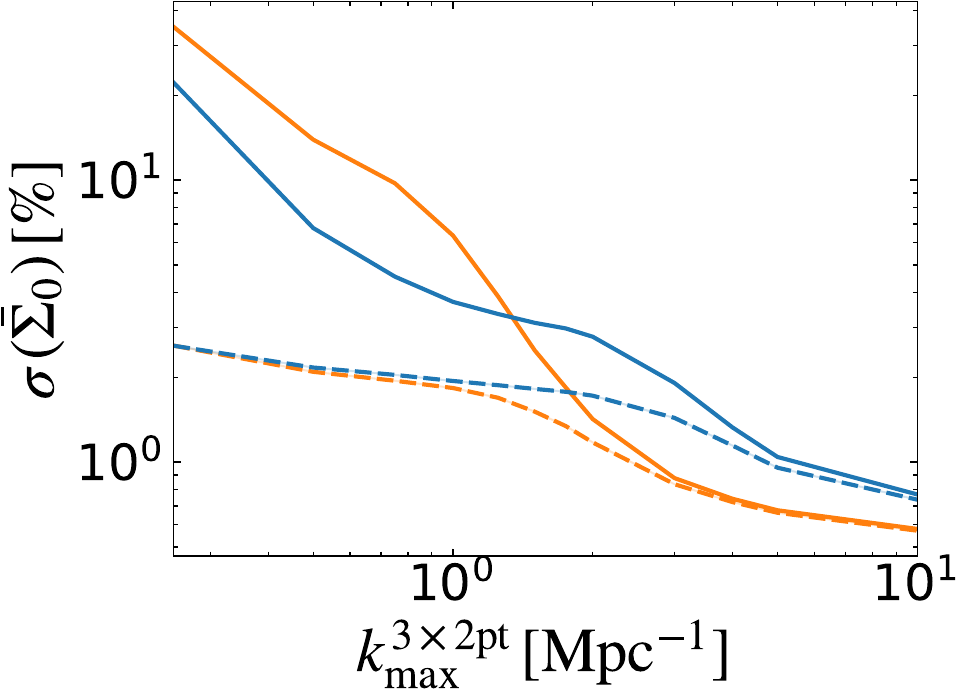} &
\includegraphics[width=0.3\textwidth]{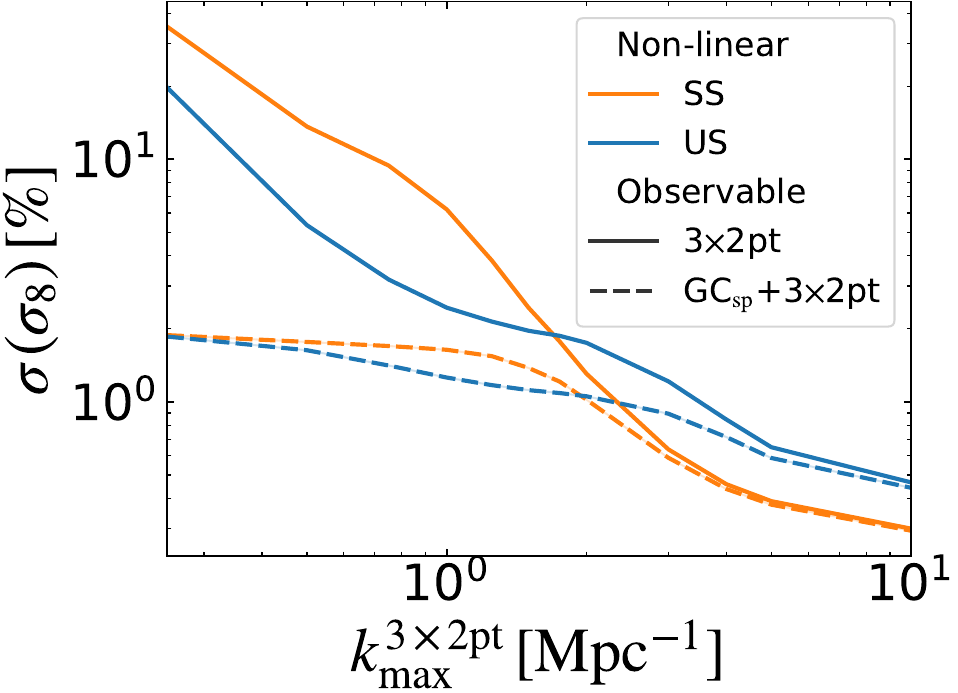}\\
\includegraphics[width=0.3\textwidth]{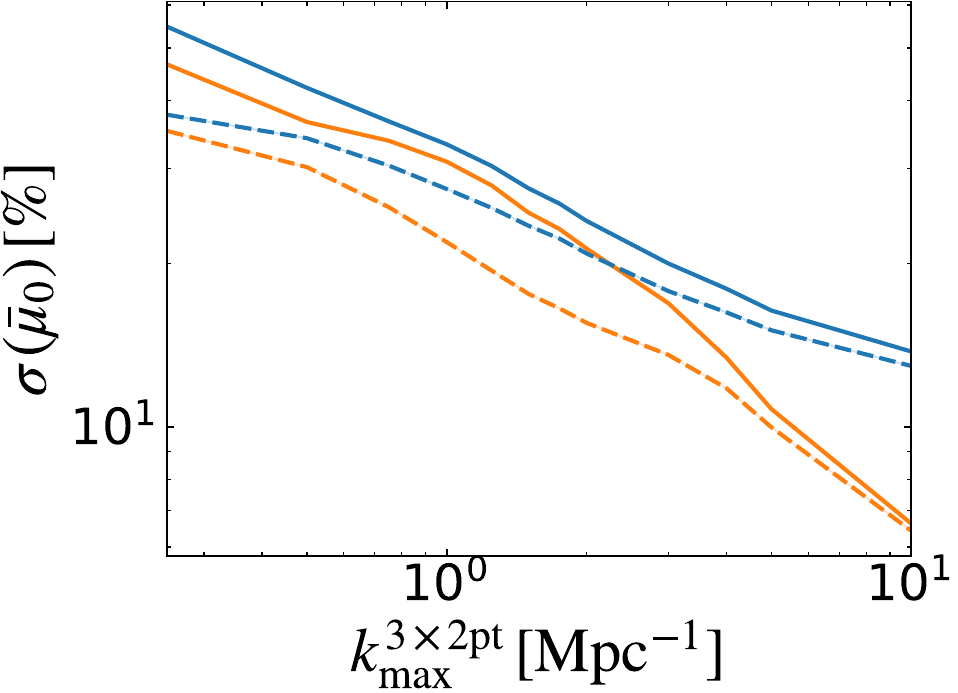}&
\includegraphics[width=0.3\textwidth]{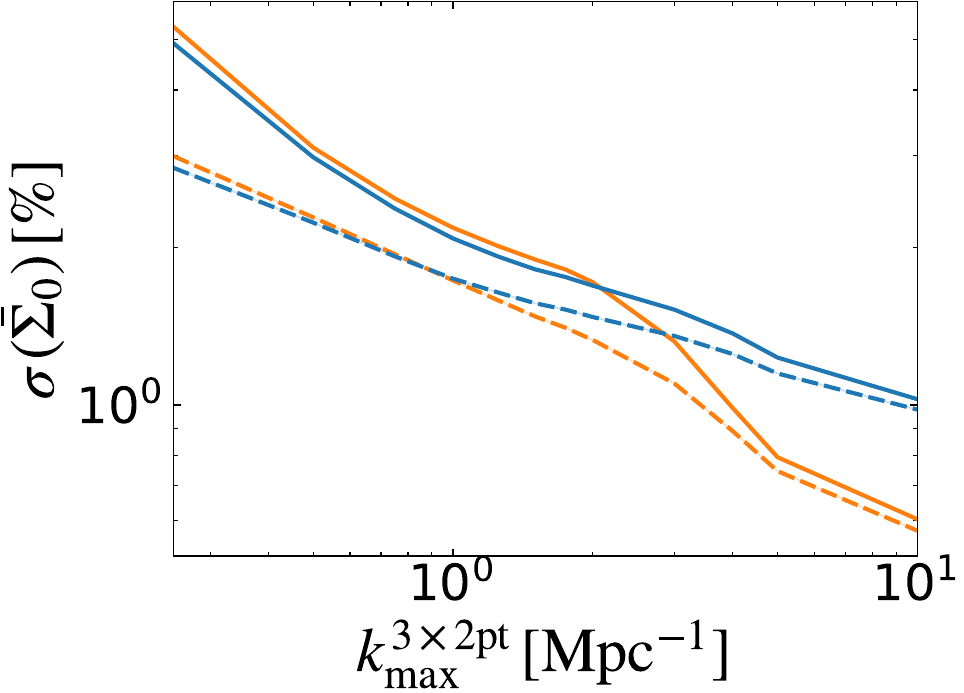}&
\includegraphics[width=0.3\textwidth]{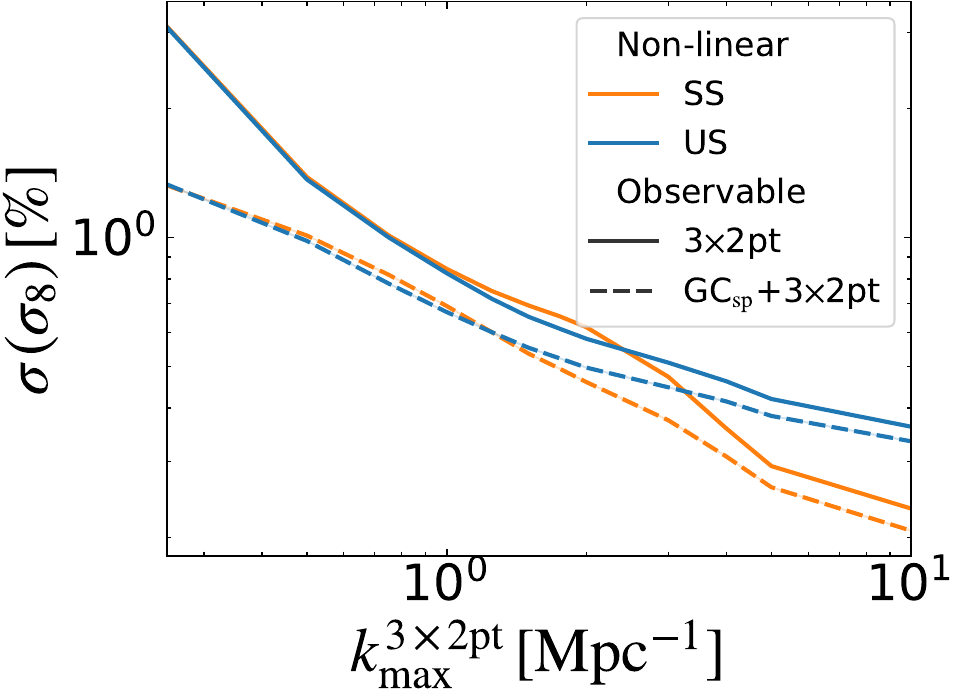}
    \end{tabular}
    \caption{3\texttimes2\,pt. statistics (solid lines) and \GCsp+3\texttimes2\,pt. statistics (dashed lines) 68\% marginal error on $\mu_{0}$, $\Sigma_{0}$, and $\sigma_8$ as a function of the $k^{\rm 3\times2pt}_{\rm max}$ value considered in the photometric analysis for PMG-1 (top row) and PMG-2 (bottom row) models.}
    \label{fig:GC_kmaxtrend}
\end{figure*}

In Fig.~\ref{fig:GC_kmaxtrend}, we illustrate the impact of the choice of $k_{\rm max}^{\rm 3\times2pt}$ on the constraints for the MG functions in both PMG-1 (top panels) and PMG-2 (bottom panels). We observe that the constraints on these parameters improve as $k_{\rm max}^{\rm 3\times2pt}$ increases. By identifying the scales at which \texttt{ReACT} results are within $1\%$ agreement with what is obtained from $N$-body simulations (see the discussion in \cref{sec:surveyspecs}), we find that $k_{\rm max}^{\rm 3\times2pt}\approx 4\,{\rm Mpc}^{-1}$ for PMG-1 and $k_{\rm max}^{\rm 3\times2pt}\approx 0.6\,{\rm Mpc}^{-1}$ for PMG-2. Consequently, the constraints on $\bar{\mu}_0$ ($\bar{\Sigma}_0$) can reach $4\%$ ($1\%$) and $6\%$ ($1\%$) for the two cases using 3\texttimes2\,pt. statistics alone.

\begin{figure*}
    \centering    
\includegraphics[width=0.35\textwidth]{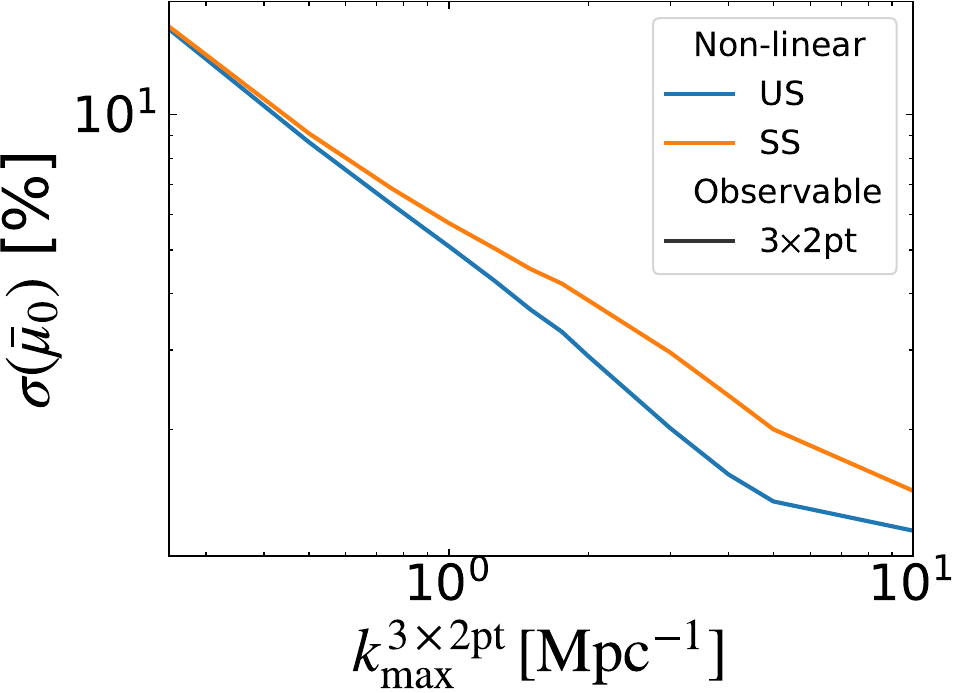} 
\includegraphics[width=0.35\textwidth]{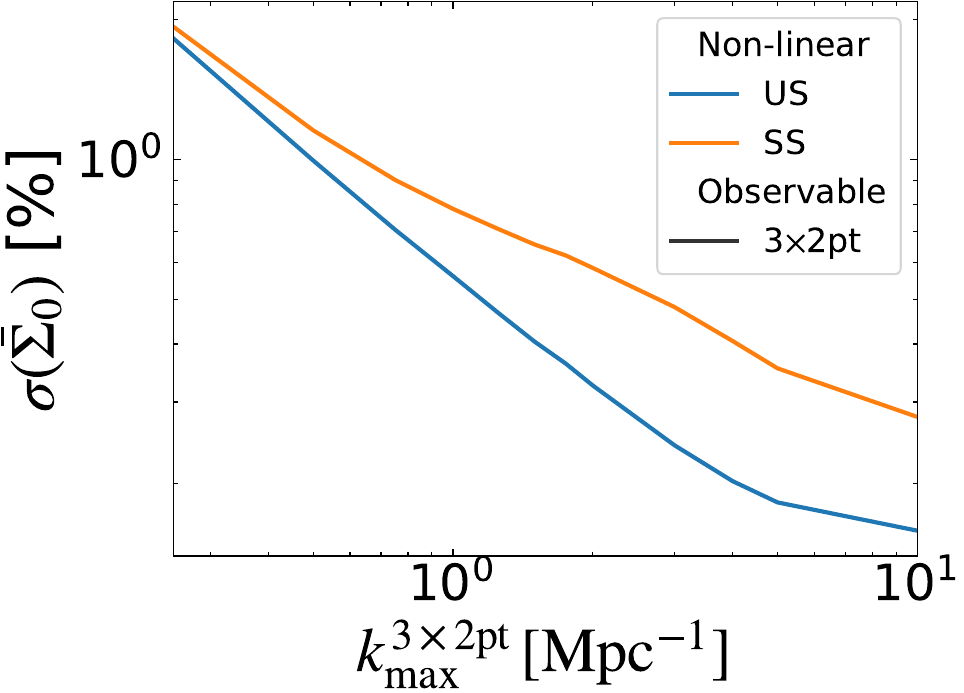}    
\caption{3\texttimes2\,pt. statistics 68\% marginal error on $\bar{\mu}_0$ (left panel), $\bar{\Sigma}_0$ (right panel) as a function of the $k^{\rm 3\times2pt}_{\rm max}$ value considered in the photometric analysis for the PMG-1 model when all other parameters are fixed to their fiducial values.}
    \label{fig:musigma_maxed}
\end{figure*}

Such an improvement is possible thanks to the inclusion of more scales, which provide more information in the analysis, particularly in the US prescription. In this case, the MG effects also enter the computation of nonlinear evolution, leading to tighter bounds than in the SS case. As highlighted in Fig.~\ref{fig:GC_kmaxtrend}, this trend holds until we delve deeper into the nonlinear regime, where reverting to GR in the SS case allows us to break the degeneracies between $\sigma_8$ and the MG parameters, resulting in tighter constraints in relation to the US case. The impact of such degeneracies is clarified in Fig.~\ref{fig:musigma_maxed} for the PMG-1 model from 3\texttimes2\,pt. statistics alone by having fixed the Fisher matrix values for all cosmological parameters except $\bar{\mu}_0$ (or $\bar{\Sigma}_0$). This procedure effectively removes the impact of parameter correlations on the final marginalised covariance matrix. Without the impact of degeneracy breaking, we expect the US to always provide stronger constraints with respect to the SS case, which is what is found in Fig.~\ref{fig:musigma_maxed}.
The substantial improvement in the constraining power of 3\texttimes2\,pt. statistics with respect to the conservative case shown in Table \ref{tab:GCresults} highlights the need to accurately model nonlinear scales, since these contribute significantly to the constraining power in these models.

Finally, we can also notice that, when combined with \GCsp, the US and SS cases provide identical or close constraints for $k_{\rm max}\lesssim2\,{\rm Mpc}^{-1}$. Below this scale, the sensitivity of \GCsp\ to $\mu_0$ and its ability to break MG degeneracies dominate the constraining power. Therefore, any variations in the 3\texttimes2\,pt. statistics analysis will yield negligible effects. However, beyond this threshold, the contribution of the 3\texttimes2\,pt. statistic becomes significant, and we can observe differences between the SS and US analyses.

\begin{table*}[h]
    \centering
    \caption{Marginalised 68\% confidence level constraints on the free parameters of the phenomenological modified gravity analysis. Values are reported as percentages of the parameters' fiducial values.}
    \label{tab:GCresults}
    \begin{tabular}{cccccccccc}
    \hline
    \rule{0pt}{2ex}
    Model & NL & Probes & $\Omega_{\rm m,0}$ & $\Omega_{\rm b,0}$ & $h$ & $n_{\rm s}$ & $\sigma_{8}$ & $\bar{\mu}_{0}$ & $\bar{\Sigma}_{0}$ \\
    \hline 
    \multirow{5}{*}{PMG-1}	&	&	\GCsp	    &  $3.8\%$  &  $16.5\%$     &  $15.0\%$  &  \;\;$7.2\%$  &  \;\;$2.6\%$  &  \;\;$53.0\%$     &  \dots  \\
    \cline{2-10}
	&	\multirow{2}{*}{SS}	&	3\texttimes2pt	&  $1.5\%$  &  $11.6\%$     &  $14.7\%$	    &  $12.1\%$     &  $35.2\%$     &  $278.5\%$        &  $35.0\%$  \\	
	&	&	\GCsp+3\texttimes2pt	            &  $1.3\%$  &  \;\;$6.9\%$  &  \;\;$6.5\%$  &  \;\;$3.0\%$  &  \;\;$1.9\%$  &  \;\;\;\;$23.7\%$  &  \;\;$2.6\%$  \\
 \cline{2-10}	
	&	\multirow{2}{*}{US}	&	3\texttimes2pt	&  $1.5\%$  &  $11.5\%$     &  $17.7\%$     &  $14.8\%$     &  $19.7\%$     &  $169.4\%$        &  $22.2\%$  \\	
	&	&	\GCsp+3\texttimes2pt	            &  $1.3\%$  &  \;\;$6.9\%$  &  \;\;$6.4\%$  &  \;\;$3.0\%$  &  \;\;$1.9\%$  &  \;\;\;\;$23.3\%$	&  \;\;$2.6\%$	\\
    \hline 
    \rule{0pt}{2ex}
     \multirow{5}{*}{PMG-2}	&	&	\GCsp	   &  $4.0\%$  &  \;\;$14.6\%$  &  \;\;$9.8\%$  &  \;\;$6.1\%$  &  \;\;$2.9\%$  &  \;\;$103.4\%$     &  \dots	\\	
     \cline{2-10}
	&	\multirow{2}{*}{SS}	&	3\texttimes2pt  &  $2.2\%$  &  $12.3\%$     &  $10.2\%$	    &  \;\;$9.3\%$	&  \;\;$3.1\%$  &  \;\;$54.7\%$	    &  \;\;$4.9\%$	\\	
	&	&	\GCsp+3\texttimes2pt                &  $1.7\%$  &  \;\;$6.7\%$  &  \;\;$4.7\%$	&  \;\;$2.7\%$	&  \;\;$1.3\%$  &  \;\;$34.4\%$	    &  \;\;$2.8\%$	\\\cline{2-10}	
	&	\multirow{2}{*}{US}	&	3\texttimes2pt  &  $2.3\%$  &  $11.2\%$     &  \;\;$9.7\%$	&  \;\;$9.0\%$	&  \;\;$3.1\%$  &  \;\;$46.6\%$     &  \;\;$5.3\%$	\\	
	&	&	\GCsp+3\texttimes2pt                &  $1.6\%$  &  \;\;$6.9\%$  &  \;\;$4.9\%$	&  \;\;$2.8\%$	&  \;\;$1.3\%$  &  \;\;$36.2\%$     &  \;\;$2.7\%$	\\	
    \hline
    \end{tabular}
\end{table*}

%--------------------------
\subsection{Effective Field Theory of dark energy and modified gravity}\label{Sec:ResultsEFT}

\begin{figure*}[ht!]
 \centering
 \includegraphics[width=0.49\linewidth]{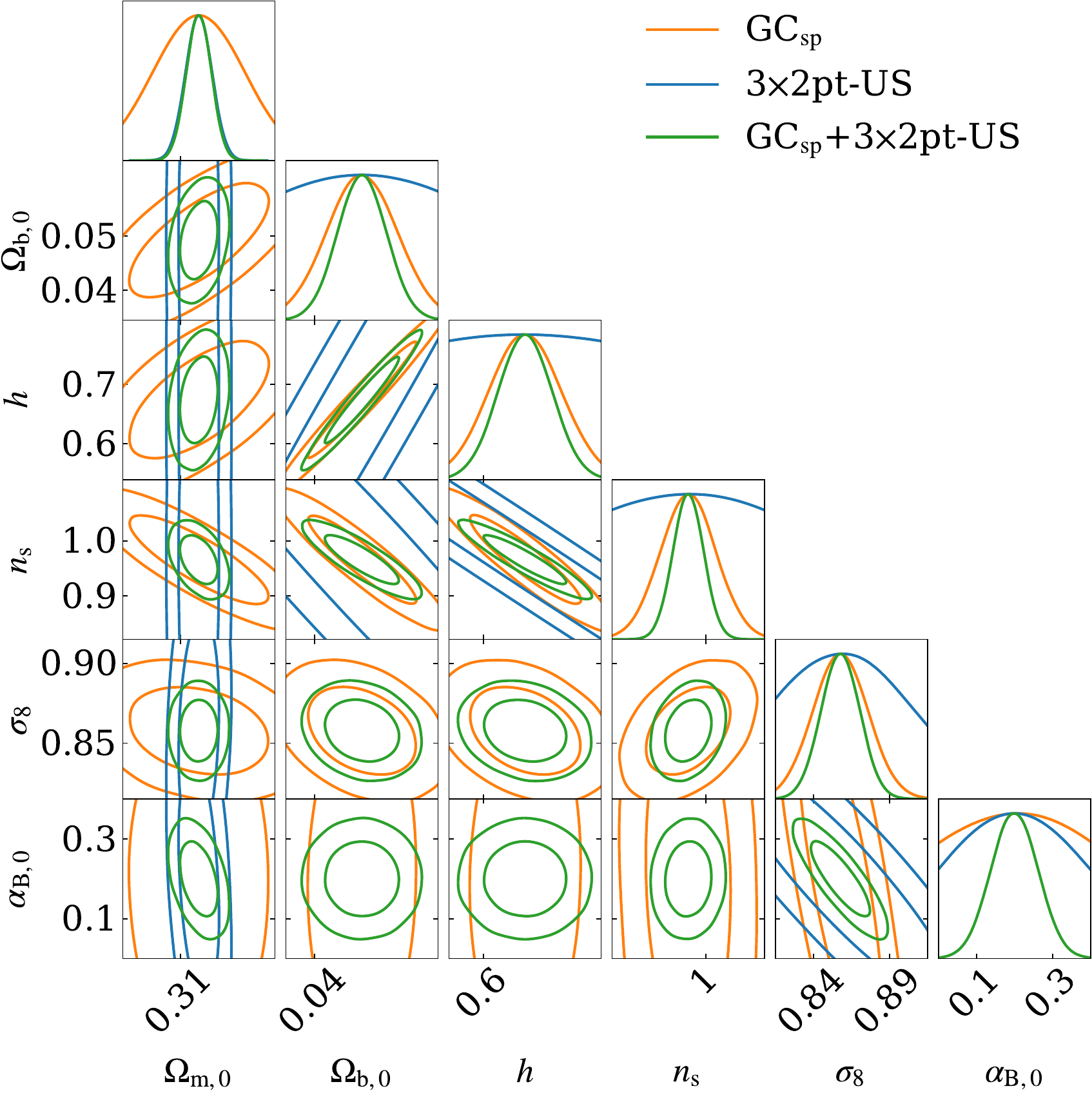}
 \includegraphics[width=0.49\linewidth]{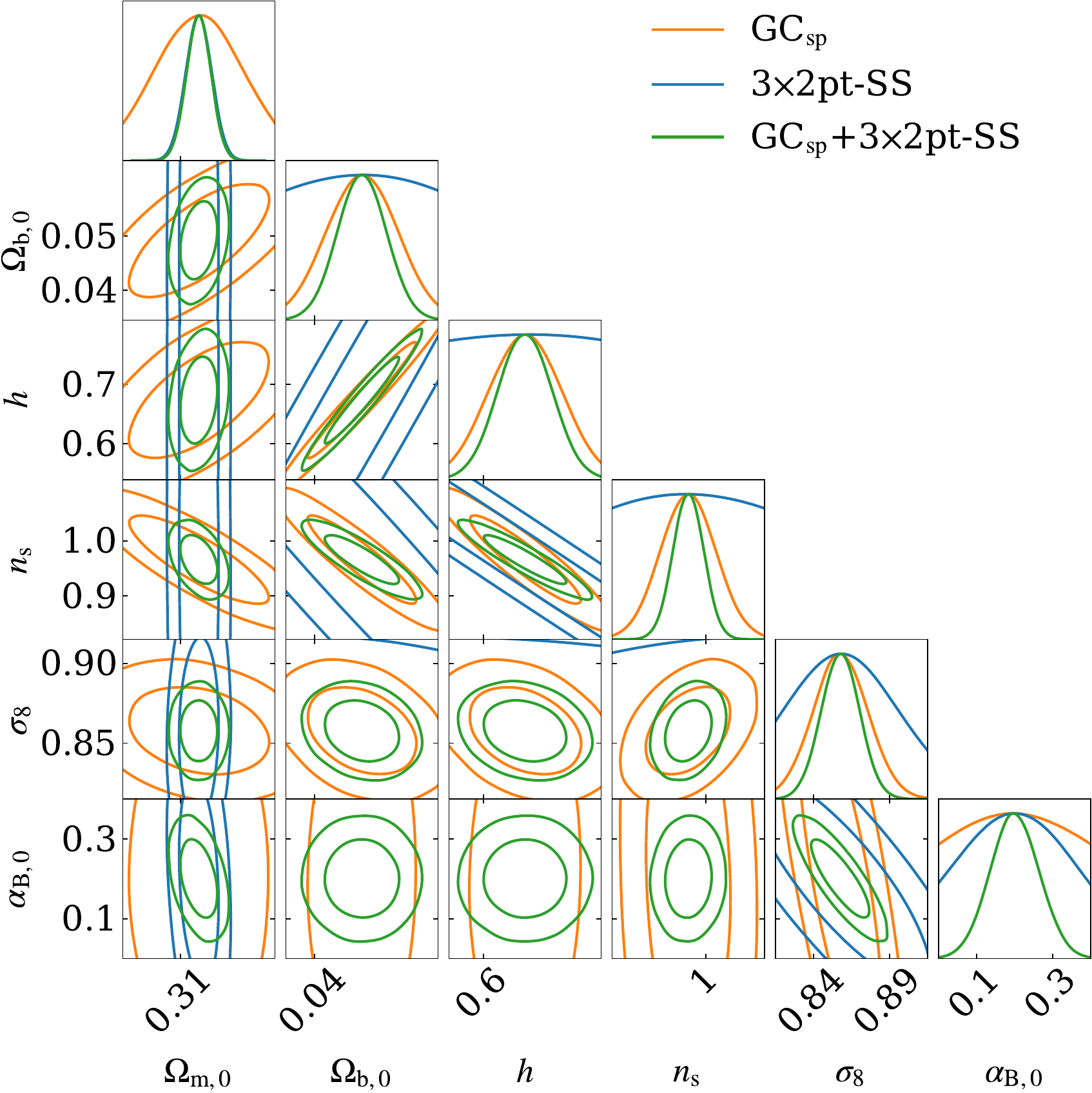}\\
 
 \caption{68.3\% and 95.4\% joint marginal error contours on the cosmological parameters for the EFT-1 model using different probe combinations: \GCsp\, (orange), 3\texttimes2\,pt. statistics (blue), and \GCsp+3\texttimes2\,pt. statistics (green). Two different nonlinear corrections have also been considered for the photometric probes: US (left panels) and SS (right panels).}
 \label{fig:ellipses-EFT1}
\end{figure*}

\begin{figure*}[ht!]
 \centering
 \includegraphics[width=0.49\linewidth]{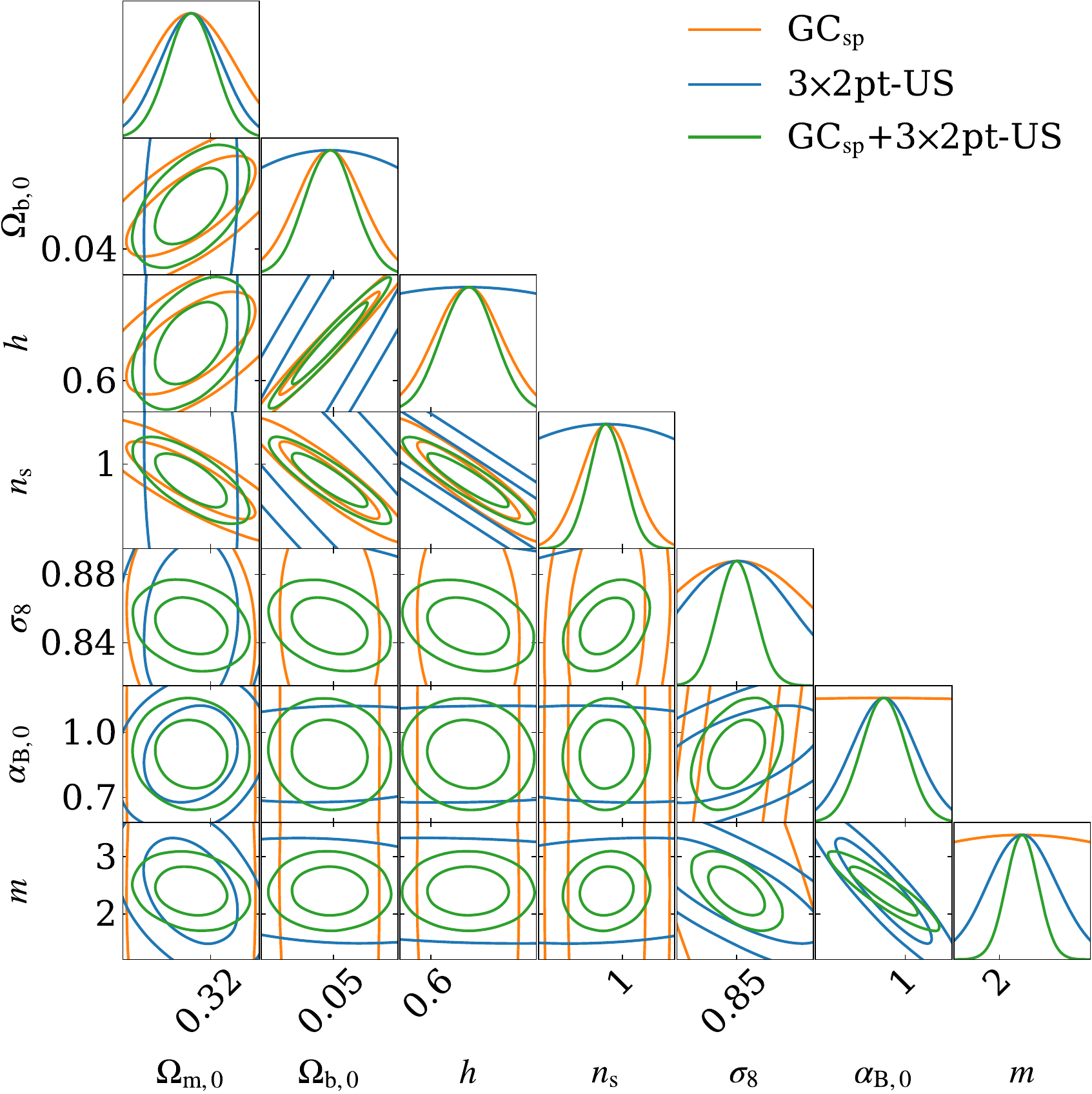}
 \includegraphics[width=0.49\linewidth]{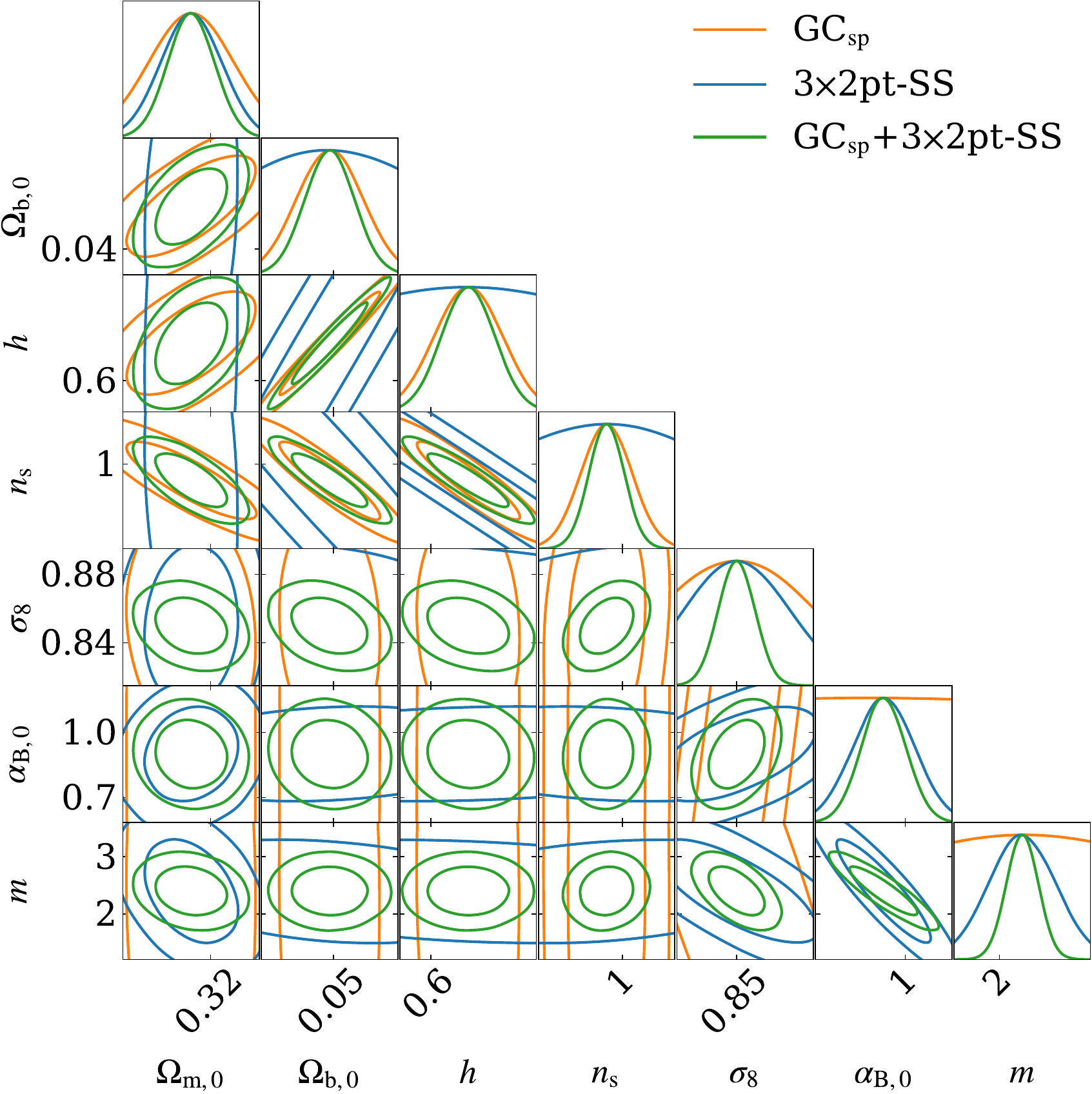}\\
 
 \caption{68\% and 95\% joint marginal error contours on the cosmological parameters for the EFT-2 model using different probe combinations: \GCsp\ (orange); 3\texttimes2\,pt. statistics (blue); and \GCsp+3\texttimes2\,pt. statistics (green). Two different nonlinear corrections have also been considered for the photometric probes: US (left panels); and SS (right panels).}
 \label{fig:ellipses-EFT2}
\end{figure*}

We present the results for the two EFT parameterisations considered, as detailed in \cref{Sec:modelEFT}. We show in Figs.~\ref{fig:ellipses-EFT1} and \ref{fig:ellipses-EFT2} the $68\%$ and $95\%$ joint marginal error contours for the cosmological and model parameters for EFT-1 and EFT-2, respectively.
The marginalised $68\%$ confidence level bounds for both models expected from the primary probes of \Euclid are reported in Table \ref{tab:EFTresults}, expressed as percentage constraints relative to the fiducial values of the parameters and accounting for the two different approaches to handling MG effects at nonlinear scales, namely SS and US.

In the EFT-1 model, the cosmological parameters $\{\Omega_{\rm m,0},\Omega_{\rm b,0},h,n_{\rm s}, \sigma_8\}$ are very well constrained using \GCsp\, ($\{3.7\%, 14.2\%,9.7\%, 5.5\%,2.1\%\}$), while they mostly degrade for the 3\texttimes2\,pt. statistics. Specifically, the constraints for $\{\Omega_{\rm b,0}, h, n_{\rm s}\}$ degrade by approximately one order of magnitude, while $\Omega_{\rm m,0}$ is the only parameter for which there is improvement. This trend holds for both SS and US. In the SS case, the error forecasts for $\Omega_{\rm b,0}$, $h,$ and $n_{\rm s}$ become $62.9\%$, $69.6\%$ and $2.9\%$, while for $\Omega_{\rm m,0}$ it reaches $1.0\%$. For the US case, we report $61.6\%$, $67.8\%$, $32.0\%$ and $1.1\%$, respectively. The constraints on $\sigma_8$ slightly degrade with respect to \GCsp\ but remain of the same order, namely $4.5\%$ and $5.7\%$ for SS and US, respectively. When considering the full combination of probes, the errors across all the cosmological parameters improve with respect to the individual probes, with no difference in the SS and US cases. For the additional parameter $\alpha_{\rm B,0}$, we find a substantial relative error for the standalone photometric and spectroscopic probes. Consequently, we find $148.5\%$ for \GCsp\ and $86.9\%$ for 3\texttimes2\,pt. statistics in the SS case, while it is poorly constrained for the US case ($100.6\%$). The full combination \GCsp+3\texttimes2\,pt. statistics greatly reduces the error in $\alpha_{\rm B,0}$ to $32.3\%$ for SS ($31.1\%$ for US), as a result of the breakdown of degeneracies produced by the full-probe combination, highlighted in the $\alpha_{\rm B,0}$-$\sigma_8$ plane shown in Fig.~\ref{fig:ellipses-EFT1}. We also note that in EFT-1 the correlations between $\alpha_{\rm B,0}$ and the cosmological parameters are unchanged between US and SS.

In EFT-2, the \GCsp\ strongly constrains the cosmological parameters $\{\Omega_{\rm m,0},\Omega_{\rm b,0},h, n_{\rm s}, \sigma_8\}$, which we find to be $3.8\%$, $14.6\%$, $9.9\%$, $5.6\%$, and $5.4\%$. When considering the 3\texttimes2\,pt. statistics probes instead, the forecast constraints notably degrade for the parameters $\{\Omega_{\rm b,0},h, n_{\rm s}\}$, reaching $53.7\%$, $60.5\%$, and $29.6\%$ for the SS case and $52.5\%$, $58.9\%$, and $28.7\%$ for the US one. The constraint on $\Omega_{\rm m,0}$ does not change significantly (around $3\%$), while the constraint on $\sigma_8$ improves (around $4\%$) for both SS and US. The full combination \GCsp+3\texttimes2\,pt. statistics yields better constraints, which are not affected by the choice of SS or US. 
EFT-2 contains two additional parameters, $\alpha_{\rm B, 0}$ and $m$, which remain unconstrained by \GCsp. However, as shown in Table \ref{tab:EFTresults}, the photometric probes have a strong constraining power over these parameters. The $\alpha_{\rm B,0}$ parameter is constrained to $16.0\%$ and $16.5\%$ for SS and US, respectively, while $m$ is constrained to $25\%$ for SS and $25.7\%$ for US. Additionally, the \GCsp+3\texttimes2\,pt. statistics combination provides further improvements: $\alpha_{\rm B,0}$ is constrained to $11.6\%$ for both SS and US, while $m$ is constrained to $11.8\%$ for both SS and US. It is evident from the discussion above that in the EFT-2 case, using the SS or US prescription does not significantly affect the results.

\begin{figure*}
    \centering
    \begin{tabular}{ccc}
\includegraphics[width=0.3\textwidth]{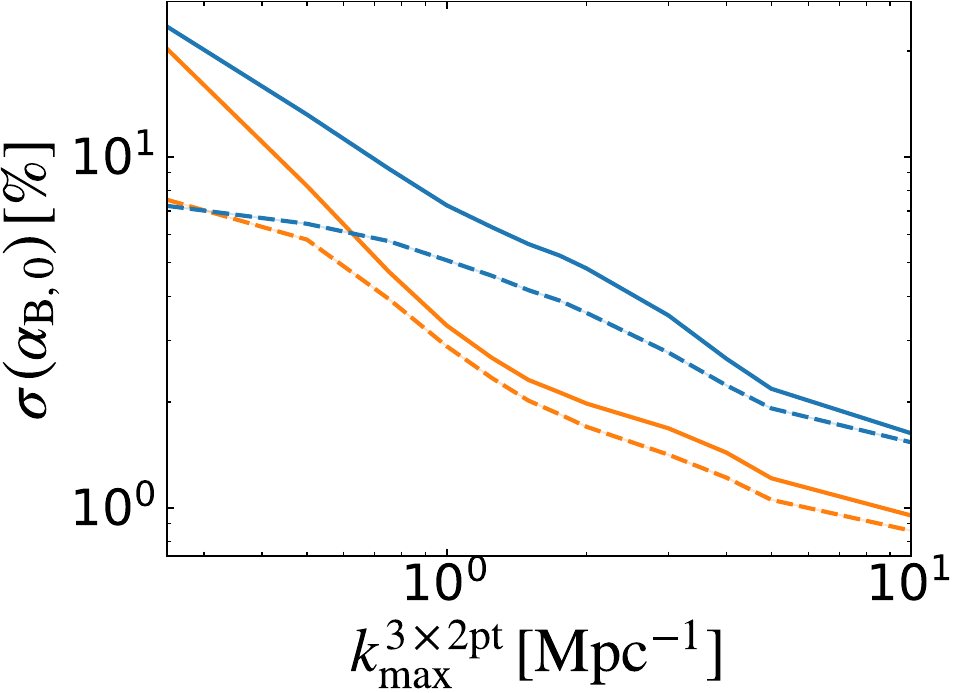} &
 &
\includegraphics[width=0.3\textwidth]{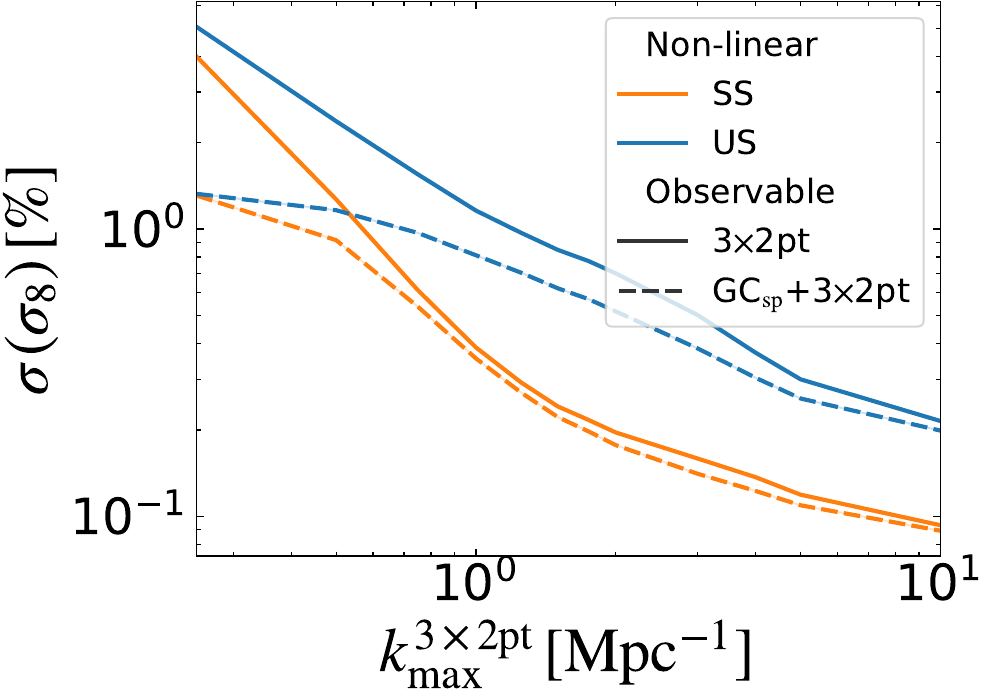} \\ 
\includegraphics[width=0.3\textwidth]{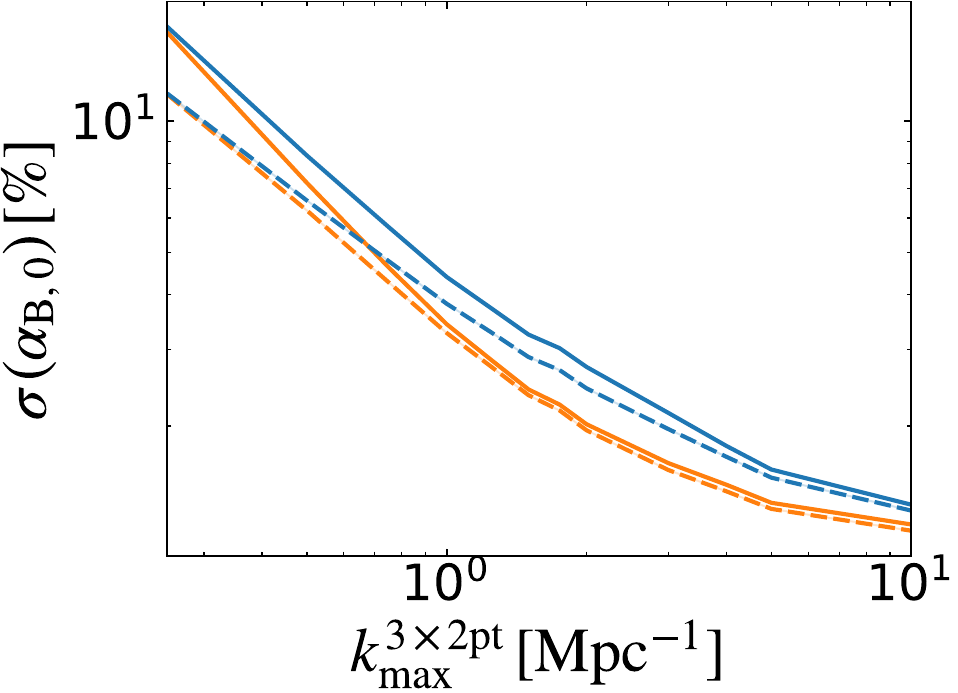} &
\includegraphics[width=0.3\textwidth]{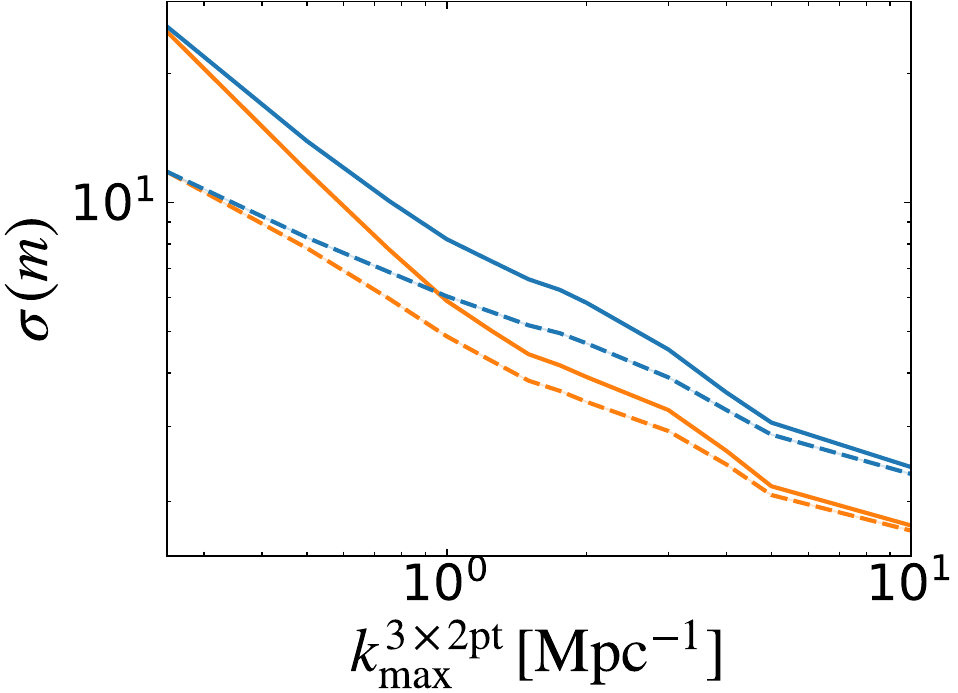}&
\includegraphics[width=0.3\textwidth]{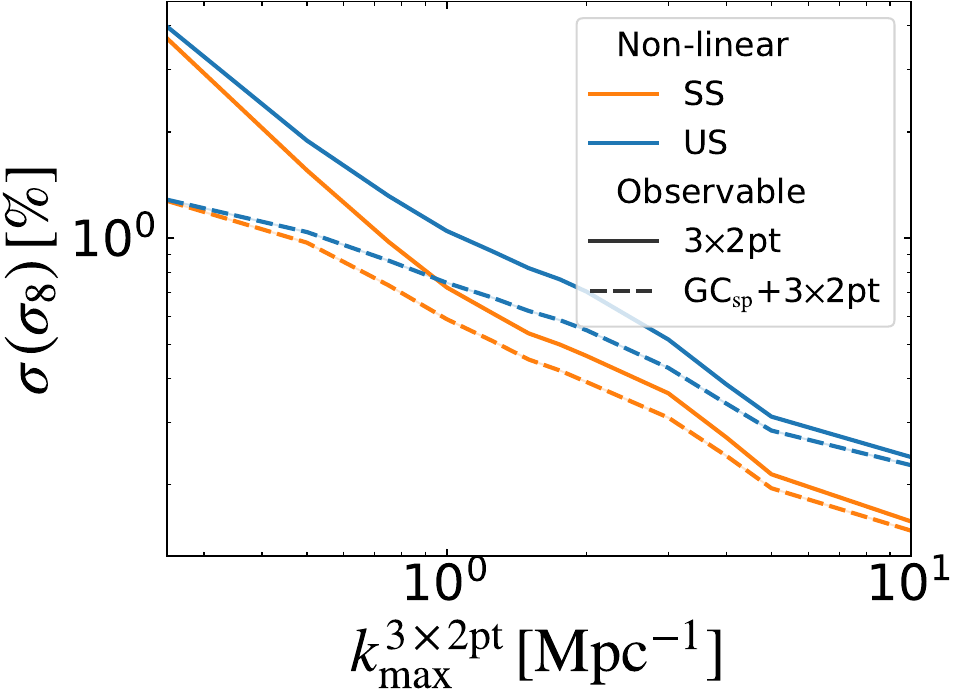}\\
    \end{tabular}
    \caption{3\texttimes2\,pt. statistics (solid lines) and \GCsp+3\texttimes2\,pt. statistics (dashed lines) 68\% marginal error on $\alpha_{\rm B,0}$, $m$, and $\sigma_8$ as a function of the $k^{\rm 3\times2pt}_{\rm max}$ value considered in the photometric analysis for EFT-1 (top row) and EFT-2 (bottom row) models.}
    \label{fig:EFT_kmaxtrend}
\end{figure*}

Finally, in Fig.~\ref{fig:EFT_kmaxtrend}, we provide the trends of the constraints of the MG parameters and $\sigma_8$ for different cut-off scales $k^{\rm 3\times2pt}_{\rm max}$ in the cases EFT-1 (top row) and EFT-2 (bottom row). In general, all constraints improve as $k^{\rm 3\times2pt}_{\rm max}$ increases, once again highlighting the benefit of incorporating more information from smaller scales. This further underscores the importance of accurately modelling the nonlinear regime. Additionally, as expected, the SS generally shows more constraining power than the US.
For the EFT-1 model, we notice that the \GCsp+3\texttimes2\,pt. statistics combination produces better constraints on $\alpha_{\rm B,0}$ compared to 3\texttimes2\,pt. statistics alone. The same is true for $\sigma_8$ with the US prescription, while for SS, both sets of probes yield nearly identical results for $k^{\rm 3\times2pt}_{\rm max}>1$ Mpc$^{-1}$.
In EFT-2, as $k^{\rm 3\times2pt}_{\rm max}$ increases the 3\texttimes2\,pt. statistics and \GCsp+3\texttimes2\,pt. statistics constraints on $\alpha_{\rm B,0}$ closely resemble one another.
On the other hand, the constraining power on the $m$ parameter from the \GCsp+3\texttimes2\,pt. statistics combination is visibly stronger than the one from 3\texttimes2\,pt. statistics alone up to larger values of $k^{\rm 3\times2pt}_{\rm max}$. Lastly, the error on the $\sigma_8$ parameter follows a similar trend to that of $m$, with the \GCsp+3\texttimes2\,pt. statistics combination providing stronger constraints with respect to the 3\texttimes2\,pt. statistics, especially for smaller $k^{\rm 3\times2pt}_{\rm max}$ choices.

Similarly to what is observed for PMG-1 in Fig.~\ref{fig:musigma_maxed}, in EFT-2 the impact of the US prescription on constraining power only becomes apparent when we remove the influence of parameter correlations by maximising the remaining cosmological parameters. In that case, the US prescription leads to smaller errors than the SS for the 3\texttimes2\,pt. statistics only due to the inclusion of the modified gravity effects in the nonlinear evolution. However, this difference is minimal and is mainly related to the slight variation in errors between SS and US in EFT-2. Figure~\ref{fig:EFT_kmaxtrend} shows that the SS case consistently provides tighter constraints on the additional EFT-2 parameters, regardless of the scale cut applied. The difference between SS and US becomes more noticeable as we include more nonlinear scales in the analysis. Including more information helps break the degeneracy between $\sigma_8$ and the other parameters in SS, underscoring once again the importance of accurately modelling the nonlinear scales.

As discussed in \cref{sec:surveyspecs} for the PMG approach, we can find the maximum scale at which our nonlinear approach can match simulations within the accuracy of $1\%$. Since there are no simulations for the EFT case, we must rely on the ones from the PMG approach, which requires finding the corresponding $\mu_{\rm mg}$ values for the EFT models. This involves mapping the EFT parameterisation into the PMG framework as detailed in \cref{Sec:theoryEFT}.
We obtain $\mu_0=0.1$ for EFT-1, indicating $k_{\rm max}^{\rm 3\times2pt}\approx 3$ Mpc$^{-1}$, and $\mu_0=0.5$ for EFT-2, corresponding to $k_{\rm max}^{\rm 3\times2pt}\approx 1$ Mpc$^{-1}$.

\begin{table*}[h]
    \centering
    \caption{Marginalised 68\% confidence level constraints on the free parameters of the effective field theory analysis. Values are reported as percentages of the parameters' fiducial values.}
    \label{tab:EFTresults}
    \begin{tabular}{cccccccccc}
    \hline
    \rule{0pt}{2ex}
    Model & NL & Probes & $\Omega_{\rm m,0}$ & $\Omega_{\rm b,0}$ & $h$ & $n_{\rm s}$ & $\sigma_8$ & $\alpha_{\rm B,0}$ & $m$ \\
    \hline 
     \multirow{5}{*}{EFT-1} &  & \GCsp       &  $3.7\%$  &  $14.2\%$  &  \;\;$9.7\%$  &  \;\;$5.5\%$  &  $2.1\%$  &  $148.5\%$  &  \dots \\ 
     \cline{2-10}
      & \multirow{2}{*}{SS} & 3\texttimes2pt &  $1.0\%$  &  $62.9\%$     &  $69.6\%$     &  $32.9\%$     &  $4.5\%$  &  \;\;$86.9\%$  &  \dots \\
      &  & \GCsp+3\texttimes2pt              &  $1.0\%$  &  \;\;$9.7\%$  &  \;\;$7.2\%$  &  \;\;$3.1\%$  &  $1.5\%$  &  \;\;$32.3\%$  &  \dots \\
      \cline{2-10}
      & \multirow{2}{*}{US} & 3\texttimes2pt &  $1.1\%$  &  $61.6\%$     &  $67.8\%$     &  $32.0\%$     &  $5.7\%$  &  $100.6\%$     &  \dots \\
      & & \GCsp+3\texttimes2pt               &  $1.0\%$  &  \;\;$9.7\%$  &  \;\;$7.2\%$  &  \;\;$3.1\%$  &  $1.5\%$  &  \;\;$31.1\%$  &  \dots \\
    \hline 
    \rule{0pt}{2ex}
     \multirow{5}{*}{EFT-2} &  & \GCsp       &  $3.8\%$  &  \;\;$14.6\%$  &  \;\;$9.9\%$  &  \;\;$5.6\%$  &  $5.4\%$  &  $301.3\%$     &  $146.0\%$ \\ 
     \cline{2-10}
      & \multirow{2}{*}{SS} & 3\texttimes2pt &  $2.7\%$  &  $53.7\%$     &  $60.5\%$     &  $29.6\%$     &  $3.7\%$  &  \;\;$16.0\%$  &  \;\;$25.0\%$ \\
      &  & \GCsp+3\texttimes2pt              &  $2.1\%$  &  \;\;$11.0\%$  &  \;\;$7.9\%$  &  \;\;$3.9\%$  &  $1.3\%$  &  \;\;$11.6\%$  &  \;\;$11.8\%$ \\
      \cline{2-10}
     & \multirow{2}{*}{US} & 3\texttimes2pt  &  $2.7\%$  &  $52.5\%$     &  $58.9\%$     &  $28.7\%$     &  $4.0\%$  &  \;\;$16.5\%$  & \;\;$25.7\%$ \\
      & & \GCsp+3\texttimes2pt               &  $2.1\%$  &  \;\;$11.0\%$  &  \;\;$7.9\%$  &  \;\;$3.9\%$  &  $1.3\%$  &  \;\;$11.6\%$  & \;\;$11.8\%$ \\
    \hline
    \end{tabular}
\end{table*}

%%%%%%%%%%%%%%%%%%%%%%
\section{Conclusions} \label{sec:conclusions}

In this paper, we have investigated how \Euclid spectroscopic and photometric probes can constrain two model-independent approaches to modifications of gravity, namely the phenomenological modified gravity (PMG) and the effective field theory of dark energy and modified gravity (EFT). For the former, we have selected a late-time parameterisation with a \lcdm\ fiducial (PMG-1) and a far-from-\lcdm\ one (PMG-2). In the case of EFT we have chosen two parameterisations for the braiding parameter, namely a linear one in the scale factor (EFT-1) and one inspired by shift-symmetric scalar tensor models (EFT-2). The forecasts are obtained through the Fisher matrix method applied individually to the photometric (WL, \GCph, and their cross-correlation), spectroscopic (\GCsp) surveys, and finally to their combination. By combining both surveys, we have shown a general and overall considerable improvement in the constraining power. This confirms the two surveys' complementary character and their combination's value in reducing the effect of nuisance parameters and successfully breaking degeneracies between cosmological parameters. 
To model the nonlinearities in the matter power spectrum for WL, we applied the halo-model reaction approach to cover two limiting cases for screening mechanisms: unscreened (US) and super-screened (SS). This methodology provided a way to assess the impact of screening mechanisms on predicted errors. These two cases were necessary given the lack of an exact prescription for modelling the nonlinear scales in the matter power spectrum in model-independent approaches.

We find that with \GCsp+3\texttimes2\,pt. statistics, \Euclid can achieve relative errors for $\bar{\mu}_0$ and $\bar{\Sigma}_0$ of about $23\%$ and $3\%$ for the PMG-1 case (this is for both US and SS) and about $34\%$ (36\%) and $3\%$ for PMG-2 in the SS (US) case. This will provide improvements by one order of magnitude with respect to currently available constraints, such as those from the \Planck mission \citep{Planck:2018vyg} or the DES survey \citep{DES:2022ccp}, while our constraints are comparable to forecasts from \cite{Casas:2022vik} using a different combination of Stage IV experiments. For the EFT-1 case with \GCsp+3\texttimes2\,pt. statistics, the relative error of $\alpha_{\rm B,0}$ is $32\%$ for the SS (31\% for US) case. 
Previous constraints on the EFT-1 parameterisation give $\sigma(\alpha_{\rm B,0}) \approx 0.05$ when fixing $\alpha_{\rm M,0} = 0$ \citep{Noller:2018wyv}, almost identical to the errors found here (recall that our fiducial value is $\alpha_{\rm B,0} = 0.2$). This similarity is likely a consequence of the specific parameterisation ($\alpha_i \propto a$) we employ in EFT-1. We expect the full probe combination considered here to yield additional improvements when considering other parametrisations, where the inclusion of galaxy cross- and auto-correlations is known to have a more substantial impact on constraints \citep{Seraille:2024beb}.
In fact, for EFT-2 with \GCsp+3\texttimes2\,pt. statistics and still for SS, we find $12\%$ relative errors for both $\alpha_{\rm B,0}$ and $m$, with similar values found for the US case. In terms of absolute uncertainties, this results in $\sigma(\alpha_{\rm B,0}) = 0.11$ and $\sigma(m) = 0.29$. Previous constraints for the EFT-2 parameterisation give $\sigma(\alpha_{\rm B,0}) = 0.3$ and $\sigma(m) = 0.4$ errors for $\alpha_{\rm B,0}$ and $m$, respectively \citep{Traykova:2021hbr}. 
\Euclid alone will strongly improve the constraint on $\alpha_{\rm B,0}$ with any combination of probes, while for $m$ the joint \GCsp+3\texttimes2\,pt. statistics will be crucial to see an improvement in this parameter.

Overall, our results highlight one of the key characteristics of \Euclid: the combination of 3\texttimes2\,pt. statistics and \GCsp\ can be used to break the degeneracies between the cosmological and MG parameters, thus significantly improving the precision with which the latter are measured. Therefore, one could think of combining other observations, both coming from \Euclid\ itself, such as galaxy clusters, and from external data, such as CMB measurements. In particular, including CMB observables in the analysis, properly accounting for their cross-correlations with 3\texttimes2\,pt. statistics, would allow us to further break down degeneracies with cosmological parameters not related only to the late-time evolution of the Universe.

Furthermore, since our results are sensitive to the scale cut adopted in the 3\texttimes2\,pt. statistics, we have shown how the constraints on these parameters change with different $k^{\rm 3\times2pt}_{\rm max}$ adopted. As a general result, we found that the higher the cut-off scale, the better the constraints. This illustrates the information gain enclosed in the nonlinear scales. We also found that, in general, US and SS have broadly similar constraints in most cases, with US having the advantage that the MG effects also enter the computation of the nonlinear evolution (resulting in tighter constraints in some cases), while SS has the advantage that in some cases it better breaks the degeneracies between the MG parameters and $\sigma_8$.

It is worth mentioning that while the approaches we adopted in this work are very powerful in describing the phenomenology of MG effects on cosmological observables, they come with an intrinsic caveat: the functions describing the MG effects in both approaches are unknown functions of time, and in the case of PMG of scale as well. Therefore, one must fix their functional forms, balancing between the need for general parameterisations and the risk of oversimplification.
Nevertheless, smooth parameterisations are, in general, enough because cosmological observables do not seem to be extremely sensitive to short time-scale variations. In light of this, one has to be very careful in extrapolating general information from constraints obtained using phenomenological approaches. A different method would be to treat the free functions in these approaches as unknown functions of $z$ (for the EFT case), and both $z$ and $k$ (for PMG), and to bundle them into a sufficiently large number of bins in their corresponding domain. However, this method would sacrifice some of the advantages discussed in the introduction.

Additionally, further developments will also be needed to overcome the two limiting cases of SS and US used in this study to properly include corrections to the matter power spectrum from nonlinear scales. This is particularly needed for the phenomenological approaches investigated here. On the one hand, it will allow us to obtain more accurate predictions for the matter power spectrum and, on the other hand, to exploit the full power of constraints from future data, which are expected to go deep into the nonlinear regime, without the limitations of current prescriptions. 
Moreover, we are missing an accurate treatment of the baryonic effects which are important for the interpretation of the smaller scales. This is a more general issue that arises in MG theories and is not limited to phenomenological frameworks. Although in this work the baryonic effects are negligible because of the conservative scale cut adopted, in studies where this cut is not employed a careful inclusion of them, or estimation of their effects, will be necessary.

In conclusion, the results we have presented provide an estimate of the precision with which \Euclid alone can test deviations from $\Lambda$CDM, refining our understanding of gravity on large scales and possibly offering a decisive test to rule out or support certain classes of alternative theories.

%%%%%%%%%%%%%%%%%%%%%%

%\FloatBarrier

%%%%%%%%%%%%%%%%%%%%%%
 \begin{acknowledgements}
% %%%%%%%%%%%%%%%%%%%%%%
% We thank \ldots. 
 \AckEC
 I.S.A. is supported by FCT through the PhD fellowship grant with ref. number 2020.07237.BD (\url{https://doi.org/10.54499/2020.07237.BD}). I.S.A. and L.A. acknowledge support from FCT research grants UIDB/04434/2020 and UIDP/04434/2020. I.S.A., L.A., N.F. and F.P. acknowledge the FCT project with ref. number PTDC/FIS-AST/0054/2021.
L.A. is supported by FCT through the PhD fellowship grant with ref. number 2022.11152.BD.
 B.B. is supported by a UKRI Stephen Hawking Fellowship (EP/W005654/2).
 M.C. acknowledges the financial support provided by the Alexander von Humboldt Foundation through the Humboldt Research Fellowship program, as well as support from the Max Planck Society and the Alexander von Humboldt Foundation in the framework of the Max Planck-Humboldt Research Award endowed by the Federal Ministry of Education and Research.
N.F. is supported by MUR through the Rita Levi Montalcini project with reference PGR19ILFGP.
 K.K. is supported by STFC grant ST/W001225/1.
M.M. acknowledges funding by ASI under agreement no. 2018-23-HH.0 and support from INFN/Euclid Sezione di Roma.
 J.N is supported by an STFC Ernest Rutherford Fellowship (ST/S004572/1).
 S.S. acknowledges the financial support provided by the Alexander von Humboldt Foundation through the Humboldt Research Fellowship program.
E.M.T is supported by ERC under the European Union's HORIZON-ERC-2022 (grant agreement no. 101076865).
D.B.T. acknowledges support from the STFC, grant numbers ST/P000592/1 and ST/X006344/1).
F.P. acknowledges partial support from the INFN grant InDark and from from the Italian Ministry of University and Research (MUR), PRIN 2022 `EXSKALIBUR – Euclid-Cross-SKA: Likelihood Inference Building for Universe's Research', Grant No.\ 20222BBYB9, CUP C53D2300131 0006, and from the European Union -- Next Generation EU.

\end{acknowledgements}

%%%%%%%%%%%%%%%%%%%%%%
%\appendix
%%%%%%%%%%%%%%%%%%%%%%

\bibliographystyle{aa}
\bibliography{Euclid, biblio}
\label{LastPage}

\end{document}